\documentclass[%
    amsmath,amssymb, 
    aps,superscriptaddress, 
    pra, %
    nofootinbib,%
    twocolumn,
    floatfix
    ]{revtex4-1} 

\usepackage{amsmath}
\usepackage{amssymb}
\usepackage{amsthm}
\usepackage{mathrsfs}
\usepackage{graphicx}
\usepackage{color}
\usepackage{bbm}
\usepackage{nicefrac}
\usepackage[shortlabels]{enumitem}

\usepackage[nosectionfonts,nonotetitle,nocosmeticdefs,nopagegeomdefs]{docnote}

\usepackage{ourdefs}

\makeatletter
\newcommand\bibalias[2]{%
  \@namedef{bibali@#1}{#2}%
}

\newtoks\biba@toks
\let\bibalias@oldcite\cite
\def\cite{%
  \@ifnextchar[{%
    \biba@cite@optarg%
  }{%
    \biba@cite{}%
  }%
}
\newcommand\biba@cite@optarg[2][]{%
  \biba@cite{[#1]}{#2}%
}
\newcommand\biba@cite[2]{%
  \biba@toks{\bibalias@oldcite#1}%
  \def\biba@comma{}%
  \def\biba@all{}%
  \@for\biba@one:=#2\do{%
    \@ifundefined{bibali@\biba@one}{%
      \edef\biba@all{\biba@all\biba@comma\biba@one}%
    }{%
      \PackageInfo{bibalias}{%
        Replacing citation `\biba@one' with `\@nameuse{bibali@\biba@one}'
      }%
      \edef\biba@all{\biba@all\biba@comma\@nameuse{bibali@\biba@one}}%
    }%
    \def\biba@comma{,}%
  }%
  \immediate\write\@auxout{\noexpand\bgroup\noexpand\renewcommand\noexpand\citation[1]{}\noexpand\citation{#2}\noexpand\egroup}%
  \edef\biba@tmp{\the\biba@toks{\biba@all}}%
  \biba@tmp%
}
\makeatother

\bibalias{Alicki79}{Alicki1979_davies}
\bibalias{BBPSSW1996}{Bennett1996PRL_purification}
\bibalias{BDSW1996}{Bennett1996PRA_MSEntglQECorr}
\bibalias{VidalC-irre}{Vidal2001_Irrev}
\bibalias{baugh2005experimental}{Baugh2005_algocool}
\bibalias{dahlsten2011inadequacy}{Dahlsten2011NJP_inadequacy}
\bibalias{del2011thermodynamic}{delRio2011Nature}
\bibalias{horodecki_reversible_2003}{Horodecki2003PRA_NoisyOps}
\bibalias{howard1997molecular}{Howard1997_molecular}
\bibalias{linden2010small}{Linden2010PRL_SmallFridge}
\bibalias{popescu2006entanglement}{Popescu2006NPhys_entanglement}
\bibalias{scovil1959maser}{Scovil1959masers}

\newcommand\suppnoteIntro{\ref{sec:appx-intro}}
\newcommand\suppnoteClarif{\ref{sec:appx-some-initial-clarifications}}
\newcommand\suppnoteSmoothEnt{\ref{sec:appx-SmoothEntropies}}
\newcommand\suppnoteApplications{\ref{sec:appx-Applications}}
\newcommand\suppnoteAltProof{\ref{sec:appx-alt-proof-lambda-maj}}
\newcommand\suppnoteFormal{\ref{sec:appx-appendixFormalLambdaMajorization}}
\newcommand\supppropDilation{\ref{prop:appx-DilationOfSubUnitalCPMsToUnitalCPMs-samesystem}}

\renewcommand{\thethm}{\Roman{thm}}

\begin{document}

\title{The Minimal Work Cost of Information Processing}
\author{Philippe Faist}
\affiliation{Institute for Theoretical Physics, ETH Zurich, Wolfgang-Pauli-Str. 27, 8093 Switzerland}
\author{Fr\'ed\'eric Dupuis}
\affiliation{Institute for Theoretical Physics, ETH Zurich, Wolfgang-Pauli-Str. 27, 8093 Switzerland}
\affiliation{Department of Computer Science, Aarhus University, IT-Parken, Aabogade 34, 8200 Denmark}
\affiliation{Faculty of Informatics, Masaryk University, Botanick\'a 68a, 612 00 Brno, Czech Republic}
\author{Jonathan Oppenheim}
\affiliation{Department for Physics and Astronomy, University College of London, Gower Street, WC1E~6BT U.K.}
\author{Renato Renner}
\affiliation{Institute for Theoretical Physics, ETH Zurich, Wolfgang-Pauli-Str. 27, 8093 Switzerland}

\date{July 7, 2015}

\begin{abstract}
  Irreversible information processing cannot be carried out without some inevitable
  thermodynamical work cost. This fundamental restriction, known as Landauer's principle,
  is increasingly relevant today, as the energy dissipation of computing devices impedes
  the development of their performance. Here we determine the minimal work required to
  carry out any logical process, for instance a computation. It is given by the entropy of
  the discarded information conditional to the output of the computation. Our formula
  takes precisely into account the statistically fluctuating work requirement of the
  logical process.  It enables the explicit calculation of practical scenarios, such as
  computational circuits or quantum measurements. On the conceptual level, our result
  gives a precise and operationally justified connection between thermodynamic and
  information entropy, and explains the emergence of the entropy state function in
  macroscopic thermodynamics.
\end{abstract}

\maketitle


Thermodynamics in essence is an information theory---its purpose is to make
statements about systems for which we only have certain partial information, such as a gas
of many particles for which only macroscopic quantities like temperature,
volume, and pressure are accessible. Following this point of view, Jaynes showed that
the entropy function derived in statistical mechanics corresponds to the information-theoretic
entropy of the gas associated with a macroscopic observer who is maximally ignorant of the
microscopic degrees of freedom~\cite{Jaynes1957PR_InfThStatMech}, resorting to Shannon's
mathematical theory of information~\cite{Shannon1948BSTJ} developed in the context of
telecommunications.

\begin{figure}
  \centering
  \includegraphics[width=87mm]{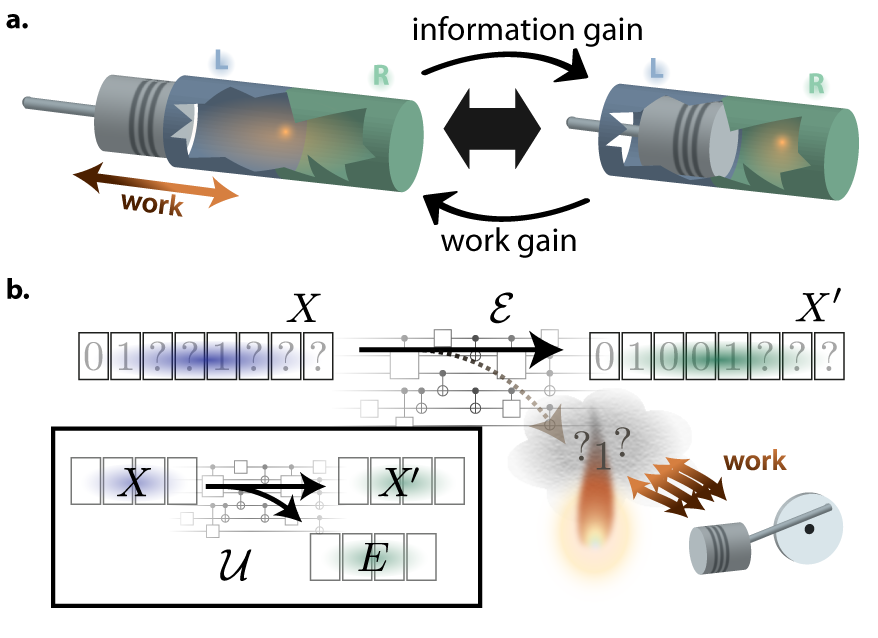}
  \caption{Work and information are related by physical processes.
    \textbf{a.}~A gas formed by a single particle can either be on the left (``L'') or the
    right (``R'') side of the cylinder (known as a Szilard
    box~\protect\cite{Szilard1929ZeitschriftFuerPhysik}). This one bit of information can be
    reversibly traded for $kT\ln2$ work by isothermally compressing the gas with a piston
    or letting the gas isothermally expand. This illustrates that discarding 1 bit of
    entropy (or uncertainty) requires $kT\ln2$ work.
    \textbf{b.}~An implementation of the logical process $\mathcal{E}$ mapping a system
    $X$ to an output $X'$ interacts with the thermal bath, may discard information and in
    general costs work. The logical process $\mathcal{E}$ may be written as part of a
    global unitary $\mathcal{U}$ on an additional hypothetical system $E$, which
    represents the discarded information (inset). Our main result states that the minimum
    work required in a physical implementation of $\mathcal{E}$ is the amount of discarded
    information, which the implementation has to dump into the environment.
  }
  \label{fig:ProcessEntropyEnvironment}
\end{figure}

When the observers have access to knowledge about microscopic quantities, such as
positions and velocities of particles in a gas, the second law of thermodynamics seems to
break down, as was illustrated by Maxwell's demon. To address this problem,
Szilard~\cite{Szilard1929ZeitschriftFuerPhysik} studied a one-particle gas that can be
located on either side of a box, left (``L'') or right (``R''), and noted that by
isothermally compressing the gas or letting the gas expand, one can trade this one bit of
information for $kT\ln2$ work, as depicted in
Fig.~\ref{fig:ProcessEntropyEnvironment}\textbf{a} (in the presence of a heat bath at
temperature $T$, and where $k$ is Boltzmann's constant).  Landauer and Bennett later
realized that the information content of data stored in a memory register, independently
of the nature of its physical representation, counts as thermodynamic entropy when
considering thermodynamical operations on that register~\cite{%
  Landauer1961_5392446Erasure,Bennett1982IJTP_ThermodynOfComp,Bennett2003_NotesLP,%
  Bennett1973IBMJRD_LogRevComp,Shizume1995_HeatGeneration,%
  Maruyama2009RMP_ColloquiumPhysMaxDemAndInf,%
  Piechocinska2000PRA,BookFeynmanLecturesOnComputation1996,%
  BookLeffRex2010,%
  Sagawa2013_InfTh}.
For example, given a bit in an unknown state, any operation that resets
it to zero must dissipate at least $kT\ln2$ heat, and thus the corresponding amount of
work must be supplied (this is known as Landauer's principle).  This fact salvages the
second law of thermodynamics and resolves the paradox of Maxwell's demon.

More recently with the advent of quantum information, efforts were made to understand the
laws of quantum thermodynamics from an information-theoretic viewpoint~\cite{%
  Lloyd2000_ultimate,%
  Plenio2001_forgetting,%
  Popescu2006NPhys_entanglement,%
  Gemmer2009_quantum,%
  Oppenheim2002PRL_thermodynamical%
}, while the increasing technological ability to control and manipulate nanoscale
systems~\cite{Hanggi2009_brownian,Baugh2005_algocool} has prompted the study of particular
operational models and frameworks, leading to characterization of the work cost of various
information-theoretic tasks such as erasure and work extraction~\cite{%
  Alicki2004_hamiltonian,%
  Janzing2006Habil,%
  Linden2010PRL_SmallFridge,%
  Dahlsten2011NJP_inadequacy,%
  delRio2011Nature,%
  Egloff2012arXiv,%
  Brandao2013_resource,%
  Aberg2013_worklike,%
  Horodecki2013_ThermoMaj,%
  Skrzypczyk2013arXiv_extracting,%
  Reeb2014NJP_improved,%
  Brandao2015PNAS_secondlaws%
}.
For a more specific review of existing results, we refer to (Appendix~\suppnoteIntro).

The aim of this work is to study thermodynamics in such generalized scenarios, where one
may have knowledge about microscopic degrees of freedom, by resorting to modern tools of
information theory~\cite{%
  PhdRenner2005_SQKD,%
  PhDTomamichel2012}. We provide a fundamental lower bound to the work cost of
a physical implementation of a logical process, discuss several examples, and illustrate
how traditional thermodynamics emerges from our micrsocopic result in the limit of
macroscopic systems.


%

\paragraph{The Framework}
We determine a general expression for the minimal amount of work needed to carry out any
given logical process $\mathcal{E}$. This can be for example an \AND{} gate or any quantum
or classical computation; most generally $\mathcal{E}$ is defined as any completely
positive, trace-preserving map from quantum states on an input Hilbert space $\Hs_X$ to
quantum states on an output Hilbert space $\Hs_{X'}$. We assume these spaces to be of
finite dimension for simplicity; note that such a space can be a subspace of an
infinite-dimensional Hilbert space in which the relevant computation or logical process
takes place. The terminology \emph{logical process} is meant to emphasize that the
mathematical object $\mathcal{E}$ only specifies for each input state the corresponding
output state, and does not prescribe its physical realization, which would consist
of a full description of a physical system including the parts of its environment that are
relevant to determine its time evolution. Note that in performing a logical process one
does not merely transform one quantum state into another, rather, the output must be
related to the input in a precisely specified way. In the case where the input is a
classical value, this means that the output depends on the particular input value
received, and not only on the distribution of inputs. This might be checked in practice,
for example, if one keeps a copy of the input as a reference system and observes the
correlations between the output and the reference system.

There are, generally, many ways of actually realizing a logical process with an actual
physical device. The device and its interactions with the environment (e.g.\@ a heat bath)
may for example be described by a Hamiltonian or a Liouvillian.
For our purposes, it is sufficient to specify the set of operations which the device is
allowed to perform as well as the associated work cost. We then optimize the work
expenditure over precisely those strategies which realize the given logical process
$\mathcal{E}$. Observe that the more permissive our framework is, the more robust our
bound will be.
In our model, we shall be allowed to implement at no work cost any trace-preserving
completely positive map which is unital, i.e. which preserves the identity operator.
Note that if we were to allow any logical process that is not unital to be performed for
free, one could flagrantly violate the second law of thermodynamics on a macroscopic
scale: in this sense, unital maps are the most permissive logical operation that we can
allow for free.
The model must also include a description of a ``battery'' that provides the energy
required to to drive the process. For this we resort to Bennett's idea of an
``information fuel tape''~\cite{%
  Bennett1982IJTP_ThermodynOfComp,%
  BookFeynmanLecturesOnComputation1996%
}: such a battery consists of a large number of qubits with a degenerate
Hamiltonian. Initially, a 
certain number $\lambda_1$ of these qubits are in the maximally mixed state, and the rest are
pure. We may freely implement any joint unital map on the system and battery. At the end
of the operation, the state of the battery consists of a possibly different number
$\lambda_2$ of qubits in a maximally mixed state, while the rest should be pure. (The
requirement that 
these $\lambda_2$ qubits be maximally mixed is not a restriction, see Methods
section.) 
We then count the amount of work consumed as
$W = kT\ln2\cdot\left(\lambda_2 - \lambda_1\right)$, which is the amount of work required
to restore the battery system into its initial state.
Indeed, a vast amount of literature has well 
underscored the correspondence between possessing a pure degenerate qubit, or storing
$kT\ln2$ work, and vice versa~\cite{Szilard1929ZeitschriftFuerPhysik,%
  Bennett1982IJTP_ThermodynOfComp,%
  BookFeynmanLecturesOnComputation1996,%
  BookLeffRex2010}.
The quantity $W$ may be negative, indicating that work can be extracted 
from the battery when restoring it to its initial state.
In addition, we assume that
the input to the logical process $\mathcal E$ is encoded in a system whose initial
Hamiltonian is degenerate. The same is assumed about the
output system at the end of the computation. Note that this does not exclude making use of
systems with nontrivial Hamiltonians during the implementation of the process.
Also, this requirement is in
practice not a limitation, as many other frameworks may be mapped to this
setting~\cite{Aberg2013_worklike,Horodecki2013_ThermoMaj,Egloff2012arXiv}; indeed the
assumption should rather be regarded as a technicality to ensure a clean way of accounting
for work.

To obtain physically relevant results, we also have to exclude overwhelmingly unlikely
events from our considerations. This is actually quite common in thermodynamics and is
usually done implicitly. For example, consider a stone lying on the ground. There is a
very small chance that by thermal fluctuation the stone spontaneously jumps in the air.
However this event is so disproportionately unlikely that in a physical theory we may
safely choose to ignore this possibility. Within our framework, we do this more
explicitly. That is, we consider a parameter that specifies the total probability of all
events we want to exclude. In the quantum regime, where events are generally not
well-defined, this idea is captured by $\epsilon$-approximations: the stone has a very
small amplitude of being found in the air, but its state is $\epsilon$-close to a state
completely located on the ground. Analogously, we study the work requirement of logical
processes that are $\epsilon$-approximations of the desired logical process. This is a
standard procedure in information theory~\cite{PhdRenner2005_SQKD,PhDTomamichel2012}, and
is justified by the fact that an $\epsilon$-approximation cannot be distinguished from the
original logical process with probability greater than $\epsilon$.

\paragraph{The Main Result}
To formulate our main claim, we represent the logical process $\mathcal E$ by its
\emph{Stinespring dilation}~\cite{BookNielsenChuang2000}. This is an isometry $\mathcal{U}$
(which can be seen as part of a unitary) that maps $X$ onto $X'$ as well as an extra
system $E$ such that the original map $\mathcal{E}$ is retrieved by ignoring $E$ (see
Fig.~\ref{fig:ProcessEntropyEnvironment}\textbf{b}).  Our main result asserts that
$W^\epsilon$, the work one needs to supply to execute the operation up to an
$\epsilon$-approximation, is lower bounded by
\begin{equation}
  \label{eq:MainResult}
  W^\epsilon \geqslant kT\ln(2)\,\Hmax^{\bar\epsilon}\left(E|X'\right)\ .
\end{equation}
The right hand side is the smooth max-entropy of $E$ conditioned on $X'$ and may be
interpreted as a measure for the irreversibility of the logical process. More precisely,
the smooth max-entropy is an information-theoretic measure defined in the Methods section,
and quantifies the uncertainty one has about $E$ when given access to $X'$. The parameters
$\epsilon$ and $\bar\epsilon$ are related by $\bar\epsilon=\sqrt{2\epsilon}$; $\epsilon$
may be chosen arbitrarily.
We stress that the system $E$ is an abstract mathematical concept used to represent
the logical map $\mathcal E$, and can be interpreted as the information discarded by the
mapping. In particular, our bound is independent of the choice of this representation.

The form of the bound~\eqref{eq:MainResult} naturally expresses our intuition that the
amount of work that needs to be provided corresponds to the amount of information that is
logically discarded, and which therefore has to be dumped into the environment. This
consideration is done from the viewpoint of the observer who has completed the
computation, and thus has access to $X'$, explaining the occurrence of the conditional
entropy. Also, if $E$ is classical, the max-entropy 
has the operational interpretation of being the amount of memory space needed to
compress the information contained in $E$ when possessing knowledge of
$X'$~\cite{Koenig2009IEEE_OpMeaning}. (In the fully quantum case, it corresponds to quantum
state merging~\cite{Horodecki2005Nat_partial,Koenig2009IEEE_OpMeaning}.)

The proof of our main result proceeds by first considering the special case in which
$\epsilon=0$. The bound one then obtains is
\begin{align}
  \label{eq:MainResultEpsEqZero}
  W^{\epsilon=0} \geqslant
  kT\ln2\cdot\log_2\znorm{\mathcal{E}\left(\Pi_X\right)}_\infty \ ,
\end{align}
where $\Pi_X$ is the projector onto the support of the input state.  This
expression proves particularly useful for calculating some simple practical examples.

The proof of this special case, and its generalization to the regime where $\epsilon>0$,
is presented in the Methods section. An alternative proof,
using techniques from majorization, is given in (Appendix~\suppnoteAltProof).

\begin{figure}
  \centering
  \includegraphics[width=87mm]{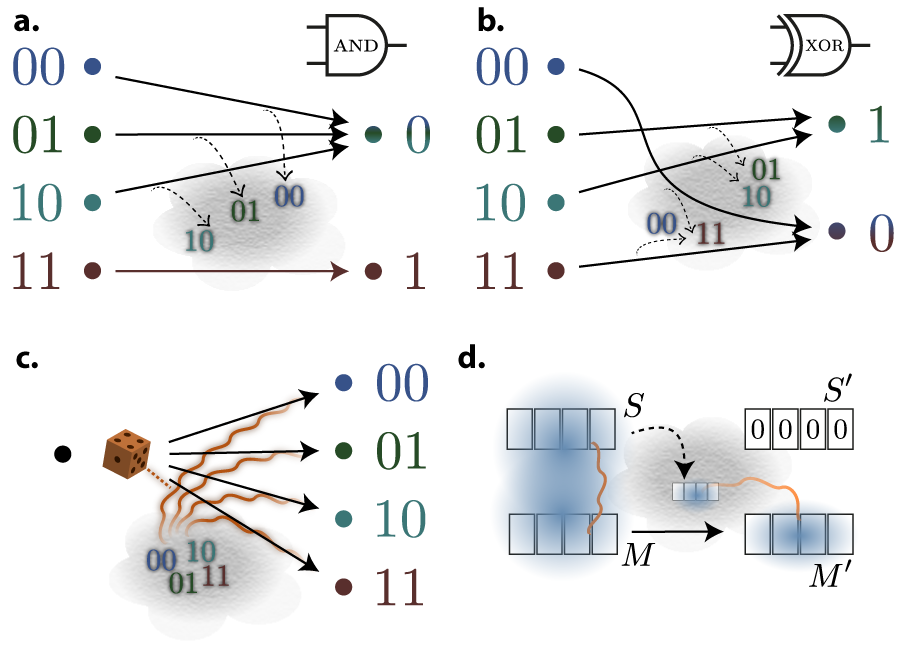}
  \caption{Examples of logical processes.
    \textbf{a.}~The \AND{} gate is one of the building blocks of computers. Our result
    implies that any successful implementation of this logically irreversible gate
    requires at least work $\log_2 3\cdot kT\ln2\approx 1.6\,kT\ln2$ due to the entropy of
    the discarded information (dotted arrows).
    \textbf{b.}~The \XOR{} gate only requires $kT\ln2$ work, as it discards less entropy
    per output event than the \AND{} gate.
    \textbf{c.}~Work can be extracted if randomness is being produced: the discarded
    information is entangled with the output (orange wavy lines), and the conditional
    entropy on the right hand side of~\eqref{eq:MainResult} is negative.
    \textbf{d.}~The erasure of a quantum system $S$ with access to a quantum memory $M$
    must transfer the content of $S$ into the system $E$ containing the discarded
    information, while preparing $S'$ in a pure state and mapping $M$ to $M'$
    identically. The corresponding minimal work cost is
    $kT\ln(2)\cdot\Hmax^\epsilon\left(S|M\right)$; this can be achieved using the
    procedure of del Rio~\emph{et al.}~\protect\cite{delRio2011Nature}. If the system is
    entangled with the memory, this quantity is negative and work may be extracted.
  }
  \label{fig:ProcessesExamples}
\end{figure}

\paragraph{Classical mappings and dependence on the logical process}
Our result, which is applicable to arbitrary quantum processes, applies to all classical
computations as a special case. Classically, logical processes correspond to stochastic
maps, of which deterministic functions are a special case.  As a simple example, consider
the \AND{} gate. This is one of the elementary operations computing devices can perform,
from which more complex circuits can be designed. The gate takes two bits as input, and
outputs a single bit that is set to \B{1} exactly when both input bits are \B{1}, as
illustrated in Fig.~\ref{fig:ProcessesExamples}\textbf{a.}

The logical process is manifestly irreversible, as the output alone doesn't allow to infer
the input uniquely. If one of the inputs is zero, then the logical process effectively has
to reset a three-level system to zero, forgetting which of the three possible inputs
\B{00}, \B{01}, or \B{10} was given; this information can be viewed as being
discarded, and hence dumped into the environment. We can confirm this
intuition with our main result, using the fact that a general classical mapping is given
by the specification of the conditional probability $p\left(x'|x\right)$ of observing $x'$
at the output if the input was $x$. Embedding the classical probability distributions into
the diagonals of quantum states, the infinity norm in
expression~\eqref{eq:MainResultEpsEqZero} becomes simply 
\begin{align}
  \label{eq:MainResultClassical}
  W^{\epsilon=0} \geqslant kT\ln2\cdot \log_2 \max_{x'} \sum_x p\left(x'|x\right)\ ,
\end{align}
where the sum ranges only over those $x$ that have a non-zero probability of occurring. In
the case of deterministic mappings $p\left(x'|x\right)\in\{0,1\}$, this corresponds to the
maximum number of input states that map to a same output state.
For the \AND{} gate,
provided all four states \B{00}, \B{01}, \B{10} and \B{11} have non-negligible probability
of occurring, there are three input states mapping to the same output state,
so~\eqref{eq:MainResultClassical} gives us simply $W^{\epsilon=0}_{\AND{}} \geqslant
\log_2 3\cdot kT\ln2\approx 1.6\,kT\ln2$.
Also, in simple examples as considered here, the expression~\eqref{eq:MainResultClassical}
is stable to considering an $\epsilon$-approximation (Appendix~\suppnoteApplications);
this quantity is thus physically justified.

Crucially, our result reveals that the minimal work requirement depends in general on the
specific logical process, and not only on the input and output states. This contrasts with
traditional thermodynamics for large systems, where the minimal work requirement of a
state transformation can always be written as a difference of a thermodynamical potential,
such as the free energy. For example, the minimal work cost of performing specifically an
\AND{} gate may differ from that of another logical process mapping an input distribution
$\left(p_{00}, p_{01}, p_{10}, p_{11}\right)$ (with $\sum_{i} p_{i}=1$) to the
distribution $\left(p'_0, p'_1\right) = \left(p_{00}+p_{01}+p_{10},\,p_{11}\right)$.
(Recall that the classical counterpart of a quantum state is a probability distribution.)
To see this, consider the \XOR{} gate, which outputs a \B{1} exactly when both inputs are
different (see Fig.~\ref{fig:ProcessesExamples}\textbf{b}). The minimal work cost
requirement of this gate, as given by~\eqref{eq:MainResultClassical}, is now only
$kT\ln2$, as in the worst case, only a single bit of information is erased (again
supposing that all four input states have non-negligible probability of occurring). Now,
suppose that for some reason, the input distribution is such that $p_{01}+p_{10}=p_{11}$,
i.e. the input \B{11} occurs with the same probability as of either \B{01} or \B{10}
appearing.
Then, the \XOR{} gate reproduces the exact same output distribution
as the \AND{} gate: in both cases, we have $p_0' = p_{00}+p_{10}+p_{01} = p_{00}+p_{11}$
and $p_1' = p_{11} = p_{01}+p_{10}$. In other words, both logical processes have the same
input and output state, yet the \XOR{} gate only requires work $kT\ln2$ compared to the
\AND{} gate, which requires $1.6\,kT\ln2$. Furthermore, we point out that this difference,
which appears small in this case, may be arbitrarily large in certain scenarios
(Appendix~\suppnoteApplications).

On the one hand, we are by definition interested in the work cost of a given logical
process, so one might have expected that this work cost should not only depend on the
input and output states. On the other hand, it might seem contradictory that the full
logical process matters even though we have fixed an input state $\sigma_X$. However, this
makes sense if we consider preparing the input state as part of a pure state on the input
system and a reference system. In this case, the logical process which is implemented
influences the (in principle detectable) correlations between the output and the reference
system, even if the reduced state on the input is the fixed state $\sigma_X$.

We emphasize that the phenomenon observed here is fundamentally different
from the notion of thermodynamic irreversibility. Here, we always consider the
optimal procedure for implementing the logical process, whereas a thermodynamically
irreversible process is in fact an ``inefficient'' physical process which could be
replaced by a more efficient, reversible one. In our framework, the
thermodynamically irreversibile processes are those physical implementations which do not
achieve the bound~\eqref{eq:MainResult}. A longer discussion with examples is provided in
(Appendix~\suppnoteClarif).

\paragraph{Work extraction}
While erasure requires work, it is well known that in a wide range of frameworks one can
in general extract work with the reverse logical process, which corresponds to
taking a register of bits which are all in the zero state and making them maximally
mixed~\cite{Szilard1929ZeitschriftFuerPhysik,Bennett1982IJTP_ThermodynOfComp}. Our
result intrinsically reproduces this fact: the Stinespring dilation
$\mathcal{U}_{X\to{}X'E}$ of a logical process which generates randomness in fact creates
entanglement between the output $X'$ and $E$ (see
Fig.~\ref{fig:ProcessesExamples}\textbf{c}). The conditional entropy
$\Hmax^\epsilon\left(E|X'\right)$ then becomes negative, such that the
bound~\eqref{eq:MainResult} allows work to be extracted.  We remark that,
even if the logical process $\mathcal{E}_{X\to X'}$ is classical, the relevant state for
the entropic term in~\eqref{eq:MainResult} is entangled, and thus all but classical; this
is due to the construction of $E$ as a purifying system for the logical process.

\paragraph{Erasure with a quantum memory and tightness of our bound}
Recently, del Rio \emph{et. al.}~\cite{delRio2011Nature} have constructed an explicit procedure
capable of resetting a quantum system $S$ to a pure state using an erasure mechanism
assisted by a quantum memory $M$, and doing so at a work cost of approximately
\begin{align}
  \label{eq:LidiasErasureCost}
  W_\text{erasure}^\epsilon \approx kT\ln2\cdot\Hmax^\epsilon\left(S|M\right)\ .
\end{align}
The approximation holds up to terms of the order of the logarithm of $\epsilon$ and
are negligible in typical scenarios (Appendix~\suppnoteApplications).

Our main result implies that their procedure is nearly optimal
(Fig~\ref{fig:ProcessesExamples}\textbf{d}). Indeed, consider the total system $X=S\otimes
M$, in the initial state $\sigma_{SM}$, with the logical process
$\mathcal{E}\left(\cdot\right) = \tr_S\left(\cdot\right)\otimes\proj0_{S'}$, denoting
symbolically with a prime the output system $S'$. (The state on $M$ remains unchanged.)
One then straightforwardly sees that the resulting joint state on $E$ and the output
$X'=S'\otimes M$ is obtained from the initial state on $S$ and $M$ by isometrically
``transferring'' the $S$ part to $E$ and replacing it by a fixed pure state. The entropy
term in our bound~\eqref{eq:MainResult} then becomes
$\Hmax^\epsilon\left(E|S'M\right)=\Hmax^\epsilon\left(E|M\right)
=\Hmax^\epsilon\left(S|M\right)$, the latter entropy being evaluated on the input state.
This matches the term in~\eqref{eq:LidiasErasureCost}.

Conversely, this optimal erasure procedure can be used to show that for any arbitrary
logical process, the minimal amount of work our result associates to it can be in
principle achieved to good approximation. Given a logical process $\mathcal{E}$ and an
input state $\sigma_X$, calculate its Stinespring dilation $\mathcal{U}_{X\to X'E}$ as
explained above, and consider an ancillary system $A_E$ of the same dimension as $E$. This
ancilla system is initialized in a pure state $\ket{0}_{A_E}$. One can then carry out a
unitary $\mathcal{U}'_{XA_E\to{}X'A_E'}$ on $X$ and $A_E$, chosen such that
\begin{align}
  \mathcal{U}'_{XA_E\to{}X'A_E'}\left(\sigma_X\otimes\proj0_{A_E}\right) =
  \mathcal{U}_{X\to X'A_E'}\left(\sigma_X\right)\ .
\end{align}
In effect, $A_E'$ impersonates the
abstract system $E$ while we perform a unitary corresponding to the Stinespring dilation
of $\mathcal{E}$ (see inset of Fig.~\ref{fig:ProcessEntropyEnvironment}\textbf{b}). This
unitary operation can be implemented at no work cost because it is reversible. The
aforementioned optimal erasure procedure can then be used to restore the ancilla $A_E'$ to
its original pure state, using the output system $X'$ as the quantum memory, at a work
cost of approximately $kT\ln(2)\cdot\Hmax^\epsilon\left(A_E'|X'\right)$. As $A_E'$
corresponds to $E$, this matches our bound~\eqref{eq:MainResult} and therefore proves its
tightness.

\begin{figure}
  \centering
  \includegraphics[width=87mm]{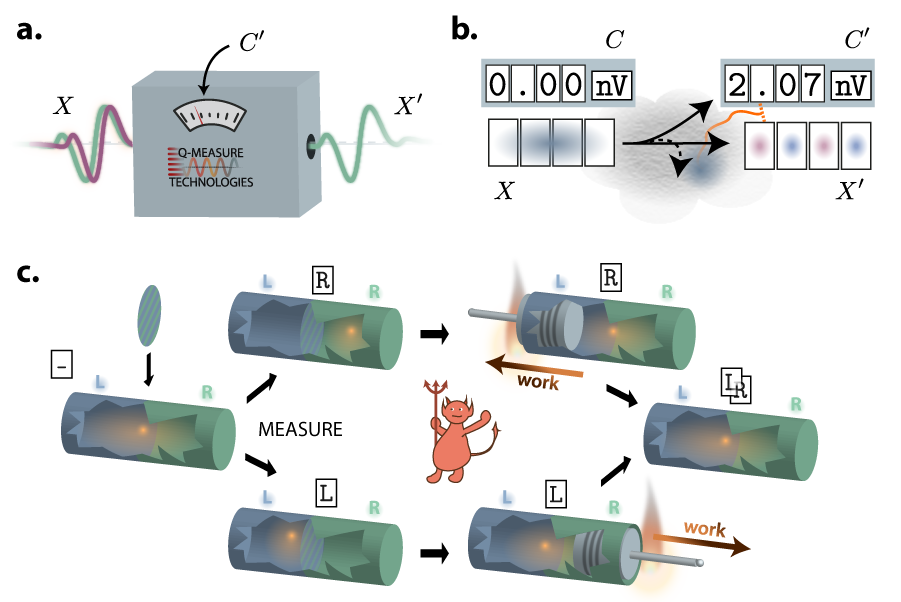}
  \caption{The work cost of quantum measurements.
    \textbf{a.}~A quantum measurement may be thought of as a device which produces a
    post-measurement state $X'$ and a classical reading $C'$ from an input state $X$.
    \textbf{b.}~The corresponding logical process maps the input system $X$ and classical
    register $C$ to a classical outcome on the output register $C'$ and a
    post-measurement state on $X'$. The initial register $C$ is prepared in a pure
    state. Our main result implies that the measurement costs no work in principle.
    \textbf{c.}~Maxwell's demon with a Szilard box, as proposed by
    Bennett~\protect\cite{Bennett1982IJTP_ThermodynOfComp}. A measurement detects on which
    side of the inserted separator the particle is, and extracts work with a piston in
    either case. The cylinder is left in its original state, apparently creating a
    perpetuum mobile with net work gain. However the measurement outcome (represented by
    ``L'' or ``R'') had to be stored in a memory register, which was initially in some
    pure state (represented by ``--'') and the work cost of resetting it to a pure state
    again compensates the work gain. The register could have been reset using the
    post-measurement state at no work cost, but the latter was consumed during work
    extraction.  }
  \label{fig:ProcessMeasurement}
\end{figure}

\paragraph{The work requirement of a quantum measurement}
The problem of determining the amount of work needed to carry out a quantum measurement
has been the subject of much literature~\cite{Sagawa2009PRL_minimal,%
  Buscemi2008PRL_global,Jacobs2012_EnergyMeas}, especially in the context of Maxwell's
demon~\cite{Bennett1982IJTP_ThermodynOfComp,%
  Bennett2003_NotesLP,%
  Earman1999_ExorcistXIVp2,%
  BookLeffRex2010}.
A quantum measurement is a logical process (depicted in Fig.~\ref{fig:ProcessMeasurement}\textbf{a})
acting on a system $X$ to be measured and a classical register $C$ initially set to a pure
state, and outputting systems $C'$ and $X'$, with $C'$ containing the
measurement result and $X'$ the quantum post-measurement state.
We will consider a projective measurement for simplicity, treating the more general case
in (Appendix~\suppnoteApplications). The logical process corresponding to the
measurement described by a complete set of projectors $\{P_i\}_i$ takes the form
\begin{align}
  \mathcal{E}_{CX\to C'X'}\left(\sigma_{CX}\right) = \sum_i
  \proj{i}_{C'}\otimes\left(P_i\,\sigma_X P_i\right)\ .
\end{align}
Our bound~\eqref{eq:MainResultEpsEqZero} for this map is at most zero (since
$\norm{\mathcal{E}\left(\proj0_C\otimes\Pi_{X}\right)}_\infty \leqslant 1$), implying that
the measurement can be carried out in principle at no work cost, as was already stated by
Bennett~\cite{Bennett1982IJTP_ThermodynOfComp}.
Note that a work cost is required if the classical register $C$ was not initially
pure~\cite{Jacobs2012_EnergyMeas}.

A related question is the work cost of erasing the information contained in the register
$C'$ after the measurement. Doing so would allow us to construct a cycle. The cost of this
erasure can be reduced using the post-measurement state as a quantum memory, by employing
the procedure presented above, to $kT\ln2\cdot\Hmax^\epsilon\left(C'|X'\right)$. But because
$C'$ and $X'$ may only be classically correlated, no work may be extracted in this
way~\cite{delRio2011Nature}. In some cases this work cost may be zero, for
example for projective measurements on a maximally mixed state
(Appendix~\suppnoteApplications). This might seem to save Maxwell's demon from
Bennett's information-theoretic exorcism which argues that the demon must pay work to
reset its memory~\cite{Bennett1982IJTP_ThermodynOfComp} (see
Fig.~\ref{fig:ProcessMeasurement}\textbf{c}). However the key point is to notice that the
demon can't use the post-measurement state to both extract work and to reset its
internal memory register.

\paragraph{Discussion}
Our main result exposes various features of thermodynamics in the
microscopic regime that are not present in the standard setting of large systems. In
particular, as argued above, the minimum work cost of a logical process cannot be given in
terms of a state function, such as the entropy or the free energy in thermodynamics.

Traditional thermodynamics is concerned with macroscopic systems, and we may retrieve this
limit by considering logical processes that consist of many individual operations. Under
appropriate independence assumptions and using typicality
arguments~\cite{Tomamichel2009IEEE_AEP}, one can show that the
average minimal work cost per process as determined by~\eqref{eq:MainResult} simply takes
the form $kT\ln(2)\cdot\left[H\left(X\right) - H\left(X'\right)\right]$, where
$H\left(X\right) = -\tr\left(\rho_X\log_2\rho_X\right)$ is the usual von Neumann entropy
(see Methods section): the minimal work requirement is now given by a function 
of state $H\left(X\right)$, and no longer depends on the logical process that maps $X$ to
$X'$ (see Methods).

\begin{figure}
  \centering
  \includegraphics[width=87mm]{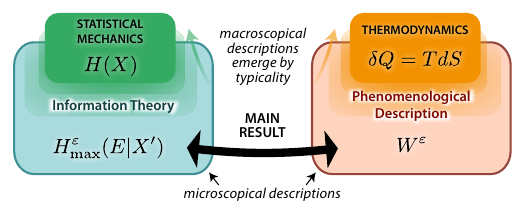}
  \caption{Relation between information-theoretic and thermodynamic quantities. Our result
    relates two quantities that depend on the microscopic details of the system: the
    information-theoretic entropy $\Hmax^\epsilon\left(E|X'\right)$ which 
    quantifies the amount of information discarded by the logical process, and the
    amount of work $W^\epsilon$ needed to carry out a logical process on the microscopic
    level.  Standard
    thermodynamics is obtained in the limit of macroscopic systems. In this limit, it
    follows from typicality arguments that the entropic measure $\Hmax^\epsilon$ converges
    to the von Neumann entropy $H(X)$, which may thus be seen as an emergent
      quantity. Furthermore, in this regime, the minimum amount of work $W^\epsilon$ used
    by a process corresponds to the heat $Q$ that is reversibly transferred to the
    environment, which in turn is related to the thermodynamic entropy, $S$, as defined by
    Clausius. Our result thus permits the identification of the information-theoretic
    entropy $H(X)$ for a macroscopic observer, i.e.\@ the entropy considered in
    statistical mechanics, with the thermodynamic entropy $S$.}
  \label{fig:DiagramEmergence}
\end{figure}

Our result thus provides the following fresh view on the macroscopic regime.
Thermodynamics can be seen as a general framework, in which the second law postulates the
existence of a state function, the thermodynamic entropy, which relates to the heat flow
in processes. Many standard results of thermodynamics follow from
that starting point. It is now the role of a microscopic theory to construct a state
function with this property, based on the microscopic dynamics of the particular
system. In textbook statistical mechanics, this construction is given for several physical
setups, such as gases or lattices; one usually considers e.g. the configuration entropy,
or an appropriately normalized Shannon or von Neumann entropy of the density of the
statistical ensemble.
Our result generalizes this construction, and clarifies when it is justified: the state
function, in general, appears whenever the inherent fluctuations due to the microscopic
stochastic nature of the process vanish by typicality.
The existence of an entropy state function is therefore not a property of the
microscopic system; it is rather an \emph{emergent quantity} that appears whenever the
full system is typical, such as in the limit of macroscopic processes
(Fig.~\ref{fig:DiagramEmergence}).

Finally, one should note that the system in consideration need not be large for the
typicality arguments to apply. For example, if one considers the work requirement of
performing many independent repetitions of a single given logical process (seen as one
big joint process), then the work requirement $W^\epsilon$ per repetition converges to the
average work requirement as calculated via statistical mechanics, even if the
individual system is small: in this case, the entropy function emerges. This further
justifies the usage of the von Neumann entropy in statistical mechanics even for small
systems.
Conversely, a large system does not necessarily display typicality; such is the case for
systems out of thermodynamic equilibrium. An explicit example is provided in
(Appendix~\suppnoteApplications).

In summary, our main result quantifies the minimal required work to perform a logical
process on the microscopic level. On the conceptual level, our result shows how, for
macroscopic systems, the information-theoretic von Neumann entropy emerges as a state
function and can thus be strictly identified with the thermodynamic entropy.

\paragraph{Acknowledgements}
The authors would like to thank Johan \AA{}berg, Francesco Buscemi, Lea Kr\"amer Gabriel,
Joe Renes, L\'\i{}dia del Rio and Paul Skrzypczyk for discussions.  PhF, FD and RR were
supported by the Swiss National Science Foundation (SNSF) through the National Center of
Competence in Research ``Quantum Science and Technology'', through grant No.\
200020-135048, and by the European Research Council through grant No.\ 258932. FD was also
supported by the SNSF through grants PP00P2-128455 and 20CH21-138799, as well as by the
German Science Foundation (grant CH 843/2-1). JO is funded by the Royal Society of
London. This work was also supported by the COST Action MP1209.

\section*{Methods}

\sloppy

\paragraph{Mathematical formulation and proof of the main result}
The task is to implement the logical process $\mathcal{E}_{X\to X'}$.
Recall the framework allows for the implementation of any unital map, i.e.
$\bar{\mathcal{E}}\left(\Ident\right)=\Ident$, to be performed on the systems at hand. We
first adapt a well-known classical result about doubly stochastic and doubly
sub-stochastic matrices~\cite{BookBhatiaMatrixAnalysis1997} to relate unital quantum
maps to so-called \emph{subunital maps}, i.e. maps $\tilde{\mathcal{E}}$ that
satisfy
$\tilde{\mathcal{E}}\left(\Ident\right) \leqslant \Ident$. Note also that the composition
of two unital maps is unital, and similarly the composition of two subunital maps is
subunital. We will need the following proposition, which we prove in
(Appendix~\suppnoteFormal) as Prop.~\supppropDilation.

\begin{prop}[Dilation of a subunital map]
  \label{prop:DilationOfSubUnitalCPMsToUnitalCPMs}
  Let $\Hs_K$ and $\Hs_L$ be finite dimensional Hilbert spaces, and let
  $\tilde{\mathcal{E}}_{K\to L}$ be a completely positive, trace-nonincreasing, subunital
  map. Then there exists finite dimensional Hilbert spaces $\Hs_Q$ and $\Hs_{Q'}$, and a
  completely positive, trace-preserving, unital map $\bar{\mathcal{E}}_{KQ\to LQ'}$ such
  that
  \begin{align}
    \label{eq:DilationOfSubUnitalCPMsToUnitalCPMs_recovercondition_Q}
      \tilde{\mathcal{E}}_{K\to L}\left(\cdot\right)
      &= \left(\Ident_{L}\otimes\bra{\mathrm f}_{Q'}\right)~
      \bar{\mathcal{E}}_{KQ\to LQ'}\bigl[ \nonumber\\
    &\hspace{10ex} \left(\Ident_{K}\otimes\ket{\mathrm i}_{Q}\right)
      \left(\cdot\right) \left(\Ident_{K}\otimes\bra{\mathrm i}_{Q}\right)
      \nonumber \\
      &\hspace{4ex}
        \bigr] ~\left(\Ident_{L}\otimes\ket{\mathrm f}_{Q'}\right) \ ,
  \end{align}
  for some pure states $\ket{\mathrm{i}}_Q$, $\ket{\mathrm{f}}_{Q'}$.
  In addition, $\dim\left(\Hs_K\otimes\Hs_Q\right) = \dim\left(\Hs_L\otimes\Hs_{Q'}\right)$.
\end{prop}

Let's now denote by $A$ the ``information battery'' system, which is the physical system
that tracks how much work we have used or extracted. The system $A$ may be as large
as we might wish (but finite), and starts in a state
$2^{-\lambda_1}\Ident_{2^{\lambda_1}}\otimes\proj{0\ldots 0}$ with some given number of
mixed qubits $\lambda_1$. The system $X$ starts in a given state $\sigma_X$, and we assume
that the Hamiltonians of $X$ and $A$ vanish at the beginning and at the end of the
physical process.

Our framework specifies that we are allowed to perform any sequence of joint unital
operations on any subsystems of $X$ and $A$. The final state on $X'\otimes A'$ should be a
product state, with the state on $A'$ of the form
$2^{-\lambda_2}\Ident_{2^{\lambda_2}}\otimes\proj{0\ldots0}$. Note that the structure
imposed on this state is not a restriction: if the final state on $A'$ is not of this
form, an additional unital map can be applied on the support of the final state on $A'$ to
replace the latter by a maximally mixed state on its support. However, this condition does
assume that there is no way to extract work while transforming a state $\rho$ to a
maximally mixed state of the same rank, or, equivalently, that the worst-case erasure cost
of a state $\rho$ is $kT\ln2\,\log_2\rank\rho$. This can usually be seen as a consequence
of the choice of framework, and is in line with the findings of
(refs~\cite{Aberg2013_worklike,Horodecki2013_ThermoMaj}). Alternatively, given a state
$\rho$, let $m$ be its rank, $p_\mathrm{min}$ its smallest non-zero eigenvalue and $\Pi$
the projector on its support. The state $\rho$ may be written as a statistical mixture of
$\frac1m\Ident_m$ with probability $m\cdot p_\mathrm{min}$ and some state $(\rho -
p_\mathrm{min}\Pi)/(1-m\,p_\mathrm{min})$ with probability $1-m\cdot p_\mathrm{min}$. In
the event where the system is prepared in the maximally mixed state of rank $m$, the work
requirement for erasure is deterministic because the state is uniform, and equals
$kT\ln2\,\log_2 m$~\cite{Szilard1929ZeitschriftFuerPhysik,%
  Bennett1982IJTP_ThermodynOfComp,%
  BookFeynmanLecturesOnComputation1996,%
  BookLeffRex2010}; it follows that the work required for erasing $\rho$ with certainty is
at least $kT\ln2\,\log_2 \rank\rho$.

Observe that our framework is equivalent to allowing the agent to perform a single unital
operation on the whole of $X$ and $A$, leaving both systems in the state $\rho_{X'}\otimes
\left(2^{-\lambda_2}\Ident_{2^{\lambda_2}}\otimes\proj{0\ldots0}\right)_{A'}$: indeed the
composition of unital maps is unital, and extending a unital map by an identity map still
yields a unital map.

Even though we have presented our results while hinting that $X$ and $X'$
represent the same system, and are thus of the same dimension, this need not be the
case: our results are valid for arbitrary finite dimensions of $X$ and $X'$. However, we
will assume that one can bring in ancillas of arbitrary finite dimension in pure states
and dispose of ancillas restored to a pure state for free. Henceforth, we will assume that
such ancillas are counted as part of the pure systems composing the work storage systems
$A$ and $A'$. (The systems $A$ and $A'$ hence need not be of same dimension.)

We must in addition require that the physical process implement the
logical process $\mathcal{E}$. Let $\ket\sigma_{XR}$ be a purification of $\sigma_X$ on a
system $R$. If one applies the physical process to $X$ while leaving $R$ untouched, then
the state on $X'\otimes R$ that results from the physical process must be equal to the
state $\rho_{X'R}$ that would result by applying the mapping
$\mathcal{E}\otimes\operatorname{id}_R$ on $\sigma_{XR}$, i.e.
$\rho_{X'R}=\mathcal{E}\otimes\operatorname{id}_R\left(\sigma_{XR}\right)$. Observe that
this constraint is equivalent to requiring the logical mapping corresponding to the
physical process to be exactly $\mathcal{E}$ on the support of $\sigma_X$, due to the
Choi-Jamio\l{}kowski isomorphism. So, even with a fixed given input state $\sigma_X$, the
full information about the mapping can be observed in the resulting state on $X'\otimes
R$, by keeping a purification of $\sigma_X$: in other words, the full information about
the mapping and the input state is one-to-one encoded in the bipartite state
$\rho_{X'R}$.

Let's now state a formal version of our problem, in the case where we don't yet consider
an $\epsilon$-approximation. The task is to find
the minimal $kT\ln2\cdot\left(\lambda_2-\lambda_1\right)$, such that there exists a
unital, trace-preserving, map $\bar{\mathcal{E}}_{XA\to X'A'}$ satisfying
\begin{align}
  \bar{\mathcal{E}}_{XA\to X'A'}\left(\sigma_{XR}\otimes
  \left(2^{-\lambda_1}\Ident_{2^{\lambda_1}}\right) \right)
  = \rho_{X'R}\otimes
  \left(2^{-\lambda_2}\Ident_{2^{\lambda_2}}\right)\ ,
\end{align}
where $\rho_{X'R} = \mathcal{E}\left(\sigma_{X'R}\right)$ and where an identity mapping on
$R$ is implicitly understood. (We henceforth omit the pure states on system $A$, i.e., the
factors ``${}\otimes\proj{0\ldots0}$'' above, for readability.)

At this point, note that whenever for given $\lambda_1$, $\lambda_2$, there is such a
unital map, then there is also a subunital map achieving the same logical process and vice
versa. Let's write this as a proposition:
\begin{prop}
  Let $\lambda_1$, $\lambda_2 \geqslant 0$, and let $\mathcal{E}$ be given. Then are
  equivalent:
  \begin{enumerate}[(i)]
  \item For a large enough $A$, and corresponding $A'$, there exists a trace-preserving
    unital map $\bar{\mathcal{E}}_{AX\to A'X'}$ such that
    \begin{multline}
      \label{eq:PropFwUnitalSubunital_Unital}
      \bar{\mathcal{E}}_{AX\to A'X'}\left(
        \sigma_{XR}\otimes\left(2^{-\lambda_1}\Ident_{2^{\lambda_1}}\right)_A \right)
      \\
      =
      \rho_{X'R}\otimes\left(2^{-\lambda_2}\Ident_{2^{\lambda_2}}\right)_{A'}
      \ ;
    \end{multline}
  \item For a large enough $B$, and large enough $B'$, there exists a trace-nonincreasing
    subunital map $\tilde{\mathcal{E}}_{XB\to{}X'B'}$ such that
    \begin{multline}
      \label{eq:PropFwUnitalSubunital_SubUnital}
      \tilde{\mathcal{E}}_{XB\to X'B'}\left(
        \sigma_{XR}\otimes\left(2^{-\lambda_1}\Ident_{2^{\lambda_1}}\right)_B \right)
      \\
      =
      \rho_{X'R}\otimes\left(2^{-\lambda_2}\Ident_{2^{\lambda_2}}\right)_{B'}
      \ .
    \end{multline}
  \end{enumerate}
\end{prop}
\begin{proof}
  The forward direction is straightforward, as a unital map is in particular subunital.
  For the converse, we will dilate the given subunital map
  $\tilde{\mathcal{E}}_{XB\to{}X'B'}$ to a 
  unital map using Prop.~\ref{prop:DilationOfSubUnitalCPMsToUnitalCPMs}, with
  $\Hs_K=\Hs_X\otimes\Hs_B$ and $\Hs_L=\Hs_{X'}\otimes\Hs_{B'}$: let $\Hs_Q$, $\Hs_{Q'}$
  and $\bar{\mathcal{E}}_{KQ\to K'Q'} = 
  \bar{\mathcal{E}}_{XBQ\to X'B'Q'}$ be given by the Proposition. Now define
  $\Hs_A=\Hs_B\otimes\Hs_Q$ and $\Hs_{A'}=\Hs_{B'}\otimes\Hs_{Q'}$.
  We would like to show that $\bar{\mathcal{E}}_{XA\to X'A'} \left(
    \sigma_{XR} \otimes \left(2^{-\lambda_1}\Ident_{2^{\lambda_1}}\right)_A
  \right) =
  \rho_{X'R}\otimes \left(2^{-\lambda_2}\Ident_{2^{\lambda_2}}\right)_{A'}$, where we have
  defined $\left(2^{-\lambda_1}\Ident_{2^{\lambda_1}}\right)_A =
  \left(2^{-\lambda_1}\Ident_{2^{\lambda_1}}\right)_B \otimes \proj{\mathrm{i}}_Q$ and
  $\left(2^{-\lambda_2}\Ident_{2^{\lambda_2}}\right)_{A'} =
  \left(2^{-\lambda_2}\Ident_{2^{\lambda_2}}\right)_{B'} \otimes \proj{\mathrm{f}}_{Q'}$
  (as pure states, $\ket{\mathrm{i}}_Q$ and $\ket{\mathrm{f}}_{Q'}$ do not alter the
  amount of work stored in the work storage systems $A$ and $A'$). Define also the
  shorthand $\bar{E}_{X'B'Q'R} := \bar{\mathcal{E}}_{XBQ\to X'B'Q'} \left(
    \sigma_{XR} \otimes \left(2^{-\lambda_1}\Ident_{2^{\lambda_1}}\right)_B \otimes
    \proj{\mathrm i}_Q
  \right)$.  By construction, and
  using~\eqref{eq:DilationOfSubUnitalCPMsToUnitalCPMs_recovercondition_Q}, we have
  \begin{multline}
    \label{eq:PropForUsEquivUnitalSubunital-calc-dilation-orig}
    \left(\Ident_{X'B'R}\otimes\bra{\mathrm f}_{Q'}\right)\, \bar{E}_{X'B'Q'R} \,
    \left(\Ident_{X'B'R}\otimes\ket{\mathrm f}_{Q'}\right)
    \\
    = \rho_{X'R}\otimes
    \left(2^{-\lambda_2}\Ident_{2^{\lambda_2}}\right)_{A'}\ .
  \end{multline}
  Since $\bar{\mathcal{E}}$ is trace-preserving, we have
  $\tr\left(\bar{E}_{X'B'Q'R}\right)=1$ and 
  \begin{multline*}
    \tr\left[\left(\Ident_{X'B'R} \otimes
        \left(\Ident-\proj{\mathrm f}\right)_{Q'}\right)\,\bar{E}_{X'B'Q'R}\right]
    \\
    = 1 -
    \tr\left[\left(\Ident_{X'B'R} \otimes \proj{\mathrm f}_{Q'}\right)\,
      \bar{E}_{X'B'Q'R}\right] = 0\ ,
  \end{multline*}
  as the expression in~\eqref{eq:PropForUsEquivUnitalSubunital-calc-dilation-orig} has
  unit trace. It follows that $\bar{E}_{X'B'Q'R}$ lies in the support of
  $\Ident_{X'B'R}\otimes\proj{\mathrm f}_{Q'}$, and
  from~\eqref{eq:PropForUsEquivUnitalSubunital-calc-dilation-orig} we conclude as requested that
  \begin{align*}
    \bar{E}_{X'A'R} =
    \rho_{X'R}\otimes\left(2^{-\lambda_2}\Ident_{2^{\lambda_2}}\right)_{A'}\ .
    \tag*\qedhere
  \end{align*}
\end{proof}

We can now characterize the allowed operations in our framework
and their work costs with the following proposition.
\begin{prop}
  Let $\sigma_X$, $\mathcal{E}_{X\to X'}$ be given. Choose system $B$ big enough and let
  be given integers $\lambda_1,\lambda_2\geqslant 0$. Then are equivalent:
  \begin{enumerate}[(i)]
  \item There exists a trace-nonincreasing subunital map
    $\tilde{\mathcal{E}}_{XB\to{}X'B'}$ such that
    \begin{multline*}
      \tilde{\mathcal{E}}_{XB\to X'B'}\left(
        \sigma_{XR}\otimes\left(2^{-\lambda_1}\Ident_{2^{\lambda_1}}\right)_B \right) \\
      =  \rho_{X'R}\otimes\left(2^{-\lambda_2}\Ident_{2^{\lambda_2}}\right)_{B'}
      \ ;
    \end{multline*}
  \item There exists a trace-nonincreasing map $\mathcal{T}$, mapping linear operators on
    $\Hs_X$ to linear operators on $\Hs_{X'}$, such that
    $\mathcal{T}_{X\to X'}\left(\Ident\right)\leqslant
    2^{-\left(\lambda_1-\lambda_2\right)}\Ident$, and
    $
    \mathcal{T}_{X\to X'}\left(\sigma_{XR}\right) = \rho_{X'R}
    $;
  \item The map $\mathcal{E}_{X\to X'}$ satisfies
    $\norm{\mathcal{E}\left(\Pi_X\right)}_\infty\leqslant
    2^{-\left(\lambda_1-\lambda_2\right)}$,
    where $\Pi_X$ is the projector onto the support of $\sigma_X$.
  \end{enumerate}
\end{prop}
\begin{proof}
  (i)$\Rightarrow$(ii): Define
  $
  \mathcal{T}_{X\to X'}\left(\cdot\right) =
  \tr_B\left[\Ident_{2^{\lambda_2}}\,
    \tilde{\mathcal{E}}_{XB\to X'B'}\left(\left(\cdot\right)
      \otimes \left(2^{-\lambda_1}\Ident_{2^{\lambda_1}}\right) \right)
    \,\Ident_{2^{\lambda_2}}\right]
  $.
  Then,
  $
  \mathcal{T}\left(\sigma_{XR}\right)
  = \tr_{B'}\left[\Ident_{2^{\lambda_2}}\,
    \left(\mathcal{E}\left(\sigma_{XR}\right)
      \otimes \left(2^{-\lambda_2}\Ident_{2^{\lambda_2}}\right) \right)
    \,\Ident_{2^{\lambda_2}}\right]
  = \rho_{X'R}
  $. Also,
  $
  \mathcal{T}\left(\Ident\right)
  = \tr_{B'}\left[\Ident_{2^{\lambda_2}}\,
    \tilde{\mathcal{E}}_{XB\to X'B'}\left(\Ident_X
      \otimes \left(2^{-\lambda_1}\Ident_{2^{\lambda_1}}\right) \right)
    \,\Ident_{2^{\lambda_2}}\right]
  \leqslant 2^{-\lambda_1} \tr_{B'}\left[
    \Ident_{2^{\lambda_2}}\,
    \tilde{\mathcal{E}}_{XB\to X'B'}\left(\Ident_{XA}\right)
    \,\Ident_{2^{\lambda_2}}
  \right]
  \leqslant 2^{\lambda_2-\lambda_1} \Ident_X
  $, because $\tilde{\mathcal{E}}$ is subunital.
  
  (ii)$\Rightarrow$(iii): We have $\mathcal{E}\left(\Pi_X\right) =
  \mathcal{T}\left(\Pi_X\right)$ because the maps are equal on the support of $\rho_X$
  (alternatively, operate $\tr_R\left[\left(\cdot\right)\rho_R^{-1}\right]$ on both sides
  of $\mathcal{T}\left(\sigma_{XR}\right) = \rho_{X'R} =
  \mathcal{E}\left(\sigma_{XR}\right)$ noting that $\rho_R=\sigma_R$);
  then because $\Pi_X\leqslant\Ident_X$, we have
  $\norm{\mathcal{E}\left(\Pi_X\right)}_\infty
  \leqslant\norm{\mathcal{T}\left(\Ident_X\right)}_\infty
  \leqslant 2^{-\left(\lambda_1-\lambda_2\right)}$.
  
  (iii)$\Rightarrow$(i): Let
  $
  \tilde{\mathcal{E}}_{XA\to X'A'}\left(\cdot\right) = \mathcal{E}\left(\tr_B\left[
      \left(\Pi_X\otimes\Ident_{2^{\lambda_1}}\right)
      \left(\cdot\right)
      \left(\Pi_X\otimes\Ident_{2^{\lambda_1}}\right)\right]\right) \otimes
  \left(2^{-\lambda_2}\Ident_{2^{\lambda_2}}\right)
  $.
  Observe that $\tilde{\mathcal{E}}$ is subunital:
  $
  \tilde{\mathcal{E}}_{XA\to X'A'}\left(\Ident_{XA}\right)
  = \mathcal{E}\left(\tr_B\left[
      \Pi_X\otimes\Ident_{2^{\lambda_1}}
    \right]\right) \otimes
  \left(2^{-\lambda_2}\Ident_{2^{\lambda_2}}\right)
  \leqslant 2^{\lambda_1-\lambda_2}\,\mathcal{E}\left(\Pi_X\right)
  \otimes\Ident_{2^{\lambda_2}}
  \leqslant \Ident_{X'A'}
  $. Also,
  $
  \tilde{\mathcal{E}}_{XA\to X'A'}\left(\sigma_{XR}\otimes
    \left(2^{-\lambda_1}\Ident_{2^{\lambda_1}}\right)
  \right)
  = \mathcal{E}\left(\sigma_{XR}\right)\otimes
  \left(2^{-\lambda_2}\Ident_{2^{\lambda_2}}\right)
  $, because the input to $\tilde{\mathcal{E}}$ is inside the support of
  $\Pi_X\otimes\Ident_{2^{\lambda_1}}$. Hence $\tilde{\mathcal{E}}$ satisfies the
  conditions of (i).
\end{proof}

With these propositions, we can calculate straightforwardly and explicitly the
minimization in the formulation of the main problem. It now reduces to the simple
question of
minimizing $\lambda_2-\lambda_1$ subject to $\norm{\mathcal{E}\left(\Pi_X\right)}_\infty
\leqslant 2^{\lambda_2-\lambda_1}$; we have thus proven~\eqref{eq:MainResultEpsEqZero}.

\paragraph{Entropic form of the bound}
Some basic facts about the smooth entropy framework are necessary to understand the rest
of this section. For a more complete introduction on the smooth entropy framework, we
refer to (Appendix~\suppnoteSmoothEnt).

An equivalent definition of the \emph{R\'enyi-zero conditional entropy}, also known as
\emph{alternative max-entropy}, for a bipartite
state $\rho_{AB}$, is given as
\begin{align}
  \label{eq:DefinitionHzero}
  \Hzero\left(A|B\right)_{\rho} = \log_2\,\norm{\tr_A\Pi_{AB}}_\infty\ ,
\end{align}
where $\Pi_{AB}$ is the projector on the support of $\rho_{AB}$.
For consistency with the standard literature, we will express our final result in terms of
the \emph{max-entropy}, which is related to the R\'enyi-zero entropy up to factors
logarithmic in $\epsilon$~\cite{PhDTomamichel2012}. The \emph{non-smooth conditional
  max-entropy} can be defined as 
\begin{align}
  \Hmax\left(A|B\right) = \max_{\omega_B} \log F^2\left(\rho_{AB},
    \Ident_A\otimes\omega_B\right)\ ,
\end{align}
where $F\left(\rho_1,\rho_2\right)=\norm{\rho_1^{1/2}\rho_2^{1/2}}_1$ is the fidelity
between two quantum states~\cite{BookNielsenChuang2000}, and where the
optimization ranges over density operators on $B$.
The \emph{smooth conditional max-entropy} is defined by ``smoothing'' the max-entropy on states
that are $\epsilon$-close to $\rho_{AB}$ in fidelity distance:
\begin{align}
  \Hmax^\epsilon\left(A|B\right)_\rho =
  \min_{\hat\rho\stackrel\epsilon\approx\rho} \Hmax\left(A|B\right)_{\hat\rho}\ ,
\end{align}
where the minimization ranges over all $\hat\rho$ such that
$F^2\left(\rho_{AB},\hat\rho_{AB}\right) \geqslant 1-\epsilon^2$.

Let's now return to our bound~\eqref{eq:MainResultEpsEqZero}. Consider the Stinespring
dilation of $\mathcal{E}$, given by an isometry $V_{X\to X'E}$ including an additional
system $E$: $\mathcal{E}\left(\cdot\right) = \tr_E\left(V_{X\to X'E}\left(\cdot\right)
  V^\dagger\right)$. Defining the pure state $\rho_{X'ER} = V \sigma_{XR} V^\dagger$ is
obviously compatible with our previous definition of $\rho_{X'R}$, as $\tr_E\rho_{X'ER} =
\mathcal{E}\left(\sigma_{XR}\right)$. It follows that $V\Pi_XV^\dagger=\Pi_{X'E}$, where
$\Pi_{X'E}$ is the projector on the support of $\rho_{X'E}$.
Recalling~\eqref{eq:DefinitionHzero}, we have
\begin{multline}
  \norm{\mathcal{E}\left(\Pi_X\right)}_\infty = \norm{\tr_E V\Pi_X V^\dagger}_\infty
  = \norm{\tr_E\Pi_{X'E}}_\infty \\= 2^{\Hzero\left(E|X'\right)_\rho}\ ,
\end{multline}
and our bound~\eqref{eq:MainResultEpsEqZero} takes the form
\begin{align}
  \label{eq:MainResultEpsEqZeroHzero}
  W^{\epsilon=0} \geqslant kT\ln2\cdot \Hzero\left(E|X'\right)_\rho\ .
\end{align}

\paragraph{Considering an $\epsilon$-approximation}
A ``smooth'' version of the result is straightforward to obtain. In this case, we allow
the actual process to not implement precisely $\mathcal{E}$, but only approximate it
well. The best strategy to detect this inexactness is to prepare $\ket\sigma_{XR}$ and send
$\sigma_X$ into the process, and then perform a measurement on $\rho_{X'R}$. To ensure
that the approximate process is not distinguishable from the ideal process with
probability greater than $\epsilon$, we require that
the trace distance between the ideal output of the process $\rho_{X'R}$ and the actual
output $\hat\rho_{X'R}$ must not exceed $\epsilon$.
We can apply our main result to the approximate process that brings $\sigma$ to
$\hat\rho$, and lower bound the work cost of that process by
\begin{align}
  W(\sigma\rightarrow\hat\rho)
  &\geqslant \Hzero\left(E|X'\right)_{\hat\rho}\cdot kT\ln(2)\nonumber\\
  &\geqslant \Hmax\left(E|X'\right)_{\hat\rho}\cdot kT\ln(2)\ ,
  \label{eq:MainResultAppliedToApprox}
\end{align}
where the second inequality is shown in (Ref.~\cite{Tomamichel2011TIT_LeftoverHashing}). This
relaxation of $\Hzero$ to $\Hmax$ is done for the sake of presentation and consistency
with other results within the smooth entropy framework. When smoothing with a parameter
$\epsilon$, there is no significant difference with this relaxation: indeed, the two
quantities are equivalent up to adjustment of the $\epsilon$ parameter and up to a
logarithmic term in $\epsilon$ (Lemma 18 of Ref.~\cite{Tomamichel2011TIT_LeftoverHashing}).

If we optimize~\eqref{eq:MainResultAppliedToApprox} over all possible
maps $\mathcal{T}$ that output such $\hat\rho_{X'R}$, we obtain a bound on the
work requirement of the $\epsilon$-approximation,
\begin{align}
  W~
  &\geqslant~ \min_{\substack{\hat\rho_{X'R}\stackrel\epsilon\approx\rho_{X'R}}}
  \Hmax\left(E|X'\right)_{\hat\rho}\cdot kT\ln(2) \nonumber\\
  &\geqslant~ \min_{\hat\rho_{X'RE}\stackrel{\bar\epsilon}{\approx}\rho_{X'RE}}
  \Hmax\left(E|X'\right)_{\hat\rho}\cdot kT\ln(2)\nonumber\\
  &= \quad \Hmax^{\bar\epsilon}\left(E|X'\right)_\rho \cdot kT\ln(2)\ ,
  \label{eq:MainResultSmoothed}
\end{align}
where the first optimization ranges over all
$\hat\rho_{X'R}$ such that the trace distance
$\frac12\norm{\hat\rho_{X'R}-\rho_{X'R}}_1\leqslant\epsilon$, and where the second
optimization ranges over all $\hat\rho_{X'RE}$ such that 
$F^2\left(\rho_{X'RE},\hat\rho_{X'RE}\right)\geqslant 1-{\bar\epsilon}^2$, with
$\bar\epsilon=\sqrt{2\epsilon}$, where
$F\left(\rho,\hat\rho\right)=\norm{\sqrt\rho\sqrt{\hat\rho}}_1$ is the
fidelity between the quantum states $\rho$ and $\hat\rho$~\cite{BookNielsenChuang2000}.

\paragraph{Macroscopic limit: many independent repetitions}
As we have seen in the introduction, considerable previous work has focused on the limit
cases where many i.i.d. systems are provided. In such a case, the process
$\mathcal{E}^{\otimes n}$ is applied on $n$ independent copies of the input
$\sigma^{\otimes n}$, and outputs $\rho^{\otimes n}$. A smoothing parameter $\epsilon>0$
is chosen freely. We may simply apply our (smoothed) main result to get an expression
for our bound on the work cost,
\begin{align}
  W \geqslant
  \Hmax^{\bar\epsilon}\left(E^{n}|X^{n}\right)_{\rho^{\otimes n}}\cdot kT\ln(2)\ ,
\end{align}
however it is known that the smooth entropies converge to the von Neumann entropy in the
i.i.d. limit~\cite{Tomamichel2009IEEE_AEP},
\begin{align}
  \lim_{\bar\epsilon\rightarrow 0}\lim_{n\rightarrow\infty} \frac1n
  \Hmax^{\bar\epsilon}\left(E^{n}|X^{n}\right)_{\rho^{\otimes n}}
  = H\left(E|X\right)_\rho\ ,
\end{align}
which allows us to simplify the expression of the work cost per particle, or per
repetition of the process, to
\begin{align*}
  H\left(E|X\right)_\rho = H\left(EX\right)_\rho - H\left(X\right)_\rho
  = H\left(X\right)_\sigma - H\left(X\right)_\rho\ ,
\end{align*}
where the last equality holds because $\rho_{EX}$ and $\sigma_X$ have the same spectrum
being both purifications of the same $\rho_R=\sigma_R$. We conclude that in the asymptotic
i.i.d. case, the work cost is simply given by the difference of entropy between the
initial and final state,
\begin{align}
  \label{eq:IIDRepetitionsBound}
  W \geqslant
  \left[H\left(\text{initial state}\right) - H\left(\text{final state}\right) \right]
  kT\ln(2)\ .
\end{align}
Here, $W$ is the average work cost per particle, or per repetition of the process. In the
case for example of many independent particles undergoing a similar, independent process,
the total work $W$ required is obtained by considering the entropy of the full system of
all particles in both terms in~\eqref{eq:IIDRepetitionsBound}.


\vspace{2cm}
\appendix
\begin{center}{\bfseries APPENDIX}\end{center}

\def\reffigurename{Fig.}

\renewcommand{\thethm}{\arabic{thm}}
\setcounter{thm}{0}


\section{Motivation. Relation of Our Result to Previous Work.}
\label{sec:appx-intro}

The relation between thermodynamics and information, and in particular between the
statistical Gibbs entropy and the information-theoretic entropy has been extensively
studied from various perspectives. We give a short overview in this section; for a rather
comprehensive discussion we suggest (Ref.~\cite{BookLeffRex2010}).

The fundamental question was raised by Maxwell in the 19th century, who imagined a
perpetuum mobile on a gas divided into two chambers, whose net effect was the reduction of
entropy of an isolated system: a small being could have knowledge of the microscopic
degrees of freedom of the gas and operate a trap door in the splitting wall, which he
could use to filter the cold molecules from the hot molecules.
Szilard~\cite{Szilard1929ZeitschriftFuerPhysik} realized that the crucial part of the
problem was that the demon accessed microscopic degrees of freedom, which are not
accessible normally in thermodynamics. He devised a thought experiment, the Szilard box
(see main text), which illustrated the reversible conversion of $kT\ln 2$ work from or
into well-defined accessible information. This suggested that the demon had to perform
work to compensate for the entropy decrease of the gas.  Scientists at the time were then
led to believing that the measurement itself was a process that had to cost work, and some
thought models were developed~\cite{Brillouin1951_Maxwells,Gabor1961_LightInf}. It was
Landauer in 1961~\cite{Landauer1961_5392446Erasure} who first associated work cost to the
\emph{logical irreversibility} of an operation, and who stated that the erasure of 1 bit
of information had to cost $kT\ln2$ work, and studied in particular the example of a
particle in a double-V shaped potential~\cite{Landauer1961_5392446Erasure,%
  Keyes1970IBMJRD_dissipation}.  Bennett showed that computations can be made completely
reversible and devised an explicit measurement apparatus which required no work. On the
other hand, resetting the demon's memory back to its original state does cost work,
effectively exorcising Maxwell's demon~\cite{Bennett1982IJTP_ThermodynOfComp,%
  Bennett1973IBMJRD_LogRevComp,%
  Bennett2003_NotesLP,%
  Bennett1988IBMJRD_NotesHistRevComp,%
  Maruyama2009RMP_ColloquiumPhysMaxDemAndInf}.  Landauer's principle has been
criticized~\cite{Earman1998_ExorcistXIVp1,%
  Earman1999_ExorcistXIVp2}, but became widely accepted as alternative proofs were
proposed~\cite{Piechocinska2000PRA,Shizume1995_HeatGeneration} (cf.
also~\cite{Jacobs2005arXiv}) and its conceptual importance
clarified~\cite{Bennett2003_NotesLP}. Various physical computational models were
explored~\cite{Landauer1971_minimal,%
  Likharev1982_limitations,%
  Landauer1984_fundamental,%
  Landauer1988_dissipation}, while general considerations relating information and physics
were discussed~\cite{Zurek1989_cost,Landauer1996_physical,Lloyd2000_ultimate,%
  Ladyman2007_LogThIrrev}. These efforts were lead in parallel to Jaynes showing the
relevance of information theory for statistical mechanics and
thermodynamics~\cite{Jaynes1957PR_InfThStatMech,%
  Jaynes1957PR_InfThStatMech2,Jaynes1965_GibbsvsBoltzman,Jaynes1992_Gibbs}.

With the development of quantum information in the last decades, importance was given to
generalizing Landauer's principle to the quantum regime~\cite{Plenio2001_forgetting,%
  Alicki2004_hamiltonian,%
  Oppenheim2002PRL_thermodynamical,%
  Janzing2000_cost,%
  Janzing2006Habil,%
  Anders2010_LPQuantDomain},
resulting in replacing the Gibbs, or Shannon, entropy with the quantum von Neumann entropy
as relevant measure of information-theoretic entropy. These studies were motivated by the
important technological advances making it possible to construct microscopic
thermodevices~\cite{Scovil1959masers,Geusic1967quatum,Alicki1979_davies,%
  Howard1997_molecular,geva1992classical,
  Hanggi2009brownian,
  allahverdyan2000extraction,
  feldmann2006lubrication,baugh2005experimental}. Efforts were also undertaken to understand the laws of
thermodynamics from an information-theoretic point of view~\cite{gemmer2009quantum,%
  popescu2006entanglement,Linden2010PRL_SmallFridge,linden2010speed,gogolin2011absence,%
  trotzky2012probing,Hutter2011_dependence}.

While most information-theoretic approaches had studied averages over many independent
repetitions of the same experiment, known as the \iid{} regime, some effort was made to
focus on single instances of information-theoretic tasks~\cite{PhdRenner2005_SQKD}, where
the natural entropy measures to consider are the \emph{smooth entropies}%
~\cite{PhdRenner2005_SQKD,Tomamichel2009IEEE_AEP,PhDTomamichel2012}. In this regime,
information erasure is also characterized using the smooth entropy
framework~\cite{delRio2011Nature,%
  Dahlsten2011NJP_inadequacy,%
  Aberg2013_worklike,%
  Horodecki2013_ThermoMaj,%
  Egloff2010MasterThesis,%
  Egloff2012arXiv}.

The frameworks that have been considered for the study of the thermodynamics of
information processing are extremely varying. While some studies have focused on explicit
construction of physical systems, such as Szilard
boxes~\cite{Szilard1929ZeitschriftFuerPhysik,Dahlsten2011NJP_inadequacy,Egloff2012arXiv}, others have
considered for example systems described by general Hamiltonians that are interacting, or
for which we allow the modification of individual
levels~\cite{Piechocinska2000PRA,%
  Alicki2004_hamiltonian,%
  Sagawa2009PRL_minimal,%
  Linden2010PRL_SmallFridge,%
  delRio2011Nature,%
  Aberg2013_worklike,%
  Skrzypczyk2013arXiv_extracting,%
  Aberg2014PRL_catalytic,%
  Reeb2014NJP_improved}.
Another very promising approach, from which the framework in this paper is largely
inspired, is based on a resource theory of thermal operations, where the Gibbs states are
for free~\cite{Horodecki2003PRA_NoisyOps,%
  Janzing2006Habil,%
  Brandao2013_resource,%
  Horodecki2013_ThermoMaj,%
  Brandao2015PNAS_secondlaws}. The two approaches are equivalent~\cite{Brandao2013_resource}.

Our result adds to the effort of relating information theory to thermodynamics, in the
form of a general formula for the minimal work requirement of a logical process. The
logical process can be any quantum physical evolution. This result may be seen as an
information-theoretic result, which expresses the minimal size of an ancillary system needed
to store the information discarded by the logical process, with a natural direct
application to thermodynamics.

As a special case, our result allows us to study the work cost of quantum measurements. 
We start by noting that in the literature discussed above there is a slight ambiguity as
to what a \emph{measurement} is exactly, and, in particular, whether the memory register
of the apparatus starts in a well-defined pure state or first needs
initialization. If we include the memory initialization process, the measurement does cost
work, whereas with a pure memory, the act of transferring information to the memory costs
no work. This fact was also emphasized by Sagawa and
Ueda~\cite{Sagawa2009PRL_minimal,Sagawa2013_InfTh,%
  Sagawa2012PRL_FluctThm}, cf. also~\cite{Jacobs2012_EnergyMeas}.

\section{Some Initial Remarks and Clarifications.}
\label{sec:appx-some-initial-clarifications}

We wish to emphasize that in this entire work, and unless otherwise stated, by ``process''
we mean to denote a \emph{logical} process, and not a thermodynamical process. 
Thermodynamical processes will come to play via the operations allowed
by our framework.

Logical
processes, or computations, are an abstract mathematical mapping of
input states to output states. For example, and \AND{} gate maps logical states \B{00},
\B{01} and \B{10} to the logical state \B{0} and \B{11} to the logical state
\B{1}. Logical processes are defined completely independently of their physical
implementation, in the same spirit of Shannon's abstraction of the unit of information. In
the most general case, a logical process is specified by a completely positive,
trace-preserving map $\mathcal{E}$.

On the other hand, the logical information has to be stored on a physical system, and any
logical operations have to be implemented through an appropriate time evolution of the
physical system in interaction with a thermal bath and some control system(s). The
specification of a logical process
$\mathcal{E}$ contains no information about how much work was actually used to perform
it---different strategies, different thermodynamical process or different levels of noise,
losses or friction might cause the physical procedure use up very different amounts of
work.

However, there is a fundamental limit on how much work will be \emph{required}
for the implementation of a logical process, or one could build a thermodynamic cycle with
net work gain. In usual thermodynamics, this can simply be calculated as the
difference in free energies between the final state and the initial state. In other words,
the free energy, a state function, acts as a potential from which one can derive the
minimal amount of work one needs in order to perform a transition from one thermodynamic
state to another.

We derive the fundamental work requirement of implementing a logical
process on the microscopic level. This is again the work cost of the \emph{best possible
  thermodynamic process} that succeeds in implementing the given logical process.
As mentioned in the main text, one of our main conclusions is that this minimal work
requirement can no longer be given by a state function. In other words, it is not possible
to define a ``generalized free energy'', a state function which would have
the property of giving the minimal work requirement of a process as a difference between
initial and final states of the computation. This is also in line with the conclusions of
Lieb and Yngvason~\cite{Lieb2013_entropy_noneq} as well as Horodecki
\etal{}~\cite{Horodecki2013_ThermoMaj}.

We wish to draw the attention of the reader to the fact that our conclusions have nothing
to do with the statement that thermodynamic work itself is not a state function. Indeed, in
standard thermodynamics and as mentioned above, the \emph{minimal} work requirement for
going from one state to another is still given by a state function, namely the free energy.

\begin{figure}
  \centering
  \includegraphics[width=87mm]{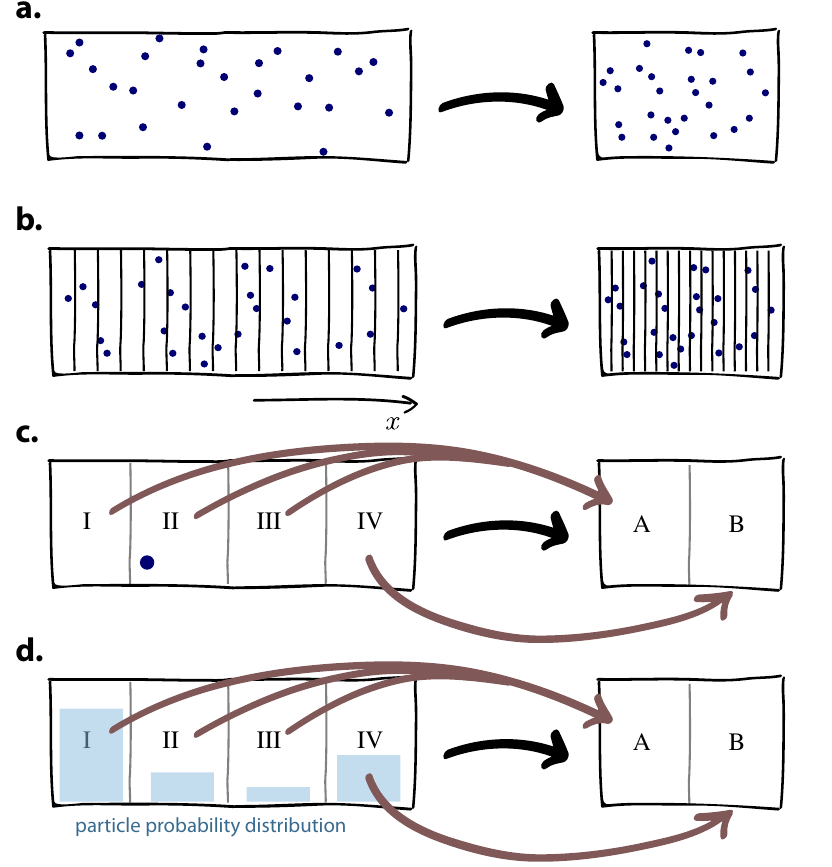}
  \caption[]{\emph{Thermodynamical} processes can be used to physically
    implement abstract \emph{logical} processes. \textbf{a.} An isothermal compression
    implements the logical process corresponding to randomizing the position of the
    particles, within half the original volume. \textbf{b.} In this logical process, the
    $X$ positions of the particles are mapped to half their value. This can be implemented
    by introducing many separators, resolving the position of the particles to a good
    enough precision, and then performing isothermal compression of each slice of gas.
    This procedure has same optimal work cost as the previous one, even though the
    \emph{logical} processes are different. \textbf{c.} A single particle gas can be used
    to model an \AND{} gate. The particle is supposed to be brought in output region A if
    it was originally in regions I, II or III, and to region B otherwise. The work cost
    fluctuates because of the probabilistic nature of the input state. \textbf{d.} If the
    probabilities that the particle initially resides within the different regions are not
    equal, the same amount of work as in the previous case is needed. However, if only the
    correct output state is to be reproduced, less work is required in some situations.
  }
  \label{fig:appx-GasLogicalThermoProcesses}
\end{figure}

In order to further clarify the relation between logical and thermodynamical processes,
consider an ideal gas of $N$ particles in a box of volume $2V$, at temperature $T$. Let's
consider bringing this gas to a new state with half the volume, given by the parameters
$(T,V,N)$ (see \reffigurename~\ref{fig:appx-GasLogicalThermoProcesses}).
The specification of a \emph{logical} process goes beyond specifying the input and output
states: indeed, one can require for example that the position of each particle be
completely randomized at the end of the process
(\reffigurename~\ref{fig:appx-GasLogicalThermoProcesses}\textbf{a}), with no correlations
between input and output; one can also require for example that a particle located at a
position $(x,y,z)$ be located at $(x/2,y,z)$ at the output
(\reffigurename~\ref{fig:appx-GasLogicalThermoProcesses}\textbf{b}). Both these logical
processes have the same input and output state.

The first logical process can be physically implemented with a simple isothermal
thermodynamical compression. The second logical process can be implemented similarly, by
using a trick: first, we insert many separators in the box, resolving the positions of the
particles to some acceptable precision, and then we perform an isothermal compression of
each of those slices of the gas independently. Then each particle originally located at
position $x$ on the $X$-axis is now located at $x/2$ at the output
and $kT\ln2$ work was expended per particle. (Should the $y$ and $z$ coordinates also be
required to be correlated between input and output, a grid of separators along the other
axis should also be inserted, while the compression is performed in the $X$ direction.)
In these simple examples, both logical
processes have the same minimal work cost (it is easy to see that the thermodynamical
processes given above are optimal, for example using our main result). This is the
illustration of our main
result in the thermodynamic limit: in an \iid{} setting, the minimal work requirement is
simply given by the difference between final and initial entropy, \emph{regardless of the
  specific logical process}. (Note that here entropy corresponds to free energy, since
there is no change in internal energy.)

Note that both logical processes could have been performed with an irreversible 
thermodynamic process, for example a fast compression followed by a thermalization (of the
whole gas in \reffigurename~\ref{fig:appx-GasLogicalThermoProcesses}\textbf{a}, or of
each slice in \reffigurename~\ref{fig:appx-GasLogicalThermoProcesses}\textbf{b}). Then additional,
irreversible work is required. However these irreversible processes are not the optimal
thermodynamical processes that carry out the requested logical processes. The expression
in our main result is given by the \emph{optimal} thermodynamic implementation of a given
logical process.

We further note that, in general, logically irreversible processes can be
implemented in a thermodynamically reversible way: a Szilard box in a completely mixed
state, for example, can be reset to a pure state by a simple reversible isothermal
process. While the thermodynamic transformation is reversible, meaning that we can recover
the initial mixed state and get back all the work invested in the erasure, the precise
logical state the memory initially was in (whether the particle was on the left or the
right side of the box) is irreversibly lost in the heat bath.%
(This reasoning does not contradict the results by Ladyman
\etal~\cite{Ladyman2007_LogThIrrev}, because the thermodynamical processes they show to
be irreversible are the thermodynamic processes $p_x$ ``conditioned'' on the particular
initial logical state $x$ of the device (using their notation).)

Consider now the example depicted in
\reffigurename~\ref{fig:appx-GasLogicalThermoProcesses}\textbf{c}: a single particle 
is in the box, and three partitions are inserted, subdividing the box into four even
regions I, II, III and IV. We wish to perform the logical process that maps a
particle in regions I, II and III to output region A, and input region IV to
region B. (Upon appropriate relabeling of the regions, this is nothing else than an
\AND{} gate.) A thermodynamical operation that would carry out this logical mapping is an
isothermal compression of the joint three first regions to one-third of their initial
volume (removing the two intermediary separators). The work cost of doing so depends on
whether the particle is located in one of the regions I, II, III or not: with
probability $\nicefrac34$, $kT\ln2\cdot\log_2 3$ work is expended. (We assume that the
isothermal process is done infinitely slowly, \ie{} quasi-statically, 
such that the fluctuations have been reduced to zero; the fluctuations of the work
cost of the logical process explained here are solely due to the probabilistic nature
of the input state or the \emph{logical} process, and not due to fluctuations of the
\emph{thermodynamical} process.)
However, with probability
$\nicefrac14$, no work is expended as those regions are empty. In this case, the work cost
is fluctuating due to the probabilistic nature of the input state. This procedure is again
the best strategy we can devise if we require the process to succeed with almost certain
probability (this is a consequence of our main result), and its worst-case work cost is
$kT\ln2\cdot\log_2 3$.

If we had considered a large number of particles, then with large probability
$\nicefrac34$ of the particles would be in regions I, II, III, and thus the work yield
would almost deterministically be the average value $\nicefrac34\cdot
kT\ln2\cdot\log_23\approx 1.2\,kT\ln2$, which one can check to be equal to the difference
between initial and final Shannon entropy. This is again in agreement with our main result
in the \iid{} regime.

Now, let's consider again the \AND{} gate with the one-particle gas, but where the
particle has different probabilities of being in the different regions, as shown in
\reffigurename~\ref{fig:appx-GasLogicalThermoProcesses}\textbf{d}. Specifically, consider the
example where 
the particle resides in region II or III with the same probability as it would be found in
region IV, \ie{} $p_\mathrm{II} + p_\mathrm{III} = p_\mathrm{IV}$. Again, we are required
to perform the logical process mapping a particle in regions I, II or III to the region A,
and mapping a particle in region IV to region B. Then, one can convince oneself that the
optimal procedure is still to isothermally compress regions I, II, and III into the volume
of region A, as before. However, as explained in the main text, it is possible to devise a
strategy that will still reproduce the correct output state, $p_\mathrm{A} = p_\mathrm{I}
+ p_\mathrm{II} + p_\mathrm{III}$, $p_\mathrm{B} = p_\mathrm{IV}$, that will actually only
cost $kT\ln2$ work. This strategy corresponds to isothermally compressing regions II and
III and merging them to region B, and isothermally compressing regions I and IV and
merging them to region A (permuting regions and moving regions around does not cost any
work, as these are 
unitary, reversible logical operations). This illustrates that the minimal work
requirement of a logical process is not fully specified by the initial and final state. In
particular, it cannot be given by a state function (such as the free energy in standard
thermodynamics).

The important fact is that in macroscopic thermodynamics, one does not care about
correlations between the input and the output, simply because for large i.i.d.\@ systems
(e.g.\ many independent particles, or large weakly interacting systems such as an ideal
gas) those correlations do not matter. This is due to the system being in a typical
microstate with overwhelming probability (see main text).  We determine, for single
quantum systems, the minimal work requirement of a logical process; our formula shows that
this requirement is not simply given by a function of state. The key point is that if one
goes to the thermodynamic limit, then our formula for the work cost of a logical process
does become a function of state. This shows that our result is of a fundamentally
different nature to the work loss of a process which is thermodynamically irreversible,
which persists in the thermodynamical limit and is due to some avoidable irreversibility.

\section{The Smooth Entropy Framework.}
\label{sec:appx-SmoothEntropies}

The literature about information-theoretic tasks, pioneered by
Shannon~\cite{Shannon1948BSTJ} has largely focused in the past on average resource costs
of asymptotically many independent repetitions of a given task, such as the average
communication rate needed to send the information output by a source generating
independent messages according to a certain distribution.

Recently, frameworks were developed in order to characterize single instances of
these tasks, such as determining how many bits are needed to compress a single message distributed according to some known distribution. The
two major approaches are the information spectrum~\cite{HanVerdu1993_approximate,%
  BookHan02_InfSpecMethods} and the smooth entropy framework~\cite{PhDTomamichel2012,%
  PhdRenner2005_SQKD}, the two approaches being closely
related~\cite{Datta2009IEEE_InfSpec}.

We will focus here on the definition of some of the smooth entropies that are needed in
this work and some of their properties. More information and proofs can be found in
(Refs.\@~\cite{PhDTomamichel2012,PhdRenner2005_SQKD,Tomamichel2010IEEE_Duality,%
  Tomamichel2011TIT_LeftoverHashing}).

In the remainder of this section, let $A$, $B$, $C$ be quantum systems and let
$\ket\rho_{ABC}$ be a pure tripartite state. We say that two states $\rho_1$ and $\rho_2$
are $\varepsilon$-close, denoted by $\rho_1\approx_\varepsilon\rho_2$, if their
\emph{purified distance} as defined in
(Ref.~\cite{Tomamichel2010IEEE_Duality,PhDTomamichel2012}) is less than or equal to
$\varepsilon$ (for normalized states, the purified distance is defined via their
fidelity~\cite{BookNielsenChuang2000}). We refer the reader to these papers for precise
definitions of the purified distance, and for comprehensive discussions about optimization
ranges over subnormalized states which will not be particularly relevant here.

\begin{thmheading}{Min and Max Entropies}
  The central quantities of the smooth entropy framework are the so-called \emph{min-} and
  \emph{max-entropies}.
  The conditional smooth min-entropy $\Hmin^\varepsilon\left(A|B\right)_\rho$
  is defined as follows:
  \begin{subequations}
    \begin{align}
      \Hmin\left(A|B\right)_{\rho|\sigma}
      &= \max\left\{
        \lambda:\; 2^{-\lambda}\Ident_A\otimes\sigma_B\geqslant\rho_{AB}
      \right\}\ ;
      \label{eq:appx-DefHminRhoSigma}\\
      \Hmin\left(A|B\right)_\rho &= \max_{\substack{\sigma_B\geqslant0\\\tr\sigma_B=1}}
      \Hmin\left(A|B\right)_{\rho|\sigma}\ ;
      \label{eq:appx-DefHminRho}\\
      \Hmin^\varepsilon\left(A|B\right)_\rho &=
      \max_{\hat\rho_{AB}\approx_\varepsilon\rho_{AB}} \Hmin\left(A|B\right)_{\hat\rho}\ .
      \label{eq:appx-DefHminRhoEpsilon}
    \end{align}
  \end{subequations}
  Similarly, the smooth conditional max-entropy $\Hmax^\varepsilon\left(A|C\right)_\rho$
  is defined by:
  \begin{subequations}
    \label{eq:appx-DefHmax}
    \begin{align}
      \Hmax\left(A|C\right)_{\rho|\sigma}
      &= \log F^2\left(\rho_{AC}, \Ident_A\otimes\sigma_C\right)\ ;
      \label{eq:appx-DefHmaxRhoSigma}\\
      \Hmax\left(A|C\right)_\rho
      &= \max_{\substack{\sigma_C\geqslant0\\\tr\sigma_C=1}}
      \Hmax\left(A|C\right)_{\rho|\sigma}\ ;
      \label{eq:appx-DefHmaxRho}\\
      \Hmax^\varepsilon\left(A|C\right)_\rho &=
      \min_{\hat\rho_{AC}\approx_\varepsilon\rho_{AC}} \Hmax\left(A|C\right)_{\hat\rho}\ .
      \label{eq:appx-DefHmaxRhoEpsilon}
    \end{align}
  \end{subequations}
  The conditional smooth entropies are invariant under local isometries. They also have
  clear operational interpretations~\cite{Koenig2009IEEE_OpMeaning}. For example, the
  min-entropy $\Hmin\left(A|B\right)$ quantifies how many bits in $A$ can be extracted
  that are uniformly random and uncorrelated to $B$; the max-entropy
  $\Hmax\left(A|C\right)$ corresponds to the amount of bits needed to send to a third
  party who has access to $C$ in order to reconstruct $A$.
\end{thmheading}

\begin{thmheading}{Duality Relation}
  The min- and max-entropy obey the so-called \emph{duality relation}.
  For $\rho_{ABC}$ pure, one has
  \begin{align}
    \Hmin^\varepsilon\left(A|B\right)_\rho = -\Hmax^\varepsilon\left(A|C\right)_\rho\ .
  \end{align}
\end{thmheading}

(The max-entropy may also be defined first by the duality relation, as originally
done~\cite{Koenig2009IEEE_OpMeaning}, and then~\eqref{eq:appx-DefHmax} becomes a theorem.)

\begin{thmheading}{Classical-Quantum states}
  A state $\rho_{AB}$ is classical-quantum (c-q) if it can be written in the form
  \begin{align*}
    \rho_{AB} = \sum_i p_i\,\proj{i}_A\otimes\rho_B^{(i)}\ ,
  \end{align*}
  for positive operators $\rho_B^{(i)}$ satisfying $\sum_i\tr\rho_B^{(i)} = 1$.

  For such states, the conditional smooth entropies
  $\Hmin^\varepsilon\left(A|B\right)_\rho$ and $\Hmax^\varepsilon\left(A|B\right)$ are
  always positive.
\end{thmheading}

\begin{thmheading}{R\'enyi-Zero Entropy}
  An additional entropy measure that will appear naturally in our calculations is the
  \emph{R\'enyi entropy of order zero}, or the \emph{R\'enyi-zero entropy}
  $\Hzero\left(A|C\right)_\rho$~\cite{PhdRenner2005_SQKD,Renyi1960_MeasOfEntrAndInf}. It
  is defined by
  \begin{align}
    \Hzero\left(A|C\right)_\rho = \max_{\sigma_C} \tr\left[\Pi_{AC}\,\sigma_C\right]
    = \norm{\tr_A\Pi_{AC}}_\infty\ ,
  \end{align}
  where $\Pi_{AC}$ is the projector onto the support of the state $\rho_{AC}$.

  The R\'enyi-zero entropy is dual to a specific variant of the min-entropy: for
  $\rho_{ABC}$ pure, one has
  \begin{align}
    \Hzero\left(A|C\right)_\rho = -\Hmin\left(A|B\right)_{\rho|\rho}\ .
  \end{align}

  The R\'enyi-zero entropy and this variant of the min-entropy have also been termed
  \emph{alternative max-entropy} and \emph{alternative min-entropy},
  respectively~\cite{Tomamichel2011TIT_LeftoverHashing}.

  When smoothed, the R\'enyi-zero entropy is closely related to the max
  entropy~\cite{Tomamichel2011TIT_LeftoverHashing}. We have on one hand
  \begin{align}
    \Hzero^\varepsilon\left(A|C\right)_\rho
    := \min_{\hat\rho\approx_\varepsilon\rho} \Hzero\left(A|C\right)_{\hat\rho}
    \geqslant \Hmax^{\varepsilon}\left(A|C\right)_\rho\ ,
  \end{align}
  and the two quantities are almost equal, up to an error term and a small adjustment
  $f(\varepsilon)$ to the smoothing parameter $\varepsilon$:
  \begin{align}
    \Hzero^{f(\varepsilon)}\left(A|C\right)_\rho
    \leqslant \Hmax^{\varepsilon}\left(A|C\right)_\rho
    \Err{+O\bigl(\log{\textstyle\frac1\varepsilon}\bigr)}\ .
  \end{align}
\end{thmheading}

\begin{thmheading}{Von Neumann entropy}
  Recall that the von Neumann entropy is defined as
  \begin{subequations}
    \begin{align}
      \HH\left(X\right)_\rho &= -\tr\left(\rho_X\log\rho_X\right)\ ;\\
      \HH\left(A|B\right)_\rho &= \HH\left(AB\right)_\rho - \HH\left(B\right)_\rho\ .
    \end{align}
  \end{subequations}
\end{thmheading}

\begin{thmheading}{Asymptotic Equipartition Property}
  The smooth entropies all converge to the von Neumann entropy in the \iid{} limit, a
  property which is known as \emph{asymptotic
    equipartition}~\cite{Tomamichel2009IEEE_AEP}.
  When considering $n$ independent copies of the same state $\rho$, and consider large
  $n$, we have:
  \begin{subequations}
    \begin{align}
      \lim_{\varepsilon\rightarrow 0} \lim_{n\rightarrow\infty}
      \frac1n\,\Hmin^\varepsilon\left(A^n|B^n\right)_{\rho^{\otimes n}}
      &= \HH\left(A|B\right)\ ;\\
      \lim_{\varepsilon\rightarrow 0} \lim_{n\rightarrow\infty}
      \frac1n\,\Hmax^\varepsilon\left(A^n|B^n\right)_{\rho^{\otimes n}}
      &= \HH\left(A|B\right)\ .
    \end{align}
  \end{subequations}

  In particular, any terms of order $\log\frac1\varepsilon$ disappear when taking the
  limit $n\rightarrow\infty$, such that the quantities
  $\Hzero^\varepsilon\left(A|B\right)_\rho$ and
  $\Hmin^\varepsilon\left(A|B\right)_{\rho|\rho}$ also obey the asymptotic equipartition
  property.
\end{thmheading}

\section{Additional Comments. Applications of our Main Result.}
\label{sec:appx-Applications}

Our main result states that in our framework, the work cost of a physical process
implementing the computation $\mathcal{E}$ exactly is lower bounded by the quantity
$W_{\mathrm{(bound)}}^{\varepsilon=0} = kT\ln2\cdot\Hzero\left(E|X'\right)_{\rho}$. In the
case where we consider an $\varepsilon$-approximation is tolerated, the bound takes the
value $W_{\mathrm{(bound)}}^{\varepsilon} =
kT\ln2\cdot\Hmax^{\bar\varepsilon}\left(E|X'\right)_{\rho}$, with
$\bar\epsilon=\sqrt{2\varepsilon}$.

In the following sections, we further discuss the implications of our result and provide
some examples.

In a slight abuse of notation, but in an effort to disencumber the mathematical
expressions, a generic superscript $\varepsilon$ on a work cost $W$ or on an entropy
measure will be understood to represent a ``smoothing'' of the quantity, in order to
account for very unlikely events. It is understood that some expressions should actually
contain variants of this quantity such as $\sqrt{2\varepsilon}$ to be technically correct,
but our calculations being relatively simple, a technically complete version should be
straightforward to obtain.

\subsection{On the Tightness of the Minimal Work Bound.}
\label{sec:appx-TightnessMainResult}

Since the bound $W_{\mathrm{(bound)}}^{\varepsilon=0}$ was obtained through a chain of
equivalences from our original framework to the expression of the bound, we know there
exists a unital map over an ``information battery'' $A$ and the system $X$ which achieves
this bound. However, it is not clear how to physically carry out a general unital
operation at no work cost (remember: unital operations were precisely chosen for
their being a very permissive framework, in order to obtain a more general bound). A
convenient special case of unital operations are noisy
operations~\cite{Horodecki2003PRA_NoisyOps}: these consist of a sequence of bringing in a
maximally mixed ancilla, performing a global joint unitary, then tracing out the ancilla.
However not all unital operations are of this
kind~\cite{Haagerup2011_factorization,Landau1993_birkhoff}. So this leaves open the
question of whether our bound is tight.

We have seen in the main text that using the method proposed by del Rio
\etal{}~\cite{delRio2011Nature}, we can construct an explicit process that carries out the
required transformation, which fails with probability less than $\varepsilon$, and which
costs work $kT\ln(2)\cdot\bigl[\Hmax^\varepsilon\left(E|X'\right)_\rho +
\Delta(\varepsilon)\bigr]$, where $\Delta(\varepsilon)$ is an error term of the order of
$\log\left(1/\varepsilon\right)$. That is, we are capable of achieving the bound up to an
error term of order $\log\left(1/\varepsilon\right)$.

In some interesting regimes, such as information coding, the error term may be
negligible. Indeed, if we want to reset $1\ \text{MB}$ of data with a
smoothing parameter of at most $\varepsilon = 10^{-10}$, then the error term is of order of
$\log\left(1/\varepsilon\right) \approx 30$ bits, which is small compared to the original
$\sim10^{7}$ bits.
However, this error term can become overwhelming when considering small systems consisting
of several qubits.

It is an open question to understand the significance of this gap, and to determine
whether the bound can be exactly achieved.
However, this type of error terms, which are clearly sublinear in the number of systems, are
widespread in information and coding theory, and are typically associated with overheads
such as encoding the word length itself, or the overhead of adapting the coding
scheme. More specifically, they invariantly appear in random constructions of protocols,
such as the one used in~\cite{delRio2011Nature}, on which our tightness proof is based.

\subsection{Simple examples: the {AND} and {XOR} gates.}
\label{sec:appx-simple-examples-and-xor}

Consider the classical \AND{} and \XOR{} gates presented in the main text. We would now
like to calculate how much work the best implementation of these gates would require. We
can apply our main result, as given by Eq.~\eqref{eq:MainResultClassical} of the
main text, to obtain
\begin{subequations}
  \label{eq:appx-WANDXOR}
  \begin{align}
    \label{eq:appx-WAND}
    W^\mathrm{(bound)}_\text{\AND} &= kT\ln2\cdot\log_2 3 \approx 1.6\, kT\ln2\ ;\\
    \label{eq:appx-WXOR}
    W^\mathrm{(bound)}_\text{\XOR} &= kT\ln2\ .
  \end{align}
\end{subequations}
(Several quantum descriptions of these classical gates are possible in quantum mechanics;
we assume for this example those which measure the input and prepare the appropriate
output.)

As long as the input distribution does not have very small eigenvalues, no eigenvalues
will be comparably small to $\varepsilon$, and all distributions that are
$\varepsilon$-close to the initial one will have same rank. Thus the
values~\eqref{eq:appx-WANDXOR} are exact also for not too large $\varepsilon$. (Note however
that this differs with the expression in Eq.~\eqref{eq:MainResult} of the main
text, because the latter was obtained with an additional relaxation of $\Hzero$ to $\Hmax$
for purposes of presentation.)

As mentioned in the main text, these gates illustrate the dependence of the minimal work
requirement on the specific computation, and not only on the input and output states.
More generally, it is worth noting that, although a specific input state is given, the
observer can still distinguish the different possible logical processes, even though they
give the same output state. Indeed, the observer can prepare a bipartite pure state on $X$
and a reference $R$, with the reduced state on $X$ matching the required input state. By
keeping this way a purification of the input state, the observer can determine exactly
which logical process was performed by appropriate measurements on the joint state
$\rho_{X'R}$ of the output and the reference (note that $\rho_{X'R}$ is then the
Choi-Jamio\l{}kowski state of the logical process).

Observe also that the value~\eqref{eq:appx-WAND} differs from the average work requirement of
the \AND{} gate, which is given by the difference in von Neumann entropy between the input
state and the output state (most previous work has focused on this regime). Assuming that
the input is uniformly random, \ie{} $\rho_X=\frac14\Ident$, then one obtains
\begin{align*}
  W_\text{\AND, avg.} = kT\ln2\cdot\left[\HH\left(X\right) - \HH\left(X'\right)\right]
  \approx 1.2\,kT\ln2\ .
\end{align*}
Additionally, the value~\eqref{eq:appx-WXOR} happens to coincide with the average work
requirement (calculated similarly) for a uniformly random input; however, if a
different input is given, the two values will differ.

\subsection{Arbitrarily large dependence on the computation, with same input and output
  states.}
\label{sec:appx-arbitrary-large-dependence}

\begin{figure}
  \centering
  \includegraphics[width=87mm]{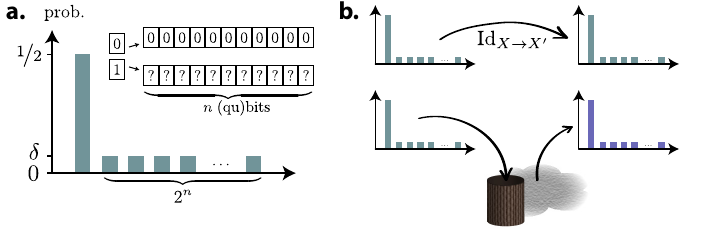}
  \caption{Examples of large, non-typical distributions.
    \textbf{a.}~The probability distribution (given by the spectrum of the state)
    of a classical system of one random qubit, along with $n$ other qubits that are all
    \B0 if the first qubit is \B0, or uniformly random otherwise.
    \textbf{b.}~Two different operations on this system may have the same input and output
    state, yet their work cost may differ arbitrarily. The first operation copies its
    input to its output (identity map), which costs no work. The second destroys the input
    and reproduces a fresh system at the output.
  }
  \label{fig:appx-DistributionCounterExample}
\end{figure}

Consider the example provided in
\reffigurename~\ref{fig:appx-DistributionCounterExample}\textbf{a},
\begin{align*}
  \rho = \frac12\bigl[\proj{0}\otimes\proj{0\ldots0}
  + \proj{1}\otimes\frac{\Ident_{2^n}}{2^n}\bigr]\ .
\end{align*}

Consider also the two logical processes depicted schematically in
\reffigurename~\ref{fig:appx-DistributionCounterExample}\textbf{b}. The first logical process
$\mathcal{E}_1$ is the identity map,
$\mathcal{E}_1 = \mathrm{id}_{X\rightarrow X'}\left(\sigma\right) = \sigma$. The
second logical process $\mathcal{E}_2$ resets its input and prepares a fresh copy
of $\rho$, \ie{} $\mathcal{E}_2\left(\sigma\right) = \tr\left(\sigma\right)\,\rho$.

First note that both computations have exactly the same input and output states. The
minimal work requirement of the identity mapping is zero, obviously, because it can be
implemented by doing nothing, or also, because it is logically reversible.  However, the
analysis is different for $\mathcal{E}_2$. If we did nothing as for $\mathcal{E}_1$, then
high correlations would remain between the input and the output, and we would not be
implementing the computation $\mathcal{E}_2$ but rather $\mathcal{E}_1$. Now the minimal
work requirement that will be needed, if we want to be almost certain that the process
succeeds, can be intuitively understood as follows: in the worst case, which happens with
probability $\nicefrac12$, the input is in the state that is almost fully mixed, and one
will first have to reset ${\sim}\,n$ bits, costing ${\sim}\,n\,kT\ln2$ work. When
preparing the output, we can decide randomly on whether to prepare the pure or the mixed
state by extracting 1 bit of work from a Szilard box. However, in the worst case with
probability $\nicefrac12$, we have to prepare the state $\ket{0\ldots 0}$, and at worst
only one bit of work can be extracted. The total worst-case work cost of this strategy is
\begin{align}
  \label{eq:appx-WorkCostToResetTheCounterExampleDistributionAndToRecreateItFromScratch}
  W^{\varepsilon\approx 0}(\mathcal{E}_2,\rho) \approx n\,kT\ln2\ .
\end{align}
This value can be calculated exactly as our example is a
special case of subsection~\ref{sec:appx-statetransfdecpl} below; it turns out to be
optimal. The approximation we made above (where we note `$\sim$' and `$\approx$') is
simply that $\log(2^n+1) \approx n$ and $n+1\approx n$. Also, we have assumed that
$\left(1-\varepsilon\right)n\approx n$.

Note that the quantity%
~\eqref{eq:appx-WorkCostToResetTheCounterExampleDistributionAndToRecreateItFromScratch}
can become arbitrarily large, as it scales with the number of qubits $n$.

This distribution might seem very artificially constructed. We however provide here an
example of a physical system which exhibits such behavior. Consider a particle detector,
which we model in the following way: as long as no particle has shown up, the detector is
initialized in a state $\ket0$. Once a particle hits the device, the state of the detector
is changed to a very disordered state $\tau$, which we may for the sake of the example
choose as a uniformly mixed state of rank $d$: $\tau = \frac1{d}\Ident_d$. Suppose we wish
to describe the state of the device, not knowing whether a particle has hit it or not. If
the probability that a particle was detected is $\nicefrac12$, then the state of the
detector is precisely $\rho$ (with $d = 2^n$).

``Erasure'' here simply means resetting the device to its initialized state: the logical
process maps the distribution $\rho$ to the pure state $\ket0$. The first logical processes
given in \reffigurename~\ref{fig:appx-DistributionCounterExample}\textbf{b} corresponds
to not doing anything to the detector. The second logical process corresponds to resetting
the detector, and then again sending a new particle in with probability $\nicefrac12$.
Note that in this case, by looking at the detector after the process, we may not know
the state of the detector at the input of the process.

\subsection{Erasure of a Quantum System Using a Quantum Memory.}%
\label{sec:appx-ErasureQuantumSystemWithMemory}%

This scenario was studied in the main text; we repeat this derivation in more detail here.

Consider the setting proposed in~\cite{delRio2011Nature}, where a system $S$ is correlated
with a system $M$ in a joint state $\sigma_{SM}$, and where our task is to erase $S$ while
preserving the reduced state on $M$ and any possible correlations of $M$ with other
systems. Formally, given a purification $\sigma_{SMR}$ of $\sigma_{SM}$, we are looking for a
process that will bring this state to the state $\rho_{SMR}=\proj0_S\otimes\sigma_{MR}$,
\ie{} we require the process to preserve $\sigma_{MR}$. In~\cite{delRio2011Nature} 
a process is proposed that performs this task at work cost
\begin{align*}
  kT\ln(2)\,\Hmax^{\varepsilon}\left(S|M\right)_\sigma
  \Err{ + O\bigl(\log{\textstyle\frac1\varepsilon}\bigr)}\ ,
\end{align*}
where $\Hmax^\varepsilon$ is the smooth max
entropy~\cite{Koenig2009IEEE_OpMeaning,PhDTomamichel2012,Tomamichel2010IEEE_Duality}.

The full process that is eventually performed can be written as
\begin{align}
  \label{eq:appx-ProcessErasureMemory}
  \mathcal{E}^\mathrm{(erasure)}_{SM\rightarrow SM}\left(\sigma\right) =
  \proj{0}_S\otimes\tr_S\left(\sigma\right)\ .
\end{align}
(It is straightforward to verify that this process preserves the reduced state
$\sigma_{MR}$.)
We can now apply our main result to this particular mapping, simply by considering $X$ to
be the joint system of $S$ and the memory $M$,  $\Hs_X=\Hs_S\otimes \Hs_M$.
Note that we have
$\rho_{SMR}=\proj{0}_S\otimes\sigma_{MR}$, purified by
$\ket{\rho}_{SMRE}=\ket0_S\otimes \ket{\rho}_{MRE}$, where $\ket\rho_{MRE}=U_{S\rightarrow
  E}\,\ket\sigma_{SMR}$ and $U_{S\rightarrow E}$ is an isometry from $S$ to $E$.

Then the bound on the work cost, including a smoothing parameter $\epsilon$, is
\begin{align}
  W &\geqslant \Hmax^\varepsilon(E|SM)_{\rho}\cdot kT\ln(2)\nonumber\\
  &= \Hmax^\varepsilon(E|M)_{\rho}\cdot kT\ln(2)\nonumber\\
  &= \Hmax^\varepsilon(S|M)_{\sigma}\cdot kT\ln(2)\ ,
\end{align}
where the first equality follows because $\rho$ is pure on $S$ and the second by
reversing the isometry $U$.
We can immediately conclude that, within our framework, any process that performs this
erasure has to cost at least $kT\ln(2)\,\Hmax^\varepsilon(S|M)_\sigma$ work.
Thus, the process proposed by del Rio \etal{} is optimal up to logarithmic
factors in $\varepsilon$. Note that if we take the memory $M$ to
be trivial \ie{} a pure state, then we are in the standard scenario of Landauer erasure on
a single system, 
and we have  $W\geqslant \Hmax^\varepsilon(S)$ which is achievable, recovering
the result of~\cite{Dahlsten2011NJP_inadequacy}.

\subsection{Coherent Preparation of a State on a System with a Memory. ``Reverse'' of a
  Logical Process.}%
One may also wonder which process is the ``reverse process'' of erasure with a quantum
memory. Specifically, starting off with a pure system $S$ and some state $\rho_M$ on a
memory $M$, one might ask how much work is needed to prepare a given bipartite state
$\sigma_{SM}$ on these systems.

The process as such is not clearly defined, as we have not specified which correlations
between the initial and final state on $M$ are to be preserved, or, equivalently, which
completely positive map $\mathcal{E}$ is to be applied for this preparation.

Let us first study the erasure mapping~\eqref{eq:appx-ProcessErasureMemory} a bit more
closely. The output state of the erasure including the reference system $R$ is given by
the state $\rho_{X'R}=\proj0_S\otimes\rho_{MR}$, where $\Hs_{X'}=\Hs_S\otimes\Hs_M$ is the
total output system.
As mentioned earlier, the joint state on $X'$ and $R$ may be interpreted as the
\emph{process matrix} of the operation $\mathcal{E}$ on $\sigma$: it can be thought of as
a joint probability distribution giving the probability that we had $\ket{k}$ at the input
and got $\ket{k'}$ at the output; also, $\rho_{X'}$ is the output state and $\rho_{R}$ is
the input state. This consideration gives us a natural way of \emph{reversing} any
process: a natural ``reverse'' process to the process $\rho_{X'R}$ is simply given by
swapping the two systems, \ie{} considering $R$ as the output and $X'$ as a purification
of the input. Let us return to the case of the erasure. First, consider the purification
of $\sigma_{SM}$ into a system $\Hs_R=\Hs_{R_S}\otimes\Hs_{R_M}$ explicitly as
\begin{align}
  \ket\sigma_{SMR_SR_M} = \sigma_{SM}^{1/2}\ket{\Phi}_{SM|R_SR_M}\ ,
\end{align}
where $\ket\Phi_{SM|R_SR_M} = \sum_{kl}\ket{k}_S\ket{k}_{R_S}\ket{l}_M\ket{l}_{R_M}$.
Now applying the erasure process on $SM$ gives us:
\begin{multline}
  \rho_{SMR_SR_M}
  = \mathcal{E}^{\mathrm{(erasure)}}_{SM\rightarrow SM} \left(
    \sigma_{SMR_SR_M} \right)\\
  = \proj0_S\otimes\tr_{S}\left(\sigma_{SMR_SR_M}\right)\ .
\end{multline}

It is thus natural, for the preparation scenario, to consider the process matrix
\begin{align}
  \rho^{\mathrm{(prep)}}_{SMR_SR_M}
  = \proj0_{R_S}\otimes\tr_{R_S}\left(\sigma_{SMR_SR_M}\right)\ .
\end{align}
(Recall that the input state is pure on $S$.)

This is obviously purified by a system $E$ which contains the traced out information,
\ie{} given an isometry $U_{R_S\rightarrow E}$,
\begin{align}
  \ket{\rho^{\mathrm{(prep)}}}_{SMR_SR_ME} = \ket0_{R_S}\otimes\left(
    U_{R_S\rightarrow E}\,\ket\sigma_{SMR_SR_M} \right)\ .
\end{align}

If we then calculate the minimal work cost of performing this process according to our
main result, we obtain
\begin{align}
  W^\mathrm{(prep)} &=
  kT\ln2\cdot\Hmax^\varepsilon\left(E|SM\right)_{\rho^\mathrm{(prep)}} \nonumber\\
  &= kT\ln2\cdot\Hmax^\varepsilon\left(R_S|SM\right)_\sigma \nonumber\\
  &= -kT\ln2\cdot\Hmin^\varepsilon\left(R_S|R_M\right)_\sigma \nonumber\\
  &= -kT\ln2\cdot\Hmin^\varepsilon\left(S|M\right)_\sigma\ .
\end{align}
(We have used the fact that $\sigma$ and $\rho$ are related by an isometry between $R_S$
and $E$, as well as the duality between min- and max-entropies.)

We notice that $kT\ln2\cdot\Hmin\left(S|M\right)_\sigma$ work can be extracted in the
reverse process of the original erasure process, which required
$kT\ln2\cdot\Hmax\left(S|M\right)_\sigma$ work. These values can be arbitrarily different;
this gap is expected as we require both processes to succeed with high probability. We
find that the gap is exactly the difference between the min- and the max-entropy, similarly
to the single-shot irreversibility between distillation rate and formation rate of
entangled pairs with LOCC operations~\cite{BBPS1996,Bennett1996PRL_purification,%
  Bennett1996PRA_MSEntglQECorr,Vidal2001_Irrev}.

\subsection{The Minimal Work Cost of a Quantum Measurement.}%
\label{sec:appx-MinimalCostQuantumMeasurement}

Quantum measurements are special cases of quantum processes, and so may also be plugged
into our main result. Note that first, we consider the measurement process to be given
access to a pure memory register to store the measurement result. We then consider the
minimal work cost of preparing the register in its pure state again for a future
measurement.

\paragraph{The Measurement Process and its Work Cost}%
Suppose that on the system $S$, in the state $\sigma$, we perform a
measurement described by a POVM $\{Q_k\}$. Each outcome, labeled by the index $k$, occurs
with probability $\tr\left(Q_k\sigma\right)$. The completely positive map associated with
this measurement is,
\begin{align}
  \label{eq:appx-MeasurementCPM}
  \mathcal{E}_{S\rightarrow S'C}\left(\sigma\right)
  = \sum_k \proj{k}_C\otimes \mathcal{E}_{S\rightarrow S'}^{(k)}\bigl(\sigma\bigr)\ ,
\end{align}
where $C$ is a classical register containing the outcome of the measurement (initially in
a pure state), and $\mathcal{E}_{S\rightarrow S'}^{(k)}\left(\cdot\right) = \sum_i
E_{i}^{(k)}\,\left(\cdot\right)\,{E_i^{(k)\,\dagger}}$ are trace-decreasing maps that map
$\sigma$ to its post-measurement state for the outcome $k$, which occurs with probability
$\tr\left(Q_k\sigma\right) = \tr\mathcal{E}_{S\rightarrow S'}^{(k)}\left(\sigma\right)$.
The Kraus operators $E_i^{(k)}$ are related to the POVM elements $Q_k$ by $\sum_i
E_i^{(k)\,\dagger}E_i^{(k)}=Q_k$.  Equation~\eqref{eq:appx-MeasurementCPM} simply expresses
that the output state of the measurement is a mixture of the possible post-measurement
states corresponding to the different outcomes $k$.  We emphasize that the register $C$
must start off in a pure state; if this is not the case (as in a purity resource
framework, for example), it should be initialized first, causing some work cost, before
performing the map $\mathcal{E}$.

We first need to calculate the Stinespring dilation of the process
$\mathcal{E}$, which is given by $\mathcal{E}_{S\rightarrow CS'}\left(\cdot\right)
= \tr_E V_{S\rightarrow ECS'}\left(\cdot\right) V^\dagger$, where the isometry
$V_{S\rightarrow ECS'}$ can be read off the operator-sum representation
of~\eqref{eq:appx-MeasurementCPM}, that is
\begin{align}
  \label{eq:appx-MeasurementCPMOperatorSum}
  \mathcal{E}_{S\rightarrow CS'} =
  \sum_{k,i} \left(\ket{k}_C\otimes E_i^{(k)}\right) \sigma
  \left(\bra{k}_C\otimes E_i^{(k)\,\dagger}\right)\ ,
\end{align}
as $V=\sum_{k,i} \ket{k,i}_E\otimes\ket{k}_C\otimes E_i^{(k)}$. This yields the
post-measurement state including $E$
\begin{align*}
  \rho_{ECS'} = V_{S\rightarrow ECS'} \;\sigma_S\; V^\dagger\ .
\end{align*}
For convenience, let $R$ be a purifying system for $\sigma_S$, i.e. let $\ket\sigma_{SR}$
such that $\tr_R\proj\sigma_{SR} = \sigma_S$. This allows us to write a full, pure,
post-measurement state $\ket\rho_{ECS'R}$ as
\begin{align}
  \label{eq:appx-MeasurementRhoECSpR}
  \ket\rho_{ECS'R} = \sum_{k\,i}\; \ket{k,i}_E\otimes\ket{k}_C\otimes
  \left( E_i^{(k)} \ket\sigma_{SR}\right)\ .
\end{align}

Our main result asserts that the minimal work cost of the
measurement~\eqref{eq:appx-MeasurementCPM} is simply given by the quantity,
\begin{align}
  \label{eq:appx-MeasurementHmaxApproxNotation}
  W_\text{meas,optimal}^\varepsilon \approx kT\ln2\cdot\Hmax^\varepsilon\left(E|CS'\right)\ ,
\end{align}
where the entropy measure is evaluated for the state $\rho_{ECS'R}$ given
by~\eqref{eq:appx-MeasurementRhoECSpR}. In the remainder of this section, all entropy measures
are implicitly evaluated on this state, unless indicated otherwise. The `$\approx$' symbol
recalls that one can only actually approximately achieve the given bound (see
Section~\ref{sec:appx-TightnessMainResult}). In the remainder of this section, when discussing
optimal work costs, we will only consider the value of our bound; it is understood that a
successful implementation is possible at a work cost close to the discussed bound in the
sense of Section~\ref{sec:appx-TightnessMainResult}. It is also implied that all work costs are
smoothed with a small but finite $\varepsilon$ parameter. In this spirit, denote the value
of the right-hand side of~\eqref{eq:appx-MeasurementHmaxApproxNotation} by
\begin{align}
  \label{eq:appx-MeasurementHmax}
  W_\text{meas} = kT\ln2\cdot\Hmax^\varepsilon\left(E|CS'\right)\ .
\end{align}

As we will see from some simple examples, the quantity~\eqref{eq:appx-MeasurementCPM} in its
most general form as presented may take any value, from a work cost to a work
yield. However, we will consider an important class of measurements: those for which
the collapse operators $\mathcal{E}_k$ don't themselves need work. In general, we know
that those processes that don't need work are sub-unital, \ie{} they satisfy
$\mathcal{E}^{(k)}\left(\Ident\right)\leqslant\Ident$. This is for example the 
case for projective measurements, or more generally if the $\mathcal{E}^{(k)}$'s only have
a single Kraus operator. We also note that any general measurement in the
form~\eqref{eq:appx-MeasurementCPM} can be written as a combination of a measurement with
collapse superoperators that have each a single Kraus operator, followed by a partial
erasure on the memory register $C$.

\begin{prop}
  Let $\mathcal{E}_{S\rightarrow CS'}$ be a measurement process of the
  form~\eqref{eq:appx-MeasurementCPM}, and assume that for all $k$,
  $\mathcal{E}^{(k)}\left(\Ident\right)\leqslant\Ident$. Then
  $W_\text{meas}$ as defined by~\eqref{eq:appx-MeasurementHmax}, satisfies
  $W_\text{meas}\leqslant 0$.
\end{prop}
\begin{proof}
  Instead of proving that the entropic quantity~\eqref{eq:appx-MeasurementHmax} is negative, we
  will show that the full measurement process $\mathcal{E}$ itself is a sub-unital
  superoperator, which we know from our framework costs no work. This is straightforward
  to see:
  \begin{multline*}
    \mathcal{E}_{S\rightarrow CS'}\left(\Ident_S\right)
    = \sum_k \proj{k}_C\otimes\mathcal{E}^{(k)}_{S\rightarrow S'}\left(\Ident_S\right) \\
    \leqslant \sum_k \proj{k}_C\otimes\Ident_{S'} \leqslant \Ident_{CS'}\ .
    \tag*{\qedhere}
  \end{multline*}
\end{proof}

\paragraph{Resetting the Memory Register Containing the Measurement Outcome}%
Let us now consider the task of resetting $C$ to a pure state, after having performed the
measurement process above.
This resetting can obviously be performed directly, with a cost given by Landauer's
principle as $\Hmax^\varepsilon\left(C\right)_\rho$, which in turn depends on the 
number of possible outcomes the measurement had (if $\varepsilon > 0$, we would only
consider the measurements that are not extremely unlikely). This procedure, however, is not optimal
if we are allowed access for example the information contained in the post-measurement
state on $S'$. Indeed, in the latter case, we may use the system $S'$ as a memory as
discussed in Sec.~\ref{sec:appx-ErasureQuantumSystemWithMemory}, and the optimal work cost is
then
\begin{align}
  \label{eq:appx-EraseMeasurementResultWorkCost}
  W_{\mathrm{reset}\ C|S'} = kT\ln2\cdot\Hmax^\varepsilon\left(C|S'\right)_\rho\ .
\end{align}
This work cost is always positive, or at best, zero, because $\rho_{CS'}$ is
classical-quantum (c-q).

We will see that this work cost may be both less and larger than $W_\mathrm{meas}$ with some
examples. Of course, this does not constitute a violation of the second law, as we will
discuss.

Additionally, we could imagine a scenario where we have kept a purification of the input
state $\ket\sigma_{SR}$ on an ancilla system $R$, in order to ``remember''  the
state of the initial system $S$. We may then of course use this system
also to reduce the work cost of erasing $C$, the latter being
\begin{align*}
  W_{\mathrm{reset}\ C|R} = kT\ln2\cdot\Hmax^\varepsilon\left(C|R\right)\ .
\end{align*}

It turns out that, if the collapse operators $\mathcal{E}^{(k)}$ all have a single Kraus
operator, this work cost is always greater than the work yield of performing the
measurement $\left(-W_\mathrm{meas}\right)$, and that the difference between both is
precisely the difference between the max and min-entropy of the system $C$ conditioned on
$R$.

\begin{prop}
  Let $\mathcal{E}_{S\rightarrow CS'}$ be a measurement process of the
  form~\eqref{eq:appx-MeasurementCPM}, with for all $k$, $\mathcal{E}^{(k)}\left(\cdot\right) =
  E_k\left(\cdot\right)E_k^\dagger$. Let $\rho_{ECS'R}$ be defined
  as in~\eqref{eq:appx-MeasurementRhoECSpR}. Then
  \begin{multline*}
    W_\mathrm{meas} + W_{\mathrm{reset}\ C|R}\\
    = kT\ln2\cdot\left[-\Hmin^\varepsilon\left(C|R\right)
      + \Hmax^\varepsilon\left(C|R\right)\right]
  \end{multline*}

  Additionally, if $\varepsilon$ is not too large (such that
  $\Hmax^\varepsilon\geqslant\Hmin^\varepsilon$~\cite{Vitanov2013_chainrules}), this
  expression is always positive,
  \begin{align*}
    W_\mathrm{meas} + W_{\mathrm{reset}\ C|R} \geqslant 0\ .
  \end{align*}
\end{prop}
\begin{proof}
  First notice that the state $\rho_{ECS'R}$ takes the form
  \begin{align*}
    \ket\rho_{ECS'R} = \sum_k \ket{k}_E\ket{k}_C\left(E_k\ket\sigma_{SR}\right)\ ,
  \end{align*}
  and in particular, $\rho$ is invariant under interchange of $E$ and $C$ systems.
  Then, using duality of the smooth entropies~\cite{Tomamichel2010IEEE_Duality}
  (see Section~\ref{sec:appx-SmoothEntropies}), we have
  $\Hmax^\varepsilon\left(E|CS'\right) = -\Hmin^\varepsilon\left(E|R\right)
  = -\Hmin^\varepsilon\left(C|R\right)$ and thus
  $W_\mathrm{meas} = -kT\ln2\cdot\Hmin^\varepsilon\left(C|R\right)$. Then recall that
  $W_{\mathrm{reset}\ C|R} = kT\ln2\cdot\Hmax^\varepsilon\left(C|R\right)$ and
  that the max-entropy is larger than the min-entropy for small $\varepsilon$.
\end{proof}

The final (pure) state on $E$, $C$, $S'$ and $R$, which is the output of the Stinespring
isometry $V$ applied on the $S$ system of $\ket\sigma_{SR}$, is still given by the
expression~\eqref{eq:appx-MeasurementRhoECSpR}.

\paragraph{Some Examples of Measurement Processes}%
Let us now focus on some examples of measurement processes, which are all special cases
of~\eqref{eq:appx-MeasurementCPM}. 
\begin{enumerate}[(I)]
\item \emph{Measurement in the computational basis $\ket0$, $\ket1$ of a single qubit in
    a maximally mixed state $\Ident_2/2$.} The measurement process then simply yields the
  output state
  \begin{align*}
    \rho_{CS'} &= \frac12\,\proj{0}_C\otimes\proj{0}_{S'}
    + \frac12\,\proj{1}_C\otimes\proj{1}_{S'}\ .
  \end{align*}
  The input state on $S$ is purified by a fully entangled state $\ket\phi_{SR}$ on
  $R$. The system $E$ also purifies the measurement process, and as given
  by~\eqref{eq:appx-MeasurementRhoECSpR},
  \begin{multline*}
    \ket\rho_{ECS'R} = \sum_k\, \ket{k}_E\ket{k}_C\,\left(\proj{k}_{S}\ket\phi_{SR}\right)
    \\
    = \frac1{\sqrt 2}\left[\ket0_C\ket0_E\ket{00}_{SR}
      + \ket1_C\ket1_E\ket{11}_{SR}\right]\ .
  \end{multline*}

  It is then evident that
  \begin{align*}
    W_\mathrm{meas} &= kT\ln2\cdot\Hmax^\varepsilon\left(E|CS'\right) = 0\ ;\\
    W_{\mathrm{reset}\ C|S'} &= kT\ln2\cdot\Hmax^\varepsilon\left(C|S'\right) = 0\ ;\\
    W_{\mathrm{reset}\ C|R} &= kT\ln2\cdot\Hmax^\varepsilon\left(C|R\right) = 0\ ,
  \end{align*}
  as the corresponding reduced states are all classically correlated.

\item \emph{Measurement of a trivial noisy POVM.}
  Consider a POVM in the extreme case where the
  state is left untouched, but a random outcome is generated according to a distribution
  $\{p_k\}$. The POVM effects are simply $Q_k=p_k\Ident$ and the post-measurement
  operators $\mathcal{E}^{(k)}\left(\sigma\right) = p_k\sigma$ are simply the
  identity superoperator weighted by the probability $p_k$.

  Intuitively, this should be no different than rolling a die, or more generally,
  generating a random outcome with a specific distribution, which is a process that can
  yield work. Indeed, based on the explicit expression of the final state
  \begin{align}
    \label{eq:appx-MeasurementNoisyPOVMRhoECSpR}
    \ket\rho_{ECS'R} =
    \Bigl( \sum_k \sqrt{p_k}\,\ket{k}_E\ket{k}_C\Bigr)\otimes\ket\sigma_{S'R}\ ,
  \end{align}
  we may express of the work cost of the measurement using some basic properties of the
  smooth entropies, presented in Section~\ref{sec:appx-SmoothEntropies}. Using the duality of
  the min- and max-entropies,
  \begin{multline*}
    W_\mathrm{meas}/(kT\ln2) = \Hmax^\varepsilon\left(E|CS'\right)
    = - \Hmin^\varepsilon\left(E|R\right) \\
    = - \Hmin^\varepsilon\left(E\right)
    = - \Hmin^\varepsilon\left(C\right)
    \leqslant \log\,\norm{\rho_C}_\infty \leqslant 0\ ,
  \end{multline*}
  because $R$ is not correlated to $E$, and $\rho$ is invariant under exchange of $E$ and
  $C$ (both these statements can be seen in~\eqref{eq:appx-MeasurementNoisyPOVMRhoECSpR}), and
  by definition the min-entropy of $C$ is given by
  $\Hmin\left(C\right) = -\log\,\norm{\rho_C}_\infty$. Also, smoothing the min-entropy can
  only increase the quantity by its definition~\eqref{eq:appx-DefHminRhoEpsilon}.

  One can also calculate, because $C$ is uncorrelated to both $S'$ and $R$,
  \begin{align*}
    W_{\mathrm{reset}\ C|S'} &= kT\ln2\cdot\Hmax^\varepsilon\left(C|S'\right)
    = kT\ln2\cdot\Hmax^\varepsilon\left(C\right)\ ; \\
    W_{\mathrm{reset}\ C|R} &= kT\ln2\cdot\Hmax^\varepsilon\left(C|R\right)
    = kT\ln2\cdot\Hmax^\varepsilon\left(C\right)\ .
  \end{align*}
  This means that the work we need to invest to reset $C$ is always larger than what we
  gain from generating the random outcome. In fact, the gap is precisely the difference
  between the max- and the min-entropy, which is the same kind of irreversibility that is
  observed between the single-shot entanglement distillation and formation cost between
  two parties~\cite{Hayden2006CommMP}.

\item \emph{Projective measurement of a pure superposition state.} One may think that
  intuitively, for the measurement to yield work, the POVM must be noisy. Surprisingly
  enough, this is not the case. Even projective measurements can yield work for specific
  input states. For example, consider the state $\ket\sigma_S = \ket+ :=
  \frac1{\sqrt 2}\left(\ket0 + \ket1\right)$. Here $R$ is a trivial system since $\sigma$
  is already pure. Now consider the usual projective measurement that measures $\sigma$ in
  the computational basis $\ket0$, $\ket1$. The final state is
  \begin{align*}
    \ket\rho_{ECS'} = \frac1{\sqrt 2}\left(\ket{0}_{E}\ket{0}_{C}\ket0_{S'}
      + \ket{1}_{E}\ket{1}_{C}\ket1_{S'}\right)\ .
  \end{align*}

  We then evidently have
  \begin{align*}
    W_\mathrm{meas} &= kT\ln2\cdot\Hmax^\varepsilon\left(E|CS'\right) = -kT\ln2\ ;\\
    W_{\mathrm{reset}\ C|S'} &= kT\ln2\cdot\Hmax^\varepsilon\left(C|S'\right) = 0\ ;\\
    W_{\mathrm{reset}\ C|R} &= kT\ln2\cdot\Hmax^\varepsilon\left(C|R\right)\\
    &\quad= kT\ln2\cdot\Hmax^\varepsilon\left(C\right) = kT\ln2\ .
  \end{align*}

  We conclude that it is possible to extract one bit of work while performing the
  measurement, and that resetting the memory register can be done at no work cost using
  $S'$ but needs one bit of work if we use the (trivial) reference system $R$.

  Note that resetting the measurement register $C$ using $S'$ costs no work. This is not
  in violation of the second law of thermodynamics: we have not returned the
  post-measurement state back to the initial state, but rather we have consumed its
  purity.

\item \emph{Measurement with erasure collapse operators.} It was noted above that if the
  collapse operators $\mathcal{E}^{(k)}$ were themselves maps that cost work, \eg{}
  erasure channels, then the measurement would also possibly cost work. It is sufficient
  to consider the following extreme example: take a single-outcome measurement, \ie{} a
  trivial measurement, with a the single collapse operator
  $\mathcal{E}_{k=0}\left(\cdot\right)=\tr\left(\cdot\right)\proj0$ being an erasure
  channel. Obviously this operation has to cost work: performing operation
  $\mathcal{E}_{S\rightarrow CS'}$ is exactly the same as performing just the erasure
  $\mathcal{E}_{k=0}$, which costs work according to our main result (which is of course
  also in line with Landauer's principle).

\item \emph{Information Gain of a Measurement.} Existing literature~\cite{Winter2004CMP,%
    Wilde2012JPA_simulating,Buscemi2008PRL_global,Berta2014TIT_gain} has studied
  and identified the amount of information that a quantum measurement provides about a
  system being measured. With the notation above, the information gain in the asymptotic,
  \iid{} regime is defined in~\cite{Buscemi2008PRL_global} as
  \begin{align}
    \label{eq:appx-MeasurementInfGainIID}
    \iota\left(\sigma_S, \mathcal{E}\right) = I\left(R:C\right)_\rho
    := \HH\left(R\right)-\HH\left(R|C\right)\ .
  \end{align}
  In our framework, information contained in a quantum system is represented by how much
  work we need in order to erase that system. Bearing this in mind, the natural way of
  defining the amount of information gained about the system using the measurement is then
  the difference in work costs of erasing $S$ \emph{before} and \emph{after} the
  measurement. Since $S$ was consumed by the measurement, this statement doesn't fully
  make sense, so we will rather consider erasing the system $R$ instead, which is a
  purification of $S$. Our take at the information gain of the measurement is then
  \begin{align*}
    \iota' = \Hmax^\varepsilon\left(R\right) - \Hmax^\varepsilon\left(R|C\right)\ .
  \end{align*}
  Notice that in the \iid{} regime where the entropies converge to the von Neumann
  entropy, this definition coincides with the previous
  one~\eqref{eq:appx-MeasurementInfGainIID}.

\end{enumerate}

\subsection{State Transformation while Decoupling from the Reference System.}%
\label{sec:appx-statetransfdecpl}
Let's return to another special case that we can derive as a corollary from our main
result. Consider the process that erases its input and prepares the required output
independently. This would occur if we required the output state to be completely
uncorrelated to the reference system $R$: $\rho_{X'R}=\rho_{X'}\otimes\rho_R$. This
corresponds to a replacement map. Any third party $R$ that would have been correlated to
the input is now completely uncorrelated to the output.

Again, we may simply apply our main result with the additional condition that
$\rho_{X'R}=\rho_{X'}\otimes\rho_R$.
In this case, the purification of $\rho_{X'R}$, $\rho_{X'RE}$, takes a special
form due to the tensor product structure, with the $E$ system split into two $E_R$ and
$E_{X'}$ systems ($E=E_R\otimes E_{X'}$),
\begin{align}
  \ket{\rho}_{X'RE} = \ket{\psi}_{X'E_{X'}}\otimes\ket{\phi}_{RE_R}\ ,
\end{align}
where $\ket{\psi}_{X'E_{X'}}$ and $\ket\phi_{RE_R}$ are purifications of $\rho_{X'}$ and
$\rho_R$, respectively.

The (smooth) lower bound on the minimal work cost $W$, given by our main result,
then reads
\begin{multline*}
  W \geqslant kT\ln2\cdot\Hmax^{\varepsilon}\left(E|X'\right)_{\rho}\\
  = kT\ln2\cdot\Hmax^{\varepsilon}\left(E_R\right)_{\ket\phi}
  + kT\ln2\cdot\Hmax^{\varepsilon}\left(E_{X'}|X'\right)_{\ket\psi} .
\end{multline*}
Now, the spectrum of $\rho_{E_R}$ is exactly the same as the spectrum
of $\rho_{R}$ by the Schmidt decomposition of $\ket\phi$. This in turn has the same
spectrum as $\sigma_X$ also by the Schmidt decomposition of $\sigma_{XR}$ and because
$\rho_R=\sigma_R$. It follows that
$\Hmax^{\varepsilon}(E_R)_{\rho}=\Hmax^{\varepsilon}(X)_\sigma$.
Also, by duality of the min- and max-entropies, we
have $\Hmax^{\varepsilon}\left(E_{X'}|X'\right)_{\ket\psi} =
-\Hmin^{\varepsilon}\left(E_{X'}\right)_\rho =
-\Hmin^{\varepsilon}\left(X'\right)_{\rho}$. In consequence,
\begin{align}
  W \geqslant kT\ln2\cdot\bigl[\Hmax^{\varepsilon}\left(X\right)_\sigma
    - \Hmin^{\varepsilon}\left(X\right)_{\rho}\bigr]\ .
\end{align}
That is, to transform a state $\sigma$ to $\rho$ while completely decorrelating $\rho$
from the input, then one has to erase $\sigma$ to a pure state (at cost
$\Hmax^{\varepsilon}\left(X\right)_\sigma$), and then prepare $\rho$ (extracting work
$\Hmin^{\varepsilon}\left(X'\right)_{\rho}$).

\subsection{Example: Erasing Part of a W State. Again, the Importance of Correlations
  Between the Input and the Output.}%
Consider the W state on a system $S$, a memory $M$ and a reference system $R$ given by
\begin{align}
  \ket{W}_{SMR} = \frac1{\sqrt 3}\left[\ket{001}+\ket{010}+\ket{100}\right]_{SMR}\ .
\end{align}
The reduced states on $SM$ and $M$ are respectively given by
$\sigma_{SM}=\frac13\proj{00}+\frac23\proj{\Psi^+}$ and
$\sigma_M=\frac23\proj{0}+\frac13\proj{1}$, where
$\ket{\Psi^+}=\frac1{\sqrt 2}\left(\ket{01}+\ket{10}\right)$. By symmetry of the W state,
the reduced state on any two or one qubit(s) have the same form.

By actions on $S$ and $M$, we would like to erase $S$, leading to the final state on $S$
and $M$ given by $\rho_{SM}=\proj{0}\otimes\sigma_M$. Let us consider two processes that
achieve this goal: the first one will preserve correlations with $R$ but will cost work,
the second will not cost work but will modify those correlations.

We may directly apply the special case above concerning the erasure of a system
conditioned on a memory: the fundamental work cost of such an erasure, if one preserves
correlations with a reference system $R$, is given by $\Hzero\left(S|M\right)_\sigma$. In
this case we have $\Hzero\left(S|M\right)_\sigma = \log\frac23\approx 0.59$ (which we
calculate below) and thus this process must cost at least this amount of work.  Because of
the small system size, we may not assert the achievability of this erasure at this work
cost (the error terms discussed in Section~\ref{sec:appx-TightnessMainResult} become
overwhelming). However we can safely exclude the possibility of performing this operation
at no work cost, a statement which suffices for our purposes here.

Observe now that both $\sigma_{SM}$ and $\sigma_M$ have the same spectrum
$\{\nicefrac23,\nicefrac13\}$. This means that there exists a unitary $U$ that performs
the erasure simply as $\proj{0}\otimes\sigma_M = U\sigma_{SM}U^\dagger$, and this unitary
process by definition does not cost any work. Note though that the correlations with $R$
are not preserved. Indeed, the unitary sends $\ket{00}$ to $\ket{01}$ and $\ket{\Psi^+}$
to $\ket{00}$, so one explicitly calculates that the state after the process is given by
$\rho_{SMR} = U\sigma_{SMR}U^\dagger =
\frac1{\sqrt3}\left[\ket{011}+\sqrt2\ket{000}\right] =
\ket0\otimes\frac1{\sqrt3}\left[\ket{11}+\sqrt2\ket{00}\right]$. We notice that the
reduced state on $M$ and $R$ is now pure and differs from initial one, given by
$\sigma_{MR}=\frac13\proj{00}+\frac23\proj{\Psi^+}$.

As before, this is an example where one can transform the input state into the output
state at no work cost a priori, but if correlations are to be preserved between the memory
and a reference system or, equivalently, if the exact erasure
process~\eqref{eq:appx-ProcessErasureMemory} is to be performed, then a physical
implementation of this operation would require work.

It remains to calculate $\Hzero\left(S|M\right)_\sigma$. Written out explicitly in the
basis $\{\ket0,\ket1\}$, the state $\sigma_{SM}$ and the projector $\Pi_{SM}$ on its
support take the form
\begin{align*}
  \sigma_{SM} = \begin{pmatrix}
    \nicefrac13 & & & \\
    & \nicefrac13 & \nicefrac13 & \\
    & \nicefrac13 & \nicefrac13 & \\
    & & & 0
  \end{pmatrix}\ ;
  \quad
  \Pi_{SM} = \begin{pmatrix}
    1 & & & \\
    & \nicefrac12 & \nicefrac12 & \\
    & \nicefrac12 & \nicefrac12 & \\
    & & & 0
  \end{pmatrix}\ .
\end{align*}
(Empty entries are zero.) We then see that
\begin{align*}
  \tr_S\Pi_{SM} = \begin{pmatrix}
    \nicefrac32 & \\
    & \nicefrac12
  \end{pmatrix}\ ,
\end{align*}
such that
\begin{align*}
  \Hzero\left(S|M\right)_\sigma = \log\,\norm{\tr_S \Pi_{SM} }_\infty = \log\frac32\ .
\end{align*}


\section{Alternative Proof Using Lambda-Majorization.}
\label{sec:appx-alt-proof-lambda-maj}

For completeness, we provide an alternative proof of our main result, based on techniques
of majorization and semidefinite programming. This proof (historically, the original one)
ventures via the study of possible state transitions, regardless of the logical process,
but then imposes that the resulting logical process be the one required.

\subsection{The Framework. Work cost or yield as generating or absorbing randomness.}

\paragraph{Framework}%
Consider a quantum mechanical system $X$ in an initial state described by the density
operator $\sigma$.  Our task is to bring the system $X$ to another state $\rho$, while
attempting to maximize some kind of notion of ``extracted'' work in the process.

We postulate a restricted set of operations as possible physical processes which we may
carry out. Throughout this paper we assume that the system starts and ends with a fully
degenerate Hamlitonian upon each application of an allowed operation. There is no further
restriction, however, on how each of the allowed operations themselves are
implemented---they might require a time-dependent Hamiltonian for example.

We first postulate two basic operations of thermodynamical nature, involving a heat bath
at temperature $T$: the erasure of a single qubit to a pure state at $kT\ln(2)$ work cost,
and the corresponding reverse process which extracts $kT\ln(2)$ work by transforming a
pure state into a fully mixed state. Here $k$ is the Boltzmann constant.
These operations are motivated by the variety of explicit physical
thermodynamical frameworks in which they can be performed, for example
using Szilard
boxes~\cite{Szilard1929ZeitschriftFuerPhysik,Dahlsten2011NJP_inadequacy} or by isothermally
manipulating energy levels of
Hamiltonians~\cite{Alicki2004_hamiltonian,delRio2011Nature,Aberg2013_worklike}. 
Crucially, we
assume the second law of thermodynamics, and require that there exist no operation that
would allow us to form a cycle for which the net effect would be the extraction of
work. This justifies that no other work extraction procedure can yield more work than
$kT\ln(2)$ from a pure qubit, or else a cycle with net work gain could be formed by
appending an erasure process, itself only costing $kT\ln(2)$.

Apart from this constraint on the set of allowed operations, it is natural to also allow
usual quantum information processing. Since our Hamiltonians are degenerate, we can allow
all global unitaries and they cost no work.  We do not need to use the fact that these
unitaries are implementable by a device operating in contact with a heat bath, since
expanding the class of allowable operations actually strengthens the bound we derive.  In
practice, one has very crude local control over the operations, and the acting agent does
not know which unitary is being implemented, however, this is actually not an obstacle for
implementation~\cite{schulman1999molecular,Brandao2013_resource}.  In addition to
unitaries, we will allow pure ancillas to be added to the system, which permits more
general computation. Crucially, ancillas will have to be exactly restored to their initial
pure state, so that it is not possible to ``hide'' a work cost in an ancilla that was left
in a mixed state.

The following framework is motivated by the above
considerations. The processes we allow are (finite) combinations of the following
elementary operations:
\begin{enumerate}[(a)]\setlength{\itemsep}{0pt}
\item Bring $n$ qubits (of the system $X$ or an ancilla $A$) from any state to a pure state (`erasure') at
  cost $n\,kT\ln 2$ work;
\item Bring $n$ qubits (of the system $X$ or an ancilla $A$) from a pure state to a fully mixed state while
  extracting $n\,kT\ln 2$ work;
\item Add and remove ancillas in a pure state at no work cost, as long as all the ancillas have been
  restored to their initial pure state when they are restored;
\item Perform arbitrary unitaries (over $X$ and any added ancillas) at no work cost.
\end{enumerate}

Operations (a) and (b) are those of thermodynamical nature, and may be
carried out in a wide range of existing frameworks as mentioned above. One may view these
operations as \emph{defining} a quantity which we call ``work''.
We note that these operations can be performed quasi-statically in a thermodynamically
reversible fashion (as long as operation (a) acts on a fully mixed state, which in fact
will turn out to be sufficient for our purposes).

On the other hand, operations (c) and (d) are purely information-theoretical. They allow us
to perform any quantum information processing circuit, since we allow pure ancillas to be
added. However, there is the condition that ``randomness'' may not be disposed of for
free, namely that ancillas have to be restored to their initial pure states at the end of
the process.

We emphasize that these operations are \emph{allowed} operations, but they are not
necessarily always optimal. For example, a pure state need not require $n\,kT\ln2$ for its
erasure, as given by operation~(a). However, any attempt to allow operation~(a) for any
state (or even just for a mixed state) at any lower cost than $n\,kT\ln2$ would result in
a macroscopic violation of the second law of thermodynamics.

\paragraph{Lambda-Majorization}%
We will now provide a simple mathematical characterization of all operations allowed
in our framework.

First, note that the operations (a)--(d) allow the use of so-called \emph{noisy
  operations}~\cite{Horodecki2003PRA_NoisyOps}, which correspond to adding an ancilla
system $N$ in a fully mixed state, performing a joint unitary, and removing the ancilla.
Specifically, a noisy operation is composed in our framework of first an operation of type
(c) (adding a pure ancilla of $n$ qubits), followed by an operation of type (b)
(extracting $n\,kT\ln 2$ work from the ancilla making it fully mixed), then one of type
(d) (performing the necessary unitary to carry out the noisy operation), and finally an
operation of type (a) (erasing the ancilla back to its pure state at a work cost $n\,kT\ln
2$). (It can be assumed without loss of generality that the ancilla is
left in a fully mixed state after the noisy operation; indeed, this is the case for the
construction of the noisy operation given by (Ref.~\cite{Horodecki2003PRA_NoisyOps}),
which is capable of performing an equivalent transformation to any other noisy
operation.)  The total process has a work balance of zero. This means that we may thus
carry out noisy operations for free within our framework and use them as building blocks
for more complex processes. In the following, we deal implicitly with the ancilla $N$ and
it should not be confused with further ancillas that will be added.

The following result by Horodecki \etal~\cite{Horodecki2003PRA_NoisyOps} relates noisy
operations to the mathematical notion of
majorization~\cite{HardyLittlewoodPolyaInequalities1952,BookBhatiaMatrixAnalysis1997,%
BookHornJohnsonMatrixAnalysis1985}.

\begin{thmheadingit}{Noisy Operations and Majorization}
  The transition on system $X$ from state $\sigma$ to
  state $\rho$ is possible by noisy operation if and only if $\sigma\succ\rho$.
\end{thmheadingit}

Majorization between two (normalized) states $\sigma\succ\rho$ captures the fact that
$\rho$ is ``more mixed'' than $\sigma$, or that the eigenvalues of $\rho$ can be written
as a ``mixture'' of the eigenvalues of $\sigma$.
Formally, majorization can be characterized by the existence of a unital,
trace-preserving completely positive map that brings $\sigma$ to
$\rho$~\cite{Uhlmann1970_shannon,Uhlmann1971_Dichtematrizen,Uhlmann1972_EdDMI,%
  Uhlmann1973_EdDMII}.
A map $\mathcal{E}$ is \emph{trace-preserving} if
$\mathcal{E}^\dagger\left(\Ident\right) = \Ident$ and \emph{unital} if
$\mathcal{E}\left(\Ident\right) = \Ident$.
\begin{prop}
  \label{prop:appx-majorizationChannelUnitalTracePreserving}
  Two positive matrices $\sigma$ and $\rho$ satisfy $\sigma\succ\rho$ if and only if there exists
  a trace-preserving, unital, completely positive map $\mathcal{E}$
  satisfying $\mathcal{E}\left(\sigma\right)=\rho$.
\end{prop}

The notion of majorization is discussed in more detail in
Section~\ref{sec:appx-appendixFormalLambdaMajorization}.

We will now provide some background insight for our new concept of lambda-majorization,
which is a generalization of majorization inspired by other majorization
variants~\cite{Joe1990JMAA,Ruch1980JMAA,Ruch1978JCP,Mead1977JCP_mixing,Horodecki2013_ThermoMaj,%
  Egloff2012arXiv}. The idea is to characterize ``how well'' a state $\sigma$ majorizes a
state $\rho$. Suppose that we have a system $X$ in state $\sigma_X$ and we want to bring
it to the state $\rho_X$, where $\sigma_X\succ\rho_X$. In this case, one can simply carry
out a noisy operation as described above.  Suppose now that we have an ancilla $A$ that is
in a fully mixed state, $\Mixedi{A}$, and suppose that we are fortunate enough for
$\sigma_X\otimes\Mixedi{A} \succ \rho_X\otimes \proj0 _A$ to also hold (for some pure
state $\ket0_A$ on $A$). Then by applying a joint noisy operation on both systems, this
would correspond to actually erasing the system $A$ ``for free'' during the transition
$\sigma\rightarrow\rho$. We could then say that the randomness of the ancilla $A$ was
``transferred'' into system $X$. We will view this type of transition as \emph{work
  extraction} on system $X$ during a transition $\sigma_X\rightarrow\rho_X$. Indeed, work
can be extracted in an initial stage of the process by starting with a pure ancilla and
making it maximally mixed; the operation described above costs no work and the ancilla can
then be restored in its pure final state.

In another situation, it might be that $\sigma_X\nsucc\rho_X$. However, in that case, for
a large enough ancilla $A$ the majorization
$\sigma_X\otimes\proj0_A\succ\rho_X\otimes\Mixedi{A}$ will hold. The corresponding noisy
operation then leaves us with a mixed ancilla that started off pure and thus requires work
to restore; we will view such a transition on system $X$ as \emph{costing work}.

Such operations can be performed within our framework,
using operations (a)--(d). In particular, the relation to work is given by elementary
erasure and work extraction (operations (a) and (b)) applied to the ancilla $A$ after the
transition to restore it to its initial state.

\begin{figure}
  \centering
  \includegraphics[width=87mm]{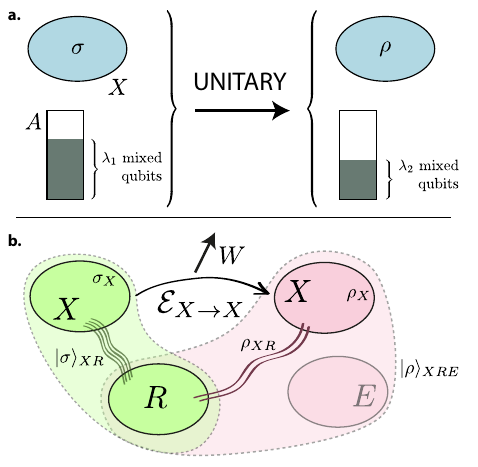}
  \caption{\textbf{a.}
    Lambda-Majorization corresponds to absorbing a certain amount of randomness
    from an ancilla during a unitary operation. The system $X$ starts in state $\sigma$,
    and the ancilla $A$ in a state with $\lambda_1$ fully mixed qubits with the remaining
    qubits pure. The goal is to devise a global unitary that will bring the system $X$ to
    the state $\rho$, while leaving the least possible number $\lambda_2$ of fully mixed
    qubits in $A$. The difference $\lambda=\lambda_1-\lambda_2$, is the work extracted by
    the process; if the value is negative, it corresponds to a work cost.
    \textbf{b.} Our main result gives a fundamental lower bound on the work cost $W$ of a process
    transforming a state $\sigma_X$ (purified by a fictitious $\ket\sigma_{XR}$) into a new state
    $\rho_{X'R}$ obtained by applying a process $\mathcal{E}_{X\rightarrow X}$. We optimize
    the work cost of lambda-majorization operations that perform the process
    $\mathcal{E}$. The lower bound to the work cost is then given by the entropy of the
    information $E$ that the process $\mathcal{E}$ has to discard (which purifies the
    state $\rho_{X'R}$), as measured by the R\'enyi-zero conditional entropy
    $\Hzero\left(E|X'\right)_{\rho}$.
  }
  \label{fig:appx-LambdaMajorizationSystemsMainResult}
\end{figure}

In general, the ancilla $A$ may start with $\lambda_1$ mixed qubits and end up with
$\lambda_2$ mixed qubits after a noisy operation; we consider in this case to have
extracted $\left(\lambda_1-\lambda_2\right)kT\ln(2)$ amount of work. This situation is
depicted in \reffigurename~\ref{fig:appx-LambdaMajorizationSystemsMainResult}\textbf{a}.
Both considerations above about work cost and work extraction are encompassed, simply
because we count the difference in the ``amount of randomness'' present in the ancilla
before and after the process. This is the idea behind the concept of lambda-majorization,
whose definition we can now state.

\begin{thmheadingit}{Lambda-Majorization}
  For two density operators $\sigma_X$, $\rho_Y$
  on two systems $X$ and $Y$,
  we will say that $\sigma_X$ \emph{$\lambda$-majorizes} $\rho_Y$, denoted by
  $\sigma_X\lambdamaj{\lambda}\rho_Y$,
  if there exists a (large enough) ancilla system $A$, as well as
  $\lambda_1,\lambda_2\geqslant 0$ with $\lambda=\lambda_1-\lambda_2$, such that
  \begin{align*}
    2^{-\lambda_1}\Ident_{2^{\lambda_1}}\otimes\sigma_X
    \succ 2^{-\lambda_2}\Ident_{2^{\lambda_2}}\otimes\rho_X\ ,
  \end{align*}
  where $2^{-\lambda_1}\Ident_{2^{\lambda_1}}$ and $2^{-\lambda_2}\Ident_{2^{\lambda_2}}$
  are fully mixed states on $\lambda_1$ (respectively $\lambda_2$) qubits of $A$, and
  where the remaining qubits of $A$ in each case are pure.%
\end{thmheadingit}

An expression for ``by how much'' a state majorizes another was originally introduced 
in~\cite{Egloff2010MasterThesis} and used in~\cite{Egloff2012arXiv}, in the context of
work extraction games from Szilard boxes. Their measure, the ``relative mixedness''
between $\sigma$ and $\rho$, corresponds to the optimal $\lambda$ such that
$\sigma\lambdamaj\lambda\rho$.

Lambda-majorization captures all the possible processes that are allowed in our
framework. Indeed, if $\sigma\lambdamaj\lambda\rho$, then one has
$2^{-\lambda_1}\Ident_{2^{\lambda_1}}\otimes\sigma\succ
2^{-\lambda_2}\Ident_{2^{\lambda_2}}\otimes\rho$ for some $\lambda_1,\lambda_2$ with
$\lambda=\lambda_1-\lambda_2$. Hence, there exists a noisy operation
(itself a combination of operations (a)--(d) with zero total work cost) that performs
the transition from $2^{-\lambda_1}\Ident_{2^{\lambda_1}}\otimes\sigma$ to
$2^{-\lambda_2}\Ident_{2^{\lambda_2}}\otimes\rho$. The $\lambda_1$ mixed qubits that we
have appended to $\sigma$ can be created by appending a large pure ancilla (operation
(c)), and using operation (b) to extract $\lambda_1\,kT\ln(2)$ work from $\lambda_1$
qubits, rendering them fully mixed. At the end of the process, after the noisy operation,
we need to restore the ancilla in a pure state; we thus
need to erase (operation (a)) the remaining $\lambda_2$ qubits,
costing $\lambda_2\,kT\ln(2)$ work. The total extracted work is then
$\left(\lambda_1-\lambda_2\right)kT\ln(2)=\lambda\,kT\ln(2)$.
Conversely, each individual operation (a)--(d), individually transforming some
state $\sigma$ into a state $\rho$ and costing work $W$, implies the lambda-majorization
$\sigma\lambdamaj\lambda\rho$ with $W = -\lambda\,kT\ln(2)$. This is clear for operations
(c) and (d). For operations (a) and (b), this follows from results derived in
Section~\ref{sec:appx-FormalLambdaMajPropertiesForQuantumStates}. Furthermore, the composition
of lambda-majorizations is again a lambda-majorization
(Section~\ref{sec:appx-appendixFormalLambdaMajorization}).

The ancilla system above may be viewed as some kind of ``information battery'', as was
proposed by Bennett~\cite{Bennett1982IJTP_ThermodynOfComp} who suggested using a blank
memory tape as ``fuel'' to extract work. In this case, the ancilla can be used as a
storage of ``purity'' (or as a storage for ``mixedness'' or ``randomness'' which we would
like to get rid of), which is increased or decreased by processes like the ones suggested
above. Equivalently, a two level system, or {\it work bit} can be
used~\cite{Horodecki2013_ThermoMaj}.

It turns out that one can characterize lambda-majorization by the existence of a
completely positive map satisfying some special normalization conditions, analogously to
Proposition~\ref{prop:appx-majorizationChannelUnitalTracePreserving}.
\begin{prop}
  \label{prop:appx-LambdaMajChannel}
  Let $\lambda\in\mathbb{R}$.
  Two normalized density matrices $\sigma_X$ and $\rho_Y$ on two systems $X$ and $Y$ satisfy
  $\sigma_X\lambdamaj{\lambda}\rho_Y$
  if and only if there exists a completely positive map $\mathcal{T}_{X\rightarrow Y}$ satisfying
  $\rho_Y = \mathcal{T}_{X\rightarrow Y}\left(\sigma_X\right)$, such that
  $\mathcal{T}_{X\rightarrow Y}^\dagger\left(\Ident_X\right)\leqslant \Ident_Y$ and
  $\mathcal{T}_{X\rightarrow Y}\left(\Ident_X\right)\leqslant 2^{-\lambda} \Ident_Y$.
\end{prop}

A map $\mathcal{T}_{X\rightarrow Y}$ that satisfies the two last conditions will be referred to as
a \emph{lambda-majorization map}.

Furthermore, although the map $\mathcal{T}$ is not directly a physical mapping (it can
be, for example, trace-decreasing), it can always be viewed as part of a unital channel
$\bar{\mathcal{E}}$, in the sense that $\mathcal{T}$ can be obtained by projection onto
specific subspaces and tracing out the ancilla $A$ of the map $\bar{\mathcal{E}}$ (see
Section~\ref{sec:appx-LambdaMajInTermsOfChannels}). In turn, unital channels are a
(strict~\cite{Haagerup2011_factorization}) superset of the noisy operations. Recall that
our task is to find a lower bound on the work cost of all possible processes allowed in
our framework, which we will do by optimizing the work cost over all processes that
perform a given state transition. However, instead of considering only the unital channels
$\bar{\mathcal{E}}$ that are noisy operations, we will relax this last condition and
consider all unital maps $\bar{\mathcal{E}}$, and thus allow the optimization to range
over all $\mathcal{T}$ that satisfy the conditions of the above proposition. This will
make our lower bound even stronger, by showing that the lower bound still holds even if we
relax somewhat the assumptions in our framework.

\subsection{The Main Result.}

\subsubsection{Formulation and Proof Sketch.}

\paragraph{Formulation of the Main Result}%
We are now ready to derive our main result. Consider a system $X$ in the state
$\sigma_X$. This system can always be purified by a reference system, $R$, in a pure joint
state $\ket\sigma_{XR}$.

Allowing actions defined by our framework on $X$, we will study the transition of this
state to a state $\rho_{XR}$, by applying a process $\mathcal{T}_{X\rightarrow X'}$. The
systems are depicted in
\reffigurename~\ref{fig:appx-LambdaMajorizationSystemsMainResult}\textbf{b}.

The task we would like to solve is the following. Given ${\sigma}_{X}$ and a logical process
$\mathcal{E}_{X\rightarrow X'}$, and given a purification $\ket{\sigma}_{XR}$ of $\sigma_X$
and an output state $\rho_{X'R}=\mathcal{E}\left(\sigma_{XR}\right)$, we would like to find
the least amount of work $W$ one has to pay for any process in our framework that
implements the action of $\mathcal{E}$ on $\sigma$. As we have seen in the previous
section, we can formulate within our framework all possible processes as
lambda-majorizations, so our task is actually to find the best $\lambda$ such that
$\sigma_{X}\lambdamaj\lambda\rho_{X'}$, with the corresponding lambda-majorization map
$\mathcal{T}$ from Prop.~\ref{prop:appx-LambdaMajChannel} satisfying
$\mathcal{T}\left(\sigma_{XR}\right)=\rho_{X'R}$.

Our main result gives an upper bound on the optimal amount of work that can be extracted
by this transition, or equivalently, a lower bound on the minimum amount of work that will
have to be paid in order to perform the transition. The main result follows directly from
following technical proposition.

We are given an input state $\sigma_X$ and a process $\mathcal{E}_{X\rightarrow X'}$. Let
$\ket\sigma_{XR}$ be a purification of $\sigma_X$, and let
$\rho_{X'R}=\mathcal{E}_{X\rightarrow X'}\left(\sigma_{XR}\right)$. Let also $\rho_{X'RE}$ be
a purification of $\rho_{X'R}$ in a system $E$.

\begin{prop}
  \label{prop:appx-MainResultHzeroEnviron}
  The $\lambda$-majorization $\sigma_{X}\lambdamaj\lambda\rho_{X'}$ holds, with the
  corresponding map $\mathcal{T}_{X\rightarrow X'}$ from
  Prop~\ref{prop:appx-LambdaMajChannel} satisfying 
  $\mathcal{T}\left(\sigma_{XR}\right)=\rho_{X'R}$, if and only if
  $\lambda\leqslant -\Hzero\left(E|X'\right)_{\rho}$.
\end{prop}
\begin{thmheadingit}{Main Result}
  Any procedure in our framework acting on system $X$ that implements the map
  $\mathcal{E}$ when given input $\sigma_X$ (or equivalently, that brings the state
  $\sigma_{XR}$ to the state $\rho_{X'R}$) has a work cost $W$ which is at least
  \begin{align}
    \label{eq:appx-MainResultWorkCost}
    W \geqslant kT\ln(2)\cdot\Hzero\left(E|X'\right)_\rho\ .
  \end{align}
\end{thmheadingit}

In other words, \emph{the minimal work cost of a process $\mathcal{E}$
mapping $\sigma$ to $\rho$ is given by the amount of (information-theoretic) entropy
discarded, and thus dumped into the environment, conditioned on the output of the
computation}. This is 
precisely the quantitative generalization to correlated
quantum systems of the original Landauer's principle~\cite{Landauer1961_5392446Erasure}.

The Main Result follows from Prop.~\ref{prop:appx-MainResultHzeroEnviron} because, as we have
noted above, lambda-majorization is equivalent to our original framework of operations
(a)--(d).

It is worth noting that instead of specifying the map $\mathcal{E}$, we may also
simply specify the output state $\rho_{X'R}$, which completely determines the process (on
the support of $\sigma_X$) since it is the Choi-Jamio\l{}kowski state
corresponding to $\mathcal{E}$ rescaled by $\sigma_X$
($\rho_{X'R}=\mathcal{E}\left(\sigma_{XR}\right)$).
One can thus understand the input to the problem to actually be a bipartite state
$\rho_{X'R}$, such that $\rho_{X'}$ is the required output, $\rho_{R}$ is the input that
will be fed into the process, and any correlations between $X'$ and $R$ specify parts of
the output that we wish be preserved and not be modified, or thermalized, by the process.

We have kept above the notation $X'$ consistently to remember that we are talking about
the output of the computation on $X$. However, $X'$ could be a different system entirely.
It could even have a different dimension than $X$, however in this case there are some
clarifications needed: whenever the system dimension increases, pure ancillas have been
brought in and haven't necessarily been restored to their pure state since they are part
of the output; however this operation need not have cost any work (in contrast to other
noisy operations resource theories, where purity is
costly~\cite{Horodecki2003PRA_NoisyOps,Brandao2013_resource}). However, whenever the
system dimension decreases, then any ancilla that was removed had to be reset to a pure
state first before being disposed of, which may have cost work. In other words, we adhere
to the convention where purity can be brought in for free, but where disposing of
randomness is costly; this is equivalent to the other approach where purity is costly but
disposing of mixed states is done for free. Our choice is \emph{a priori} arbitrary but
possesses the advantage of well integrating into our mathematical framework with simple
mathematical descriptions in terms of subunital maps and weak sub-majorization (see
Section~\ref{sec:appx-appendixFormalLambdaMajorization}).

The full proof of Prop.~\ref{prop:appx-MainResultHzeroEnviron} is provided in
Section~\ref{sec:appx-DerivationMainResultSDP}. We provide the general idea of the proof in the
following.

\paragraph{Proof Sketch of the Main Result}%
The main idea of the proof is to write the optimization problem as a semidefinite program
for the variables $\alpha=2^{-\lambda}\geqslant 0$, $T_{XX'}\geqslant 0$ (the
Choi-Jamio\l{}kowski representation of $\mathcal{T}_{X\rightarrow X'}$). Let
$\left(\cdot\right)^{t_X}$ denote the partial transpose operation on $X$. Consider the
state transformation $\sigma\to\rho$. An upper bound on the
extracted work $\lambda$ in the lambda-majorization $\sigma\lambdamaj\lambda\rho$, while
ensuring that the map $\mathcal{T}$ from Prop.~\ref{prop:appx-LambdaMajChannel} performs the
same logical operation as $\mathcal{E}$, is given
by the following semidefinite program (see \cite{barvinok-convexity,watrousnotes} for a
introduction to SDPs in a style similar to what we use here.):

\begin{flushleft}
  \begin{subequations}
    \begin{SDP*}{\columnwidth}
      \SDPprimal
      \SDPmin \alpha \\
      \SDPst \\
      \mathcal{T}_{X\rightarrow X'}\left(\Ident_X\right) \leqslant \alpha \Ident_{X'} \\
      \mathcal{T}^\dagger_{X\leftarrow X'}\left(\Ident_{X'}\right) \leqslant \Ident_X
      \\
      \mathcal{T}_{X\rightarrow X'}(\sigma_{XR}) = \rho_{X'R}\ .
    \end{SDP*}
  \end{subequations}
\end{flushleft}
\vspace{3mm}

The optimal value $\alpha = 2^{\Hzero(E|X')_{\rho}}$ is achieved (see
Section~\ref{sec:appx-DerivationMainResultSDP}) by the completely positive map 
$\mathcal{T}_{X\rightarrow X'}=\tr_E\left[V_{X\rightarrow
    X'E}\,(\cdot)\,V^\dagger\right]$, 
where $V_{X\rightarrow X'E}$ is the partial isometry with minimal support relating
$\sigma_{XR}$ to $\rho_{X'ER}$ (both being purifications of the same $\sigma_R=\rho_R$).

While it is clear from the formulation of our problem that $\mathcal{T}$ is already
completely determined on the support of $\sigma_X$ (expressed by the condition
$\mathcal{T}\left(\sigma_{XR}\right) = \rho_{X'R}$), the optimization over $\mathcal{T}$ is
done in order to (at least formally) find the optimal action on the complement of the
support of $\sigma_X$.

Also, the formulation of a lambda-majorization problem as a
semidefinite program is a more general toolbox that could be used in the case where the
mapping is not completely determined and where arbitrary additional semidefinite
conditions can be imposed at will. For example, instead of fixing the process
with $\mathcal{T}\left(\sigma_{XR}\right)=\rho_{X'R}$, one may have instead required that
$\mathcal{T}\left(\sigma_X\right) = \rho_{X'}$ for given $\sigma_X$ and $\rho_{X'}$, not
specifying and optimizing over what happens to correlations between the input and the
output (or, equivalently, one could optimize over $\rho_{X'R}$ with fixed reductions
$\rho_{X'}$ and $\rho_R$). In that case, the semidefinite program can be used to obtain
bounds to the optimal value. This also implies that the ``relative mixedness''
introduced in~\cite{Egloff2012arXiv} can be formulated as a semidefinite program.
However it is not clear if the result in this case can be written in terms of an entropy
measure.

\subsubsection{Proof of the Main Result. Formulation as a Semidefinite Program.}
\label{sec:appx-DerivationMainResultSDP}

Let $\Hs_X$ be a quantum system in the state $\sigma_X$. Let $\Hs_R$
be an additional quantum system and let $\ket\sigma_{XR}$ be a purification of
$\sigma_X$.

Suppose we want to perform the computation $\mathcal{E}_{X\rightarrow X'}$ on system $X$,
bringing its initial state $\sigma_{XR}$ into a given state $\rho_{X'R}$ with a
lambda-majorization. Here $\rho_{X'R}$ is not necessarily pure; giving the joint state
with $R$ allows us to specify which correlations we want to preserve, equivalently
specifying the computation $\mathcal{E}$ on the support of $\sigma_X$. The task is then
the following.

\begin{thmheading}{Task}
  Find the best (maximal) $\lambda$, such that there exists a completely positive,
  $2^{-\lambda}$-subunital, trace-nonincreasing map $\mathcal{T}_{X\rightarrow X'}$
  satisfying $\mathcal{T}_{X\rightarrow X'}(\sigma_{XR}) = \rho_{X'R}$.
\end{thmheading}

In other words, we would like to find the trace nonincreasing map that satisfies
$\mathcal{T}_{X\rightarrow X'}(\sigma_{XR})=\rho_{X'R}$, that has the smallest possible
$\norm{\mathcal{T}_{X\rightarrow X'}\left(\Ident_X\right)}_\infty$.

This problem can be formulated as a semidefinite program in terms of the variables
$\alpha\geqslant 0$ (defined as $\alpha=2^{-\lambda}$) and $\mathcal{T}_{X\rightarrow X'}$
(through its Choi-Jamio\l{}kowski map $T_{XX'}\geqslant 0$), and the dual
variables $\omega_{X'}\geqslant 0$, $X_X\geqslant0$ and $Z_{X'R}=Z_{X'R}^\dagger$.

\begin{flushleft}
  \begin{subequations}
    \begin{SDP}{\columnwidth}
      \SDPprimal
      \SDPmin \alpha \\
      \SDPst \\
      \label{eq:appx-TXXpAlphaSubUnital}
      \tr_X\left[T_{XX'}\right] \leqslant \alpha \Ident_{X'}
      \SDPdualvar{\omega_{X'}} \\
      \label{eq:appx-TXXpTraceNonIncr}
      \tr_{X'}\left[T_{XX'}\right] \leqslant \Ident_X \SDPdualvar{X_{X}} \\
      \label{eq:appx-TXXpRhoToRhoHat}
      \tr_{X}\left[T_{XX'}\sigma_{XR}^{t_X}\right] = \rho_{X'R} \SDPdualvar{Z_{X'R}}
    \end{SDP}
  \end{subequations}\\[3mm]
\end{flushleft}\vspace{1mm}
\begin{flushleft}
  \begin{subequations}
    \begin{SDP}{\columnwidth}
      \SDPdual
      \SDPmax \\
      \nonumber \tr\left(Z_{X'R}\,\rho_{X'R}\right) - \tr X_{X} \\
      \SDPst \\
      \label{eq:appx-TXXpDualTrOmega}
      \tr\omega_{X'} \leqslant 1\\
      \label{eq:appx-TXXpDualWierdZ}
      \tr_R\left[ \sigma^{t_X}_{XR}\, Z_{X'R} \right] \leqslant \Ident_X\otimes\omega_{X'} +
      X_X\otimes\Ident_{X'}\ .
    \end{SDP}
  \end{subequations}
\end{flushleft}
\vspace{3mm}

Note that since the map does not act on $\sigma_R$, we must necessarily have
$\sigma_R=\rho_R$. Let $E$ be a system that purifies the output state as
$\rho_{X'RE}$. As two purifications with the same reduced state on $R$, the two states
$\sigma_{XR}$ and $\rho_{X'RE}$ must be related by an isometry $V_{X\rightarrow X'E}$ as
$\rho_{X'RE} = V_{X\rightarrow X'E}\,\sigma_{XR}\,V^\dagger$. Note that this is an
equivalent construction of the Stinespring dilation of the original computation
$\mathcal{E}_{X\rightarrow X'}$.
We can choose here
$V_{X\rightarrow X'E}$ to be a partial isometry such that $VV^\dagger=\hat\Pi_{X'E}$,
the projector on the support of $\rho_{X'E}$, and $V^\dagger V=\Pi_X$, the projector on
the support of $\sigma_X$.

Now, define $\mathcal{T}$ by its Stinespring dilation
\begin{align}
  \mathcal{T}_{X\rightarrow X'}\left(\cdot\right) = \tr_E\left[V_{X\rightarrow
      X'E}\,\left(\cdot\right)\,V^\dagger\right]\ ,
\end{align}
and let $\alpha=\norm{\mathcal{T}\left(\Ident_X\right)}_\infty$. In fact, $\mathcal{T}$ is
simply a projection on the support of $\sigma_X$ followed by the mapping $\mathcal{E}$.
We will show that this
choice of variables is feasible and optimal, and will derive a more explicit value of
$\alpha$.

Condition~\eqref{eq:appx-TXXpAlphaSubUnital} is satisfied by definition
and~\eqref{eq:appx-TXXpTraceNonIncr} because $V$ is a partial isometry. Also, verifying
condition~\eqref{eq:appx-TXXpRhoToRhoHat},
\begin{multline}
  \mathcal{T}_{X\rightarrow X'}\left(\sigma_{XR}\right) = \tr_E\left[V_{X\rightarrow
      X'}\,\sigma_{XR} V^\dagger\right] = \tr_E\rho_{X'RE} \\
  = \rho_{X'R}\ .
\end{multline}

Now calculate
\begin{multline}
  \label{eq:appx-TXXpValuePrimal}
  \alpha =
  \norm{\mathcal{T}\left(\Ident_{X}\right)}_\infty = \norm{\tr_E VV^\dagger}_\infty =
  \norm{\tr_E \hat\Pi_{X'E}}_\infty\\
  = \max_{\tau_{X'}} \tr\left[\hat\Pi_{X'E}\,\tau_{X'}\right]
  = 2^{\Hzero(E|X')_{\rho}}\ .
\end{multline}

We will now show that this value is optimal by exhibiting a solution to the dual problem
that achieves the same value. Let $\omega_{X'}=\tau_{X'}$ be the optimal $\tau_{X'}$ for
the definition of $\Hzero(E|X')$ as in~\eqref{eq:appx-TXXpValuePrimal}, let
$Z_{X'R}=\sigma_R^{-1}\otimes\omega_{X'}$ and let $X_X=0$. This choice is feasible since
condition~\eqref{eq:appx-TXXpDualTrOmega} is automatically satisfied and
condition~\eqref{eq:appx-TXXpDualWierdZ} becomes
\begin{align}
	\tr_R\left[\sigma^{t_X}_{XR}\,Z_{X'R}\right]
        &=
	\tr_R\left[\sigma^{t_X}_{XR}\cdot\rho_R^{-1}\otimes\omega_{X'}\right]\nonumber\\
	&= \tr_R\left[\Phi^{t_X}_{X|R}\otimes\omega_{X'}\right]\nonumber\\
        &=
	\Pi^{t_X}_X\otimes\omega_{X'}\leqslant\Ident_X\otimes\omega_{X'}\ ,
\end{align}
where $\Phi_{X|R}$ is an unnormalized maximally entangled state on the supports of
$\sigma_X$ and $\sigma_R$. Let $\rho_{X'RE}$ and $V_{X\rightarrow X'E}$ be defined as
before. The value achieved by this choice of dual variables is then
\begin{align}
  &\tr\left[Z_{X'R}\,\rho_{X'R}\right]
  =
  \tr\left[\sigma_R^{-1}\otimes\omega_{X'}\cdot\rho_{X'R}\right]\\
  &\qquad=
  \tr\left[\sigma_R^{-1}\otimes\omega_{X'}\cdot V_{X\rightarrow X'E}\, \sigma_{XR}
    V^\dagger\right] \\
  &\qquad= \tr\left[\omega_{X'} \cdot V_{X\rightarrow X'E} \,\Phi_{X|R} V^\dagger\right]
  \nonumber\\
  &\qquad=
  \tr\left[\omega_{X'} \hat\Pi_{X'E}\right] = 2^{\Hzero(E|X')_{\rho}}\ .
\end{align}

From this, we conclude that the optimal $\lambda$ for this problem is
\begin{align}
  \lambda_\Lt{opt} = -\Hzero(E|X')_{\rho}\ . 
\end{align}
where $\rho_{X'RE}$ is a purification of $\rho_{X'R}$.

\section{Formal Tools From Majorization and Lambda-Majorization}
\label{sec:appx-appendixFormalLambdaMajorization}

\subsection{Preliminaries and Main Definition}

Let $\Hs_X$, $\Hs_Y$ be two subspaces of a finite-dimensional Hilbert space $\Hs_Z$, and let
$\Hs_A$, $\Hs_B$ be two subspaces of a finite-dimensional Hilbert space $\Hs_C$. Let $d_{(\cdot)}$
denote the dimensions of the various Hilbert spaces $\Hs_{(\cdot)}$ and specifically let
$d=d_Z=\dim\Hs_Z$. Denote by $\LOps(\Hs)$ the set of linear hermitian operators on $\Hs$, by
$\POps(\Hs)$ the set of positive semidefinite operators on $\Hs$, and by $\SOps(\Hs)$ those
operators in $\POps(\Hs)$ that have unit trace. Let also $\lambda_i(\rho)$ denote the $i$-th
eigenvalue of $\rho$ (the order doesn't matter), and $\lD_i(\rho)$ denote the $i$-th
eigenvalue of $\rho$ taken in decreasing order.

Majorization is discussed in detail in
(Refs.\@~\cite{BookBhatiaMatrixAnalysis1997,BookHornJohnsonMatrixAnalysis1985,BookMarshall2010Inequalities}).

\begin{thmheadingit}{Majorization} A matrix $\sigma\in\POps(\Hs_Z)$ is said to {\em majorize} $\rho\in\POps(\Hs_Z)$,
denoted by $\sigma\succ\rho$, if for all $k$, $\sum_{i=1}^k \lD_i(\sigma) \geqslant \sum_{i=1}^k \lD_i(\rho)$,
and if $\tr\sigma=\tr\rho$.
\end{thmheadingit}

The notion of majorization defines a (partial) order relation on $\POps(\Hs_Z)$. When considering the
set of density matrices $\SOps(\Hs_Z)$, there is a ``least'' element: the fully mixed state,
$\frac{1}{d}\Ident_Z$.

\begin{thmheadingit}{Weak Submajorization} A matrix $\sigma\in\POps(\Hs_X)$ is said to {\em weakly submajorize}
$\rho\in\POps(\Hs_Y)$, denoted by $\sigma\succ_w\rho$, if for all $k$,
$\sum_{i=1}^k \lD_i(\sigma) \geqslant \sum_{i=1}^k \lD_i(\rho)$.
\end{thmheadingit}

Remark that if $\sigma,\rho\in\SOps(\Hs_Z)$, then the concept of weak submajorization is equivalent to regular
majorization simply because the traces of these matrices are already equal to unity.

\begin{thmheadingit}{Doubly Stochastic Matrix} A $d\times d$ matrix $S$ is {\em doubly stochastic} if
$S_i^{~j}\geqslant 0$, $\sum_i S_i^{~j} = 1~\forall\,j$ and $\sum_j S_i^{~j} = 1~\forall\,i$.
\end{thmheadingit}

\begin{thmheadingit}{Doubly Substochastic Matrix} A $n\times m$ matrix $B$ is {\em doubly substochastic} if
$B_i^{~j}\geqslant 0$, $\sum_i B_i^{~j} \leqslant 1~\forall\,j$ and $\sum_j B_i^{~j} \leqslant 1~\forall\,i$.
\end{thmheadingit}

The following theorem is due to Hardy, Littlewood and P\'olya~\cite{HardyLittlewoodPolyaInequalities1952}.

\begin{thm}[Hardy, Littlewood, and P\'olya, 1929]
  \label{prop:appx-MajDblStochEquiv}
  Let $\sigma,\rho \in \POps(\Hs_Z)$. Then $\sigma\succ\rho$ if and only if there exists a $d\times d$
  doubly stochastic matrix $S_i^{~j}$ such that $\lambda_i(\rho) = \sum_j S_i^{~j}\lambda_j(\sigma)$ .
\end{thm}

A similar theorem is obtained for weak submajorization and doubly substochastic
matrices~\cite{BookBhatiaMatrixAnalysis1997}.

\begin{prop}
  \label{prop:appx-WeakMajDblSubstochEquiv}
  Let $\sigma \in \POps(\Hs_X)$ and $\rho\in\POps(\Hs_Y)$. Then $\sigma\succ_w\rho$ if and only if there
  exists a $d_Y \times d_X$ doubly substochastic matrix $B_i^{~j}$ such that
  $\lambda_i(\rho) = \sum_j B_i^{~j}\lambda_j(\sigma)$.
\end{prop}

Majorization defines a partial order on states and has a ``smallest'' element, the fully mixed
state. Also, a pure state majorizes any other state.

\begin{prop}
  \label{prop:appx-MajPreservedDirectSumsTensorProducts}
  Majorization is preserved by direct sums and tensor products, \ie{} if $\sigma\succ \rho$
  and $\sigma'\succ \rho'$, then $\sigma\oplus\sigma'\succ\rho\oplus\rho'$ and
  $\sigma\otimes\sigma'\succ\rho\otimes\rho'$. The same holds for weak submajorization.
\end{prop}

A proof for the direct sum of two vectors can be found
in~\cite[Cor. II.1.4]{BookBhatiaMatrixAnalysis1997}. We provide here an alternative proof
along with the tensor product case.

\begin{proof}
  Let $S_i^{\,j}$ and $S_i'^{\,j}$ be doubly stochastic matrices such that
  $\lambda_i(\rho) = \sum_j S_i^{\,j}\lambda_j(\sigma)$ and
  $\lambda_i(\rho') = \sum_j S_i'^{\,j}\lambda_j(\sigma')$. Then $S\oplus S'$ is
  also doubly stochastic and satisfies $\lambda_i(\rho\oplus\rho') = \sum_j (S\oplus
  S')_i^{\,j}\lambda_j(\sigma\oplus\sigma')$, because the vectors of eigenvalues of the
  direct sum are simply the direct sums of the individual vector of eigenvalues. This
  shows that $\sigma\oplus\sigma'\succ\rho\oplus\rho'$.

  Analogously, $S\otimes S'$ satisfies $\lambda_{ii'}(\rho\otimes\rho') =
  \lambda_i(\rho)\lambda_{i'}(\rho') =
  \sum_{jj'} S_i^{\,j}\lambda_j(\sigma)\,S_{i'}^{\,j'}\lambda_{j'}(\sigma')
  = \sum_{jj'} (S\otimes S')_{ii'}^{\,jj'}\,\lambda_{jj'}(\sigma\otimes\sigma')$. $S\otimes S'$ is
  doubly stochastic, $\sum_{ii'} (S\otimes S')_{ii'}^{\,jj'} = \sum_{ii'}
  S_i^{\,j}S_{i'}^{\,j'} = 1$ and $\sum_{jj'} (S\otimes S')_{ii'}^{\,jj'} = \sum_{jj'}
  S_i^{\,j}S_{i'}^{\,j'} = 1$.

  The same proof holds for doubly substochastic matrices, so majorization may be replaced
  by weak submajorization in the proposition.
\end{proof}

We are now all set for a formal definition of lambda-majorization.

Let $\lambda\in\mathbb{R}$ and let $\lambda_1,\lambda_2\geqslant 0$ such that
$\lambda = \lambda_1-\lambda_2$ and $2^{\lambda_1}$, $2^{\lambda_2}$ are integers.
(The case when $2^\lambda$ is irrational will be discussed later.) Take $\Hs_C$ of size greater
than both $2^{\lambda_1}$ and $2^{\lambda_2}$ and let $\Hs_A$ and $\Hs_B$ be subspaces 
of $\Hs_C$ of respective dimensions $2^{\lambda_1}$ and $2^{\lambda_2}$.

\begin{thmheadingit}{Lambda-Majorization} For $\sigma\in\POps(\Hs_X)$ and
  $\rho\in\POps(\Hs_Y)$, we say that $\sigma$ {\em $\lambda$-majorizes} $\rho$, denoted by
  $\sigma\lambdamaj\lambda\rho$, if there exists such $\lambda_1$, $\lambda_2$ such that
  $2^{-\lambda_1}\Ident_A\otimes\sigma \succ_w 2^{-\lambda_2}\Ident_B\otimes\rho$.
  Here $\Ident_A$, $\Ident_B$ are the projectors onto the respective subspaces $\Hs_A$ and
  $\Hs_B$ embedded in $\Hs_C$, of respective dimensions $2^{\lambda_1}$,
  $2^{\lambda_2}$. Likewise, $\sigma$ and $\rho$ are considered as living in
  $\Hs_Z$ by padding them with zero eigenvalues as necessary.
\end{thmheadingit}

We have assumed here that $2^\lambda$ is rational. If $2^\lambda$ is irrational, we say
that $\sigma$ $\lambda$-majorizes $\rho$ if for all rational $2^{\lambda'}$ with
$\lambda'<\lambda$, then $\sigma\lambdamaj{\lambda'}\rho$.

The following proposition guarantees that the definition above does not depend on the exact values of $\lambda_1$
and $\lambda_2$ but only on their difference. This is the same as saying that a fully mixed state cannot act
as a catalyst.

\begin{prop}
  \label{prop:appx-MajWithExtraIdentities}
  For any $\sigma,\rho\in\POps\left(\Hs_Z\right)$, and for any $n$, we have
  $\sigma\succ_w\rho$ if and only if
  $\sigma\otimes\Mixedin{n}\succ_w\rho\otimes\Mixedin{n}$.
\end{prop}
\begin{proof}
  If $\sigma\succ_w\rho$, then the majorization passes over the tensor product, and thus proves the claim.
  Conversely, if $\sigma\otimes\Mixedin{n}\succ_w\rho\otimes\Mixedin{n}$, then in particular,
  for any $k\leqslant d$,
  \begin{align}
    \sum_{i=1}^{n\cdot k} \lD_i(\Mixedin{n}\otimes\sigma) \geqslant
    \sum_{i=1}^{n\cdot k} \lD_i(\Mixedin{n}\otimes\rho)\ .
  \end{align}
 ($d$ is the maximum rank of $\sigma$ or $\rho$.)
 But $\lD_{in}(\Mixedin{n}\otimes\sigma) = \frac1n \lD_i(\sigma)$ and thus
 \begin{align}
   \sum_{i=1}^{k} \lD_i(\sigma) \geqslant \sum_{i=1}^k \lD_i(\rho)\ . \tag*{\qedhere}
 \end{align}
\end{proof}

The following proposition is a direct consequence of the definition of lambda-majorization, and just states
that you can move around randomness into or out of the ancillas in the definition of
lambda-majorization.

\begin{prop}
  \label{prop:appx-LambdaMajMoveIdentitiesAround}
  For any $\sigma\in\POps(\Hs_X)$, $\rho\in\POps(\Hs_Y)$, and for any $\lambda\in\mathbb{R}$, $n> 0$, we have
  \begin{align*}
    {\textstyle\frac1n}\Ident_n\otimes\sigma\lambdamaj\lambda\rho  ~\Leftrightarrow~
    \sigma\lambdamaj{\lambda+\log n} \rho
  \end{align*}
  and
  \begin{align*}
    \sigma\lambdamaj{\lambda-\log n} \rho  ~\Leftrightarrow~
    \sigma\lambdamaj\lambda {\textstyle\frac1n}\Ident_n\otimes\rho\ .
  \end{align*}
\end{prop}

Similarly to Thm.~\ref{prop:appx-MajDblStochEquiv} and to
Prop.~\ref{prop:appx-WeakMajDblSubstochEquiv}, it is possible to characterize
lambda-majorization by the existence of a matrix relating the vector of eigenvalues that
satisfies some specific normalization conditions.

\begin{prop}
  \label{prop:appx-LambdaMajTikFormal}
  Let $\sigma\in\POps(\Hs_X)$ and $\rho\in\POps(\Hs_Y)$. Then $\sigma\lambdamaj\lambda\rho$ if
  and only if there exists a $d_Y \times d_X$ matrix $T_i^{~k}$ such that
  $\lambda_i(\rho) = \sum_k T_i^{~k} \lambda_k(\sigma)$, satisfying
  $T_i^{~k}\geqslant 0$,
  $\sum_i T_i^{~k} \leqslant 1$, and
  $\sum_k T_i^{~k} \leqslant 2^{-\lambda}$ .
\end{prop}

\begin{proof}[Proof of Prop.~\ref{prop:appx-LambdaMajTikFormal}]
  Suppose $2^{-\lambda_1}\Ident_A\otimes\sigma\succ_w 2^{-\lambda_2}\Ident_B\otimes\rho$
  with $\lambda=\lambda_1-\lambda_2$. Then there exists a doubly substochastic matrix
  $S_{bi}^{~ak}$ such that
  \begin{align*}
    \lambda_{bi}\bigl(2^{-\lambda_2}\Ident_B\otimes\rho\bigr)
    = \sum_{ak} S_{bi}^{~ak}\, \lambda_{ak}\bigl( 2^{-\lambda_1}\Ident_A\otimes\sigma\bigr)\ ,
  \end{align*}
  with $S_{bi}^{~ak}\geqslant 0$, $\sum_{bi} S_{bi}^{~ak} \leqslant 1$ and $\sum_{ak}
  S_{bi}^{~ak}\leqslant 1$. (Indices $a$ and $b$ refer to the mixed ancillas of respective
  sizes $2^{\lambda_1}$ and $2^{\lambda_2}$. Since we are considering weak
  submajorization, we can safely ignore all zero eigenvalues and consider only the
  subspaces (of different sizes on the left and right hand side of the majorization) on
  which $\sigma$, $\rho$, $\Ident_A$ and $\Ident_B$ have support, as in
  Prop.~\ref{prop:appx-WeakMajDblSubstochEquiv}.)

  Now we have
  \begin{align*}
    \lambda_i\bigl(\rho\bigr) &= \sum_b \lambda_{bi}\bigl(2^{-\lambda_2}\Ident_B\otimes\rho\bigr) \\
    &= \sum_{a\,b\,k} S_{bi}^{~ak}\,\lambda_{ak}\bigl(2^{-\lambda_1}\Ident_A\otimes\sigma\bigr) \\
    &= \sum_k \left(\sum_{ab}2^{-\lambda_1}\,S_{bi}^{~ak}\right)\,\lambda_k\left(\sigma\right)\ ,
  \end{align*}
  so one can define
  \begin{align*}
    T_i^{~k} = \sum_{ab} 2^{-\lambda_1}\,S_{bi}^{~ak}\ ,
  \end{align*}
  which fulfills $\lambda_i\bigl(\rho\bigr) = \sum_k T_i^{~k}\,\lambda_k\bigl(\sigma\bigr)$. Because $S$ is
  doubly substochastic, and using the fact that indices $a$ (resp. $b$) range to $2^{\lambda_1}$
  ($2^{\lambda_2}$), the matrix $T$ satisfies
  \begin{align*}
    \sum_i T_i^{~k} = \sum_{i\,a\,b} 2^{-\lambda_1} S_{bi}^{~ak}
    = \sum_a 2^{-\lambda_1}\sum_{bi} S_{bi}^{~ak} \leqslant 1\ ,
  \end{align*}
  as well as
  \begin{multline*}
    \sum_k T_i^{~k} = \sum_{k\,a\,b} 2^{-\lambda_1}S_{bi}^{~ak}
    = \sum_b 2^{-\lambda_1}\,\sum_{ak}S_{bi}^{~ak}\\
    \leqslant \sum_b 2^{-\lambda_1} = 2^{-\lambda}\ .
  \end{multline*}
  Additionally, $T_i^{~k}\geqslant 0$ because $S_{bi}^{~ak}\geqslant 0$.

  Conversely, suppose that a matrix $T_i^{~k}$ exists, with $T_i^{~k}\geqslant 0$,
  $\sum_i T_i^{~k} \leqslant 1$, $\sum_k T_i^{~k} \leqslant 2^{-\lambda}$, and
  $\lambda_i(\rho) = \sum_k T_i^{~k} \lambda_k(\sigma)$. Let $\lambda_1,\lambda_2$ such that
  $\lambda=\lambda_1-\lambda_2$ and such that $2^{\lambda_1}$, $2^{\lambda_2}$ are integers.
  Then let $S_{bi}^{~ak} = 2^{-\lambda_2}T_i^{~k}$ for all $a$, $b$. Then $S_{bi}^{~ak} \geqslant 0$ and
  $S$ satisfies
  \begin{equation*}
    \sum_{ak} S_{bi}^{~ak} = 2^{-\lambda_2}\sum_{ak}T_i^{~k} 
    \leqslant 2^{-\lambda_2} \Bigl(\sum_a 1\Bigr)\; 2^{-\lambda} = 1\ ,
  \end{equation*}
  as well as
  \begin{equation*}
    \sum_{bi} S_{bi}^{~ak} = 2^{-\lambda_2} \sum_{bi} T_i^{~k} 
    \leqslant 2^{-\lambda_2}\Bigl(\sum_b 1\Bigr) = 1\ .
  \end{equation*}
  The required weak submajorization for the desired lambda-majorization is provided by this doubly
  substochastic matrix,
  \begin{align*}
    &\lambda_{bi}\bigl(2^{-\lambda_2}\Ident_B\otimes\rho\bigr)
    = 2^{-\lambda_2}\lambda_i\bigl(\rho\bigr)
    = 2^{-\lambda_2} \sum_k T_i^{~k} \lambda_k\bigl(\sigma\bigr) \\
    &\qquad= 2^{-\lambda_2} \sum_k
    T_i^{~k}\sum_a\lambda_{ak}\bigl(2^{-\lambda_1}\Ident_A\otimes\sigma\bigr) \\
    &\qquad= \sum_{ak} S_{bi}^{~ak}\lambda_{ak}\bigl(2^{-\lambda_1}\Ident_A\otimes\sigma\bigr)\ .\tag*{\qedhere}
  \end{align*}
\end{proof}

\subsection{Formulation of Lambda-Majorization in Terms of Maps}
\label{sec:appx-LambdaMajInTermsOfChannels}

Majorization can also be characterized in terms of unital, trace-preserving completely
positive
maps~\cite{Uhlmann1970_shannon,Uhlmann1971_Dichtematrizen,Uhlmann1972_EdDMI,Uhlmann1973_EdDMII}.

\begin{prop}
  \label{prop:appx-GeneralMajorizationChannelUnitalTracePreserving}
  Two positive semidefinite matrices $\sigma$ and $\rho$ satisfy $\sigma\succ\rho$ if and
  only if there exists a trace-preserving, unital, completely positive map $\mathcal{E}$
  satisfying $\mathcal{E}\left(\sigma\right)=\rho$.
\end{prop}

Similarly, one can prove an analogous characterization of weak submajorization. The proof of this proposition
will be given later.
\begin{prop}
  \label{prop:appx-WeakSubMajorizationSubUnitalCPM}
  Let $\sigma\in\POps(\Hs_X)$ and $\rho\in\POps(\Hs_Y)$. Then $\sigma\succ_w\rho$ if and only if
  there exists a completely positive map
  $\mathcal{E}_{X\rightarrow Y} : \LOps(\Hs_X)\rightarrow\LOps(\Hs_Y)$
  such that $\mathcal{E}_{X\rightarrow Y}\left(\sigma\right) = \rho$, with $\mathcal{E}$
  satisfying $\mathcal{E}_{X\rightarrow Y}\left(\Ident_X\right)\leqslant \Ident_Y$ and
  $\mathcal{E}_{X\rightarrow Y}^\dagger\left(\Ident_Y\right)\leqslant \Ident_X$.
\end{prop}

Let's say that $\mathcal{E}_{X\rightarrow Y}$ is {\em subunital} if
$\mathcal{E}_{X\rightarrow Y}\left(\Ident_X\right)\leqslant \Ident_Y$. Then the two conditions on the
structure of the map $\mathcal{E}_{X\rightarrow Y}$ in the above proposition require the map
to be subunital and trace-nonincreasing.

A subunital trace-nonincreasing completely positive map can always be seen as part of a
unital, trace-preserving completely positive map on a larger Hilbert space. This is
analogous of the result that doubly substochastic matrices are submatrices of stochastic
matrices~\cite{BookBhatiaMatrixAnalysis1997}.

In the following, let $\Ident_{X\to Z}$ (resp. $\Ident_{Y\to Z}$) denote the canonical
embedding isometry, and define $\Ident_{Z\to X}$ (resp. $\Ident_{Z\to Y}$) as the
canonical projection partial isometry. Let also $\Ident_X$ (resp. $\Ident_Y$) be the
projector onto the subspace $\Hs_X$ (resp. $\Hs_Y$) within $\Hs_Z$.

\begin{prop}
  \label{prop:appx-DilationOfSubUnitalCPMsToUnitalCPMs}
  Let $\mathcal{E}_{Z\rightarrow Z}$ be a unital, trace-preserving completely
  positive map. Let $\Hs_X$ and $\Hs_Y$ be two subspaces of $\Hs_Z$. Then the map
  $\mathcal{E}'_{X\rightarrow Y}\left(\cdot\right) =
  \Ident_{Z\to Y}\mathcal{E}_{Z\to Z}\left(\Ident_{X\to
      Z}\left(\cdot\right)\Ident_{Z\to X}\right)\Ident_{Y\to Z}$ is subunital and
  trace-nonincreasing.

  Conversely, let $\Hs_X$ and $\Hs_Y$ be two finite-dimensional Hilbert spaces and let
  $\mathcal{E}'_{X\rightarrow Y}$ be any trace-nonincreasing,
  subunital completely positive map. Then there exists a finite-dimensional Hilbert space
  $\Hs_Z$ with isometries $\Ident_{X\to Z},\,\Ident_{Y\to Z}$ and partial isometries
  $\Ident_{Z\to X},\,\Ident_{Z\to Y}$, and a completely positive, unital, trace-preserving
  map $\mathcal{E}_{Z\to Z}$, such that
  \begin{align}
    \label{eq:appx-DilationOfSubUnitalCPMsToUnitalCPMs_recovercondition}
    \mathcal{E}'_{X\rightarrow Y}\left(\cdot\right) =
    \Ident_{Z\to Y}\, \mathcal{E}_{Z\to Z}\left(
      \Ident_{X\to Z} \left(\cdot\right) \Ident_{Z\to X}
    \right)\Ident_{Y\to Z}\ .
  \end{align}

  The space $\Hs_Z$ and map $\mathcal{E}_{Z\to Z}$ may be chosen as
  $\Hs_Z=\Hs_X\oplus\Hs_Y$ and
  \begin{align}
    \hspace*{1em}&\hspace*{-1em}
    \mathcal{E}_{Z\rightarrow Z}\left(\cdot\right)
    ={}  \nonumber\\
    &\ \Ident_{Y\to Z}\, \mathcal{E}'_{X\rightarrow Y}\left(
      \Ident_{Z\to X} \left(\cdot\right) \Ident_{X\to Z}
    \right) \Ident_{Z\to Y}
    \nonumber\\
    &+\ \Ident_{X\to Z}\, \mathcal{E}'^\dagger\left(
      \Ident_{Z\to Y} \left(\cdot\right) \Ident_{Y\to Z}
    \right) \Ident_{Z\to X} \nonumber\\
    &+\ \Ident_{Y\to Z}\,\sqrt{G_Y}\,\Ident_{Z\to Y} \left(\cdot\right) \Ident_{Y\to Z}\,
    \sqrt{G_Y}\, \Ident_{Z\to Y} \nonumber\\
    &+\ \Ident_{X\to Z}\,\sqrt{H_X}\,\Ident_{Z\to X} \left(\cdot\right) \Ident_{X\to Z}\,
    \sqrt{H_X}\,\Ident_{Z\to X}\ ,
    \label{eq:appx-DilationOfSubUnitalCPMsToUnitalCPMs_expression}
  \end{align}
  where $G_Y = \Ident_Y - \mathcal{E}'_{X\to Y}\left(\Ident_X\right)$ and
  $H_X = \Ident_X - \mathcal{E}'^\dagger\left(\Ident_Y\right)$.

  The space $\Hs_Z$ may also be chosen to be any space bigger than $\Hs_X\oplus\Hs_Y$.
\end{prop}

In order to generalize this concept to our lambda-majorization, let's introduce the concept of an
$\alpha$-subunital map. These generalize the notion of subunital maps to arbitrary normalizations.

\begin{thmheading}{$\alpha$-subunital Maps} We'll call a map $\mathcal{T}_{X\rightarrow Y}$
{\em $\alpha$-subunital} if it satisfies $\mathcal{T}_{X\rightarrow Y}(\Ident_X) \leqslant \alpha \Ident_Y$.
\end{thmheading}

\begin{prop}[Composition of $\alpha$-subunital maps]
  \label{prop:appx-CompositionOfAlphaSubunitalMaps}
  Let $\Hs_W\in\Hs_Z$ be another subspace of $\Hs_Z$ in addition to $\Hs_X$ and $\Hs_Y$, and
  let $\mathcal{T}_{X\rightarrow Y}$,
  $\mathcal{T}'_{Y\rightarrow W}$ be completely positive, trace-nonincreasing maps. Assume
  that $\mathcal{T}_{X\rightarrow Y}$ is
  $\alpha$-subunital and that $\mathcal{T}'_{Y\rightarrow W}$ is $\beta$-subunital. Then their composition
  $\left[\mathcal{T}'\circ\mathcal{T}\right]_{X\rightarrow W}$ is $\:\alpha\cdot\beta\,$-subunital.
\end{prop}
\begin{proof}[Proof of Prop.~\ref{prop:appx-CompositionOfAlphaSubunitalMaps}]
  The composition of $\mathcal{T}_{X\rightarrow Y}$ and $\mathcal{T}'_{Y\rightarrow W}$ is
  trace-nonincreasing,
  \begin{align*}
    \mathcal{T}^\dagger\left(\mathcal{T}'^\dagger\left(\Ident_W\right)\right) \leqslant
    \mathcal{T}^\dagger\left(\Ident_Y\right) \leqslant \Ident_X\ .
  \end{align*}

  Their composition is also $\alpha\cdot\beta\,$-subunital,
  \begin{align*}
    \mathcal{T}'_{Y\rightarrow W}\left(\mathcal{T}_{X\rightarrow
        Y}\left(\Ident_X\right)\right)
    \leqslant \mathcal{T}'_{Y\rightarrow W}\left(\alpha\,\Ident_Y\right)
    \leqslant \alpha\beta\,\Ident_W\ . \tag*\qedhere
  \end{align*}
\end{proof}

\begin{remark}
  \label{remark:AlphaSubUnitalCompWithProj}
  Let $V_{X\to Y}$ be a partial isometry. Then the map $\left(\cdot\right)_X
  \longrightarrow V_{X\to Y}\nodag \left(\cdot\right)_X V^\dagger_{X\leftarrow Y}$ is
  trace-nonincreasing and subunital.

  In particular, if $\mathcal{T}_{Z\rightarrow Z}$ is a $2^{-\lambda}$-subunital,
  trace-nonincreasing map, then $\mathcal{T}_{X\rightarrow Y}$, defined by
  $\mathcal{T}_{X\rightarrow Y}(\cdot) =
  \Ident_{Z\to Y}\mathcal{T}_{Z\rightarrow Z}\left( \Ident_{X\to Z} \left(\cdot\right)
    \Ident_{Z\to X}\right) \Ident_{Y\to Z}$,
  is also a trace-nonincreasing $2^{-\lambda}$-subunital map.
\end{remark}

\begin{proof}[Proof of Prop.~\ref{prop:appx-DilationOfSubUnitalCPMsToUnitalCPMs}]
  The remark proves the first part of the proposition.
  To prove the converse, we will show that the
  expression~\eqref{eq:appx-DilationOfSubUnitalCPMsToUnitalCPMs_expression} satisfies the
  conditions of the claim. Notice first that the channel $\mathcal{E}_{Z\to Z}$ is equal
  to its own adjoint, i.e. $\mathcal{E}_{Z\to Z}=\mathcal{E}^\dagger_{Z\to Z}$. Moreover,
  the map is unital:
  \begin{align}
    \mathcal{E}_{Z\rightarrow Z}\left(\Ident_Z\right) &= {} 
    \Ident_{Y\to Z} \left(\Ident_Y-G_Y\right) \Ident_{Z\to Y} \nonumber\\
    &\quad+\ \Ident_{X\to Z} \left(\Ident_X-H_X\right) \Ident_{Z\to X} \nonumber\\
    &\quad+\ \Ident_{Y\to Z} \,G_Y\, \Ident_{Z\to Y} \nonumber\\
    &\quad+\ \Ident_{X\to Z} \,H_X\, \Ident_{Z\to X} \nonumber\\
    &= \Ident_{Y} + \Ident_X = \Ident_Z\ ,
  \end{align}
  which makes it automatically trace-preserving, the map being its own adjoint.
  Condition~\eqref{eq:appx-DilationOfSubUnitalCPMsToUnitalCPMs_recovercondition} follows from
  the definition of $\mathcal{E}_{Z\to Z}$
  in~\eqref{eq:appx-DilationOfSubUnitalCPMsToUnitalCPMs_expression}.

  If we choose $\Hs_Z$ to be any space larger than $\Hs_X\oplus\Hs_Y$, we can adapt the
  definition of $\mathcal{E}_{Z\to Z}$
  in~\eqref{eq:appx-DilationOfSubUnitalCPMsToUnitalCPMs_expression} and add a term
  $\Ident_{X\oplus Y}^\perp\,\left(\cdot\right)\,\Ident_{X\oplus Y}^\perp$ (the operator
  $\Ident_{X\oplus Y}^\perp$ projects onto the orthogonal subspace to $\Hs_X\oplus\Hs_Y$
  in $\Hs_Z$). In this case $\mathcal{E}_{Z\to Z}$ would still satisfy all the required
  properties.
\end{proof}

We now have all the necessary tools to prove
Proposition~\ref{prop:DilationOfSubUnitalCPMsToUnitalCPMs} of the Main Text.

\begin{prop}[Proposition~\ref{prop:DilationOfSubUnitalCPMsToUnitalCPMs} of the
  Main Text]
  \label{prop:appx-DilationOfSubUnitalCPMsToUnitalCPMs-samesystem}
  Let $\Hs_K$ and $\Hs_L$ be finite dimensional Hilbert spaces, and let
  $\mathcal{E}'_{K\to L}$ be a completely positive, trace-nonincreasing, subunital map.
  Then there exists finite dimensional Hilbert spaces $\Hs_Q$ and $\Hs_{Q'}$, and a
  completely positive, trace-preserving, unital map ${\mathcal{E}}_{KQ\to LQ'}$ such
  that
  \begin{multline}
    \label{eq:appx-DilationOfSubUnitalCPMsToUnitalCPMs_recovercondition_Q}
    \mathcal{E}'_{K\to L}\left(\cdot\right)
    =
    \left(\Ident_{L}\otimes\bra{\mathrm f}_{Q'}\right)\ \cdot\\
    {\mathcal{E}}_{KQ\to LQ'}\left[ \left(\Ident_{K}\otimes\ket{\mathrm i}_{Q}\right)
      \left(\cdot\right)
      \left(\Ident_{K}\otimes\bra{\mathrm i}_{Q}\right)
    \right] \\
    \cdot\ \left(\Ident_{L}\otimes\ket{\mathrm f}_{Q'}\right)
    \ ,
  \end{multline}
  for some pure states $\ket{\mathrm{i}}_Q$, $\ket{\mathrm{f}}_{Q'}$.
  In addition, $\dim\left(\Hs_K\otimes\Hs_Q\right) = \dim\left(\Hs_L\otimes\Hs_{Q'}\right)$.
\end{prop}

\begin{proof}
  Apply the converse in Prop.~\ref{prop:appx-DilationOfSubUnitalCPMsToUnitalCPMs} to dilate the
  subunital map $\mathcal{E}'_{K\to L}$ to a unital map $\mathcal{E}_{Z\to Z}$ (take
  $\Hs_X=\Hs_K$ and $\Hs_Y=\Hs_L$).
 
  We may choose $\Hs_Z=\Hs_X\oplus\Hs_Y\oplus\Hs_{W}$, where $\Hs_W$ is a space whose
  dimension we haven't yet fixed. We would like the space $\Hs_Z$ to factorize as both
  \begin{align}
    \Hs_Z &= \Hs_K\otimes\Hs_Q\ ;\\
    \Hs_Z &= \Hs_L\otimes\Hs_{Q'}\ ,
  \end{align}
  for some systems $\Hs_Q$ and $\Hs_{Q'}$.
  A necessary and sufficient condition for that is
  \begin{align}
    \dim\left(\Hs_Z\right) &=
    \dim\left(\Hs_K\right) \times \dim\left(\Hs_Q\right)\ ; \\
    \dim\left(\Hs_Z\right) &=
    \dim\left(\Hs_L\right) \times \dim\left(\Hs_{Q'}\right)\ ,
  \end{align}
  where $\dim\left(\Hs_Z\right) = \dim\left(\Hs_X\right) + \dim\left(\Hs_Y\right) +
  \dim\left(\Hs_W\right)$.  We thus fix the dimension of $\Hs_Z$ to be
  \begin{align}
    \dim\left(\Hs_Z\right) =
    m\times \operatorname{lcm}\left(\dim\left(\Hs_K\right),
      \dim\left(\Hs_{L}\right)\right)\ ,
  \end{align}
  where $\operatorname{lcm}\left(\cdot,\cdot\right)$ designates the least common multiple
  of its arguments and $m$ is some integer chosen such that $\dim\left(\Hs_Z\right)
  \geqslant \dim\left(\Hs_K\right) + \dim\left(\Hs_L\right)$.  We have now fixed
  $\dim\left(\Hs_Z\right)$, and this in turn implicitly fixes the dimension of $\Hs_W$.

  This choice of $\Hs_Z=\Hs_K\otimes\Hs_Q=\Hs_L\otimes\Hs_{Q'}$ fixes the embedding
  isometries $\Ident_{K\to Z}$, $\Ident_{L\to Z}$, and projective partial isometries
  $\Ident_{Z\to K}$, $\Ident_{Z\to L}$, as
  \begin{align*}
    \Ident_{X\to Z} &=\Ident_{X\to K}\otimes\ket{\mathrm{i}}_Q\ ; &
    \Ident_{Y\to Z} &=\Ident_{Y\to L}\otimes\ket{\mathrm{f}}_{Q'}\ ; \\
    \Ident_{Z\to X} &=\Ident_{K\to X}\otimes\bra{\mathrm{i}}_Q\ ; &
    \Ident_{Z\to Y} &=\Ident_{L\to Y}\otimes\bra{\mathrm{f}}_{Q'}\ ,
  \end{align*}
  for some fixed states $\ket{\mathrm i}_Q$, $\ket{\mathrm f}_{Q'}$.

  The right hand side of~\eqref{eq:appx-DilationOfSubUnitalCPMsToUnitalCPMs_recovercondition_Q}
  is
  \begin{align*}
    &\left(\Ident_{L}\otimes\bra{\mathrm f}_{Q'}\right)\ \cdot\\
    &\qquad \mathcal{E}_{KQ\to LQ'}\left[ \left(\Ident_{K}\otimes\ket{\mathrm i}_{Q}\right)
      \left(\cdot\right)
      \left(\Ident_{K}\otimes\bra{\mathrm i}_{Q}\right)
    \right] \\
    &\qquad\qquad\cdot\ \left(\Ident_{L}\otimes\ket{\mathrm f}_{Q'}\right)
    \\
    &= \Ident_{Z\to Y} \ \mathcal{E}_{Z\to Z}\left[ \Ident_{X\to Z}
      \left(\cdot\right)
      \Ident_{Z\to X}
    \right]
    \  \Ident_{Y\to Z}
    \\
    &= \mathcal{E}'_{X\to Y}\left(\cdot\right)\ .
  \end{align*}
  We have used the above definition of the embedding and projection (partial) isometries,
  and applied~\eqref{eq:appx-DilationOfSubUnitalCPMsToUnitalCPMs_recovercondition}. This
  proves condition~\eqref{eq:appx-DilationOfSubUnitalCPMsToUnitalCPMs_recovercondition_Q}.

  Note that any choice of $\mathcal{E}_{KQ\to LQ'}$ which would satisfy the conditions of
  the Proposition would require that $\dim\left(\Hs_K\otimes\Hs_Q\right) =
  \dim\left(\Hs_L\otimes\Hs_{Q'}\right)$. Indeed, a unital, trace-preserving completely
  positive map may only exist if the dimensions of the input and output spaces match: with
  $\mathcal{E}\left(\Ident_{KQ}\right) = \Ident_{LQ'}$ and $\mathcal{E}$ trace-preserving
  we have that $\tr\left(\Ident_{KQ}\right)=\tr\Ident_{LQ'}$ and thus that the dimensions
  are equal.
\end{proof}

\begin{proof}[Proof of Prop.~\ref{prop:appx-WeakSubMajorizationSubUnitalCPM}]
  By the weak submajorization condition, if $\tr\rho\neq\tr\sigma$, we must have $\tr\rho < \tr\sigma$. Consider
  an extension space
  $\Hs_{Y'}\in\Hs_Z$ (consider a larger $\Hs_Z$ if necessary) in which we extend $\rho$ by many small
  eigenvalues such that $\tr \rho_{Y\oplus Y'} = \tr\sigma$, while still having $\sigma \succ_w \rho_{Y\oplus Y'}$.
  Now we have a (regular) majorization, $\sigma\succ\rho_{Y\oplus Y'}$, and can apply
  Prop.~\ref{prop:appx-GeneralMajorizationChannelUnitalTracePreserving}.

  The obtained map, $\mathcal{E}_{Z\rightarrow Z}$, is then unital and
  trace-preserving. It can be restricted by projecting the input onto $\Hs_X$ and the
  output onto $\Hs_Y$,
  \begin{align*}
    \mathcal{E}_{X\rightarrow Y}(\cdot)
    = \Ident_Y\; \mathcal{E}_{Z\rightarrow Z}\,\bigl(\Ident_X \left(\cdot\right) \Ident_X \bigr)\; \Ident_Y\ .
  \end{align*}
  This restricted operator, by Remark~\ref{remark:AlphaSubUnitalCompWithProj}, is a valid
  trace-nonincreasing subunital map 
  (take $\lambda=0$).

  Conversely, if $\mathcal{E}_{X\rightarrow Y}$ is a subunital trace-nonincreasing
  completely positive map with $\mathcal{E}_{X\rightarrow Y}\left(\sigma\right)=\rho$,
  then one can dilate it with Proposition~\ref{prop:appx-DilationOfSubUnitalCPMsToUnitalCPMs}
  to a unital, trace-preserving completely positive map $\mathcal{E}_{Z\rightarrow Z}$
  such that $\Ident_Y\,\mathcal{E}_{Z\rightarrow Z}\left(\sigma\oplus 0_Y\right)\Ident_Y =
  \rho$. Note also that the map
  $\left(\cdot\right)\mapsto\Ident_Y\left(\cdot\right)\Ident_Y+\Ident_X\left(\cdot\right)\Ident_X$
  is a pinching~\cite[p.~50, Prob.~II.5.5]{BookBhatiaMatrixAnalysis1997}, so we have
  $\sigma\oplus 0_Y
  \succ\mathcal{E}_{Z\rightarrow Z}\left(\sigma\oplus 0_Y\right)
  \succ\Ident_X\mathcal{E}_{Z\rightarrow
    Z}\left(\sigma\oplus0_Y\right)\Ident_X+\Ident_Y\mathcal{E}_{Z\rightarrow
    Z}\left(\sigma\oplus0_Y\right)\Ident_Y
  \succ_w\Ident_X\mathcal{E}_{Z\rightarrow Z}\left(\sigma\oplus0_Y\right)\Ident_X=\rho$. The 
  last weak submajorization is because some eigenvalues were left out.
\end{proof}

In the same way as lambda majorization can be characterized with differently normalized
doubly substochastic maps, it can also be characterized in terms of a differently
normalized subunital map.

\begin{prop}
  \label{prop:appx-LambdaMajorizationTMap}
  Let $\sigma\in\POps(\Hs_X)$, $\rho\in\POps(\Hs_Y)$ and $\lambda\in\mathbb R$. Then $\sigma\lambdamaj\lambda\rho$
  if and only if there exists a completely positive map
  $\mathcal T_{X\rightarrow Y} : \LOps(\Hs_X) \rightarrow \LOps(\Hs_Y)$ such
  that $\mathcal{T}_{X\rightarrow Y}(\sigma) = \rho$, that is $2^{-\lambda}$-subunital and trace-nonincreasing.
\end{prop}

\begin{proof}[Proof of Prop.~\ref{prop:appx-LambdaMajorizationTMap}]
  {\bf ``$\boldsymbol\Rightarrow$''.} \hspace*{1.5mm}
  Assume first that $2^{-\lambda_1}\Ident_A\otimes\sigma \succ_w
  2^{-\lambda_2}\Ident_B\otimes\rho$, with   $\Hs_A$, $\Hs_B$ (of respective sizes
  $2^{\lambda_1}$ and $2^{\lambda_2}$) being subsystems of an ancilla system $\Hs_C$, with
  $\lambda=\lambda_1-\lambda_2$.

  By Prop.~\ref{prop:appx-WeakSubMajorizationSubUnitalCPM}, there exists a subunital
  trace-nonincreasing completely positive map $\mathcal{E}_{AX\rightarrow BY}$, such that
  \begin{align}
    \label{eq:appx-LambdaMajTMap_proof_ECZmap}
    \mathcal{E}_{AX\rightarrow BY}(2^{-\lambda_1}\Ident_A\otimes\sigma)
    = 2^{-\lambda_2}\Ident_B\otimes\rho\ .
  \end{align}

  Now let the map $\mathcal{T}$ be defined by 
  \begin{align}
    \label{eq:appx-LetTMapFunctionOfSubUnitalMap}
    \mathcal{T}_{X\rightarrow Y}\left(\cdot\right) = \tr_B\left[\mathcal{E}_{AX\rightarrow
        BY}\left(2^{-\lambda_1}\Ident_A\otimes\left(\cdot\right)\right)\right]\ .
  \end{align}

  This map is trace-nonincreasing,
  \begin{multline*}
    \mathcal{T}^\dagger_{X\leftarrow Y}\left(\Ident_Y\right)
    = 2^{-\lambda_1}\tr_A\left[\mathcal{E}_{AX\leftarrow
        BY}^\dagger\left(\Ident_{BY}\right)\right]\\
    \leqslant 2^{-\lambda_1}\tr_A\left(\Ident_{AX}\right) = \Ident_X\ ,
  \end{multline*}
  and $2^{-\lambda}$-subunital,
  \begin{multline*}
    \mathcal{T}_{X\rightarrow Y}\left(\Ident_X\right)
    = 2^{-\lambda_1}\tr_{B}\left[\mathcal{E}\left(\Ident_{AX}\right)\right]
    \leqslant 2^{-\lambda_1}\tr_B\Ident_{BY}\\
    = 2^{-\lambda}\Ident_Y\ .
  \end{multline*}
  The map $\mathcal{T}$ brings $\sigma$ to $\rho$,
  \begin{multline*}
    \mathcal{T}_{X\rightarrow Y}\left(\sigma_X\right)
    = \tr_B\left[ \mathcal{E}\left(2^{-\lambda_1}\Ident_A\otimes\sigma_X\right) \right] \\
    = \tr_B\left( 2^{-\lambda_2}\Ident_B\otimes\rho_Y \right) = \rho_Y\ ,
  \end{multline*}
  so that $\mathcal{T}$ satisfies all the claimed properties.

  {\bf ``$\boldsymbol\Leftarrow$''.} \hspace*{1.5mm}
  To prove the converse, assume that a trace-nonincreasing, $2^{-\lambda}$-subunital map $\mathcal{T}_{X\rightarrow Y}$
  exists, such that $\mathcal{T}_{X\rightarrow Y}(\sigma) = \rho$.

  Choose $\lambda_1$, $\lambda_2$ such that $\lambda=\lambda_1-\lambda_2$ and such that $2^{\lambda_1}$,
  $2^{\lambda_2}$, are integers. (Again, in case $2^\lambda$ is
  irrational, approximate $2^\lambda$ arbitrarily well by rational numbers $2^{\lambda'}$.)
  Choose $\Hs_C$ large enough to contain two subspaces $\Hs_A$ and $\Hs_B$ of respective
  dimensions $2^{\lambda_1}$ and $2^{\lambda_2}$. Let
  \begin{align}
    \label{eq:appx-LetSubUnitalMapFunctionOfTMap}
    \mathcal{E}_{AX\rightarrow BY}\left(\cdot\right)
    = 2^{-\lambda_2}\Ident_B\otimes\mathcal{T}_{X\rightarrow
      Y}\left(\tr_A\left(\cdot\right)\right)\ .
  \end{align}

  This map is trace-nonincreasing,
  \begin{multline*}
    \mathcal{E}^\dagger\left(\Ident_{BY}\right)
    = 2^{-\lambda_2}\Ident_A\otimes\mathcal{T}^\dagger\left(\tr_B\Ident_{BY}\right)\\
    = 2^{-\lambda_2}\Ident_A\otimes\mathcal{T}^\dagger\left(2^{\lambda_2}\Ident_Y\right)
    \leqslant \Ident_{AX}\ ,
  \end{multline*}
  and subunital,
  \begin{multline*}
    \mathcal{E}\left(\Ident_{AX}\right)
    = 2^{-\lambda_2}\Ident_B\otimes\mathcal{T}\left(\tr_A\Ident_{AX}\right)\\
    = 2^{-\lambda_2}\Ident_B\otimes\mathcal{T}\left(2^{\lambda_1}\Ident_{X}\right)
    \leqslant \Ident_{BY}\ ,
  \end{multline*}
  since $\lambda=\lambda_1-\lambda_2$ and $\mathcal{T}$ is $2^{-\lambda}$-subunital. Also,
  \begin{multline*}
    \mathcal{E}\left(2^{-\lambda_1}\Ident_A\otimes\sigma_X\right) =
    2^{-\lambda_2}\Ident_B\otimes\mathcal{T}\left(\tr_A\left(2^{-\lambda_1}\Ident_A\otimes\sigma_X\right)\right)\\
    = 2^{-\lambda_2}\Ident_B\otimes\mathcal{T}\left(\sigma_X\right)
    = 2^{-\lambda_2}\Ident_B\otimes\rho_Y\ .
  \end{multline*}
  By Prop.~\ref{prop:appx-WeakSubMajorizationSubUnitalCPM}, we eventually have
  \begin{align*}
    2^{-\lambda_1}\Ident_A\otimes\sigma_X\succ_w 2^{-\lambda_2}\Ident_B\otimes\rho_Y\ .
    \tag*{\qedhere}
  \end{align*}

\end{proof}

\begin{remark}
  A trace-nonincreasing, $2^{-\lambda}$-subunital completely positive map
  $\mathcal{T}_{X\rightarrow Y}$
  can always be written as in Eq.~\eqref{eq:appx-LetTMapFunctionOfSubUnitalMap} for a
  sub-unital trace-nonincreasing completely positive map $\mathcal{E}_{AX\rightarrow BY}$,
  which itself can always be written as projections of a unital map
  $\mathcal{E}_{CZ\rightarrow CZ}$ (see text of the previous proof, and
  Prop.~\ref{prop:appx-DilationOfSubUnitalCPMsToUnitalCPMs}).

  Conversely, for any unital map $\mathcal{E}_{CZ\rightarrow CZ}$ with
  $\mathcal{E}\left(2^{-\lambda_1}\Ident\otimes\sigma_X\right)=2^{-\lambda_2}\Ident\otimes\rho_Y$, in
  particular for any noisy operation in our framework, the map $\mathcal{T}$ obtained by
  Eq.~\eqref{eq:appx-LetTMapFunctionOfSubUnitalMap} is trace-nonincreasing and
  $2^{-\lambda}$-subunital.
\end{remark}

In particular, for our purposes of optimizing $\lambda$ over all possible processes of our
framework with an additional condition to the map carrying out the process (namely to
preserve correlations between our system $X$ and the reference system $R$), we may
impose that condition directly on the map $\mathcal{T}$ to obtain an upper bound on
$\lambda$.

\subsection{Properties for quantum states}
\label{sec:appx-FormalLambdaMajPropertiesForQuantumStates}

We will consider in this section some useful properties of lambda-majorization in the case
where we consider normalized states $\sigma$, $\rho$. Here, weak majorization
automatically implies (regular) majorization because $\tr\sigma=\tr\rho=1$.

In this section, let $\sigma\in\SOps(\Hs_X)$ and $\rho\in\SOps(\Hs_Y)$.

\begin{prop}[Lambda-Majorizing a Pure State]
  \label{prop:appx-lambdaMajDOpsPureState}
  For any pure state $\ket 0\in\Hs_Z$, we have $\sigma\lambdamaj\lambda\ketbra00$ if and only
  if $\rank\sigma\leqslant 2^{-\lambda}$ (obviously $\lambda$ has to be negative or zero).
  Equivalently, $\sigma\succ\frac1n\Ident_n$ if and only if $\rank\sigma\leqslant n$.
\end{prop}

\begin{proof}[Proof of Prop.~\ref{prop:appx-lambdaMajDOpsPureState}]
  Assume first that  $\sigma\lambdamaj\lambda\ketbra00$. Here $\Hs_Y$ is the one-dimensional space
  spanned by $\ket0$, and take $\Hs_X$ the subspace on which $\sigma$ has its support. By
  Prop.~\ref{prop:appx-LambdaMajTikFormal} there exists a single-row matrix
  $T_i^{~k}$ satisfying $T_i^{~k}\geqslant 0$, $\sum_i T_i^{~k} = T_{i=1}^{~k} \leqslant 1~\forall k$,
  $\sum_k T_i^{~k} \leqslant 2^{-\lambda}$ such that
  $1 = \lambda_{i=1}(\ketbra00) = \sum_k T_{i=1}^{~k} \lambda_k(\sigma)$. We also have $\lambda_k(\sigma)\neq 0$
  because $\sigma$ has nonzero eigenvalues in $\Hs_X$. Then
  $\sum_k T_{i=1}^{~k} \lambda_k(\sigma) = 1 = \sum_k\lambda_k(\sigma)$ implies $T_{i=1}^k = 1~\forall k$.
  That is, the condition $\sum_k T_{i=1}^{~k} \leqslant 2^{-\lambda}$ forces $T_{i=1}^{~k}$ to have at most
  $2^{-\lambda}$ elements, \ie{} the rank of $\sigma$ may not exceed $2^{-\lambda}$.

  The converse holds because any state majorizes a uniform state of the same rank.
\end{proof}

\begin{prop}[Condition on Support Sizes for Lambda-Majorization]
  \label{prop:appx-lambdaMajDOpsConditionRanks}
  If $\sigma\lambdamaj\lambda\rho$, then $\rank\sigma\leqslant 2^{-\lambda}\rank\rho$. 
\end{prop}

\begin{proof}[Proof of Prop.~\ref{prop:appx-lambdaMajDOpsConditionRanks}]
  Notice that $\rho\succ\frac1{\rank\rho}\Ident_{\rank\rho}$, and thus
  $\sigma\lambdamaj\lambda\frac1{\rank\rho}\Ident_{\rank\rho}$. Then, by
  Prop.~\ref{prop:appx-LambdaMajMoveIdentitiesAround} we have
  \begin{equation*}
    \sigma\lambdamaj{\lambda-\log\rank\rho} \ketbra00\ ;
  \end{equation*}
  it remains to apply Prop.~\ref{prop:appx-lambdaMajDOpsPureState}.
\end{proof}

\begin{prop}[Being Lambda-Majorized by a Pure State]
  \label{prop:appx-LambdaMajorizedDOpsByPureState}
  Let the state $\rho$ have maximum eigenvalue $\lambda_\mathrm{max}(\rho)$.
  For any pure state $\ket0$, we have $\ketbra00 \lambdamaj\lambda \rho$ if and only if
  $\lambda_\mathrm{max}(\rho) \leqslant 2^{-\lambda}$. Equivalently, $\frac1n\Ident_n\succ\rho$ if and
  only if $\lambda_\mathrm{max}(\rho) \leqslant \frac1n$.
\end{prop}

\begin{proof}[Proof of Prop.~\ref{prop:appx-LambdaMajorizedDOpsByPureState}]
  Let $T_i^{~k}$ be as in Prop.~\ref{prop:appx-LambdaMajTikFormal}. Note here $k$ only takes value
  $1$, because we consider $\Hs_Y$ being the one-dimensional space spanned by $\ket0$. Then
  $\lambda_i(\rho) = \sum_k T_i^{~k} \lambda_k(\ketbra00) = T_i^{~k=1}$ and thus
  $T_i^{~k}=\lambda_i(\rho)$. Then $2^{-\lambda} \geqslant \sum_k T_i^{~k} = T_i^{k=1} = \lambda_i(\rho)$ for
  all $i$. In particular, $2^{-\lambda}\geqslant \lambda_\mathrm{max}(\rho)$.

  Conversely, if $\lambda_\mathrm{max}(\rho) \leqslant 2^{-\lambda}$, then let $T_i^{~k=1}=\lambda_i(\rho)$. This
  matrix $T$ satisfies the conditions in Prop.~\ref{prop:appx-LambdaMajTikFormal} and thus
  $\ketbra00\lambdamaj\lambda\rho$.
\end{proof}

\subsection{Optimal Lambda Majorization for Normalized States and Relation to Single-Shot Entropy Measures}

Define the {\em absorbed randomness} (or {\em relative mixedness}~\cite{Egloff2012arXiv})
of a transition from $\sigma$ to $\rho$ as the maximal amount of randomness
that you can get rid of, or the minimal amount of randomness that you have to generate, in
a noisy operation process:
\begin{align}
  \label{eq:appx-AbsorbedRandomnessDef}
  R(\sigma\rightarrow\rho) = \sup\, \bigl\{ \lambda : \sigma \lambdamaj{\lambda} \rho \, \bigr\}\ .
\end{align}

Recent work has shown that this measure is relevant for the amount of extractable work
of processes acting on arrays of Szilard boxes~\cite{Egloff2012arXiv}.

The absorbed randomness has some tight relations to single-shot entropy measures, which we
present here. These are reformulations of results shown
in~\cite{Dahlsten2011NJP_inadequacy,Egloff2010MasterThesis}.

\begin{prop}
  \label{prop:appx-RBoundsHminmax}
  The absorbed randomness defined above satisfies the following bounds.
  \begin{align*}
    \Hmin(\rho)-\Hzero(\sigma) \leqslant R(\sigma\rightarrow\rho) \leqslant \Hzero(\rho) - \Hzero(\sigma) \ .
  \end{align*}
\end{prop}

\begin{prop}
  \label{prop:appx-RpurerhoAndRsigmapure}
  If $\ket0$ denotes any pure state, then the following relations hold:
  \begin{align}
    \label{eq:appx-Rpurerho}  R(\ket 0 \rightarrow \rho) &= \Hmin(\rho)\ ,\\
    \label{eq:appx-Rsigmapure}  R(\sigma \rightarrow \ket 0) &= -\Hzero(\sigma)\ .
  \end{align}
\end{prop}

Similar explicit values can be obtained in the case where either the initial state or the target state is mixed.
\begin{prop}
  \label{prop:appx-RmixedrhoAndRsigmamixed}
  If $\Mixedin{n}$ denotes the fully mixed state on $\log n$ qubits, then:
  \begin{align}
    \label{eq:appx-Rmixedrho} R(\Mixedin{n} \rightarrow \rho) &= \Hmin(\rho) - \log n\ , \\
    \label{eq:appx-Rsigmamixed} R(\sigma \rightarrow \Mixedin{n}) &= \log n - \Hzero(\sigma)\ .
  \end{align}
\end{prop}

\begin{proof}[Proof of Prop.~\ref{prop:appx-RBoundsHminmax}]
  {\em Lower bound:} Let $\lambda_1 = \Hmin(\rho) = -\log \lambda_\mathrm{max}(\rho)$ and
  $\lambda_2 = \Hzero(\sigma) = \log\,\rank\sigma$. By Proposition~\ref{prop:appx-LambdaMajorizedDOpsByPureState},
  we have $2^{-\lambda_1}\Ident_{2^{\lambda_1}}\succ\rho$ and by Proposition~\ref{prop:appx-lambdaMajDOpsPureState},
  $\sigma\succ 2^{-\lambda_2}\Ident_{2^{\lambda_2}}$. The majorization carries over to the tensor product,
  $2^{-\lambda_1}\Ident_{2^{\lambda_1}}\otimes\sigma\succ 2^{-\lambda_2}\Ident_{2^{\lambda_2}}\otimes\rho$,
  and $\lambda_1 - \lambda_2$ is a valid maximization candidate for~\eqref{eq:appx-AbsorbedRandomnessDef}.

  {\em Upper bound:} Let $\lambda=R(\sigma\rightarrow\rho)$ satisfying $\sigma\lambdamaj\lambda\rho$.
  Proposition~\ref{prop:appx-lambdaMajDOpsConditionRanks} immediately yields
  $2^{\lambda}\leqslant\frac{\rank\rho}{\rank\sigma}$, and
  \begin{equation*}
    R(\sigma\rightarrow\rho) = \lambda \leqslant \log\rank\rho - \log\rank\sigma\ .
  \end{equation*}
  Recalling the definition of the R\'enyi-0 entropy $\Hzero(\sigma)=\log\rank\sigma$ yields the
  required upper bound.
\end{proof}
\begin{proof}[Proof of Prop.~\ref{prop:appx-RpurerhoAndRsigmapure}]
  Equation~\eqref{eq:appx-Rsigmapure} follows from the bounds of Proposition~\ref{prop:appx-RBoundsHminmax}, which become tight
  in this special case.
  Equality~\eqref{eq:appx-Rpurerho} is a direct consequence of Prop.~\ref{prop:appx-LambdaMajorizedDOpsByPureState}.
\end{proof}
\begin{proof}[Proof of Prop.~\ref{prop:appx-RmixedrhoAndRsigmamixed}]
  The bounds of Proposition~\ref{prop:appx-RBoundsHminmax} become tight for~\eqref{eq:appx-Rsigmamixed}.
  Equality~\eqref{eq:appx-Rmixedrho} is again a consequence of Prop.~\ref{prop:appx-LambdaMajorizedDOpsByPureState},
  recalling
  Prop.~\ref{prop:appx-LambdaMajMoveIdentitiesAround} which allows us to write $\ketbra00 \lambdamaj{\lambda+\log n} \rho$
  instead of $\Mixedin{n} \lambdamaj\lambda \rho$.
\end{proof}



%
\makeatletter
\def\pdfstartlink@attr{attr{/Border[0 0 0 [1 5] ]/H/I/C[0 1 1]}}%
\renewcommand \doibase [0]{http://dx.doi.org/}%
\newcommand \Doi [0]{\begingroup \@sanitize@url \@Doi }%
\newcommand \@Doi [1]{\endgroup\@@startlink{\doibase#1}\@@Doi}%
\def\gobbledotblock , {}
\let\@@Doi\relax
\newcommand \@@Doi [1]{\textcolor{docnotelinkcolor}{\gobbledotblock#1}\@@endlink}%
\newcommand\myhrefnoop[2]{\gobbledotblock#2}
\makeatother

\setlength\bibsep{5pt}
\bibliographystyle{naturemagdoi}
\bibliography{\jobname.bibolamazi}

\end{document}